\documentclass[12pt]{article}
\pdfoutput=1
\usepackage{amsmath,amssymb,epsfig,ulem,color}
\usepackage[bookmarks=false]{hyperref}
\allowdisplaybreaks  

\def\bm#1{\mbox{\boldmath$#1$\unboldmath}}
\def\sgn{\mbox{sgn}}
\def\Mkk{M_{\rm KK}}
\def\delslash{\rlap{\hspace{0.02cm}/}{\partial}}

\newcommand{\ord}{{\cal O}}
\newcommand{\ie}{{\it i.e.}}
\newcommand{\eg}{{\it e.g.}}
\newcommand{\etc}{{\it etc.}}
\newcommand{\quotes}[1]{``#1''}

\begin{document}
\normalem

\begin{titlepage}

\begin{flushright}
  MZ-TH/10-18\\
  May 23, 2010 \\
\end{flushright}

\vspace{15pt}
\begin{center}
  \Large\bf
  The Custodial Randall-Sundrum Model: \\
  From Precision Tests to Higgs Physics
\end{center}

\vspace{5pt}
\begin{center}
  {\sc S.~Casagrande$^{a}$, F.~Goertz$^{b}$,
    U.~Haisch$^{b}$, M.~Neubert$^{b,c}$ and T.~Pfoh$^{b}$}\\
  \vspace{10pt} {$^{a}$ \sl Excellence Cluster Universe,
    Technische Universit\"at M\"unchen \\
    D-85748 Garching, Germany \\} \vspace{5pt} 
    {$^{b}$ \sl Institut f\"ur Physik (THEP),
    Johannes Gutenberg-Universit\"at\\
    D-55099 Mainz, Germany \\} \vspace{5pt} 
    {$^{c}$ \sl Institut f\"ur Theoretische Physik, 
    Ruprecht-Karls-Universit\"at Heidelberg\\
    Philosophenweg 16, D-69120 Heidelberg, Germany}
\end{center}

\vspace{10pt}
\begin{abstract}\vspace{2pt} \noindent
  We reexamine the Randall-Sundrum (RS) model with enlarged gauge
  symmetry $SU(2)_L\times SU(2)_R\times U(1)_X\times P_{LR}$ in the
  presence of a brane-localized Higgs sector. In contrast to the
  existing literature, we perform the Kaluza-Klein (KK) decomposition
  within the mass basis, which avoids the truncation of the KK
  towers. Expanding the low-energy spectrum as well as the gauge
  couplings in powers of the Higgs vacuum expectation value, we obtain
  analytic formulas which allow for a deep understanding of the
  model-specific protection mechanisms of the $T$ parameter and the
  left-handed $Z$-boson couplings. In particular, in the latter case
  we explain which contributions escape protection and identify them
  with the irreducible sources of $P_{LR}$ symmetry breaking. We
  furthermore show explicitly that no protection mechanism is present
  in the charged-current sector confirming existing model-independent
  findings. The main focus of the phenomenological part of our work is
  a detailed discussion of Higgs-boson couplings and their impact on
  physics at the CERN Large Hadron Collider.  For the first time, a
  complete one-loop calculation of all relevant Higgs-boson production
  and decay channels is presented, incorporating the effects stemming
  from the extended electroweak gauge-boson and fermion sectors.
\end{abstract}

\vfill
\end{titlepage}

\tableofcontents
\newpage

\section{Introduction}
\label{sec:intro}

Precision experiments in the last two decades have elevated the
Standard Model (SM) of particle physics from a promising description
to a provisional law of nature, tested as a quantum field theory at
the level of one percent or better. Despite its triumphs the SM is not
an entirely satisfactory theory, however, because it has various
theoretical shortcomings. In particular, the gauge hierarchy problem,
\ie, the instability of the electroweak scale $\Lambda_W$ under
radiative corrections, has spurred the imagination of many theorists
and led to the development of a plethora of models of physics beyond
the SM that envision new phenomena at or not far above the TeV
scale. A particularly appealing proposal for stabilizing the
electroweak scale, featuring one compact extra dimension with a
non-factorizable anti-de~Sitter (AdS$_5$) metric, is the
Randall-Sundrum (RS) model \cite{Randall:1999ee}, which by virtue of
the AdS/CFT correspondence \cite{Maldacena:1997re, Gubser:1998bc,
  Witten:1998qj} can be thought of as dual to a strongly coupled
four-dimensional (4D) CFT. With two three-branes acting as the
boundaries of the warped extra dimension, the AdS$_5$ background
generates an exponential hierarchy of energy scales, so that the
natural scale at one orbifold fixed point (the ultra-violet (UV)
brane) is much larger than at the other (the infra-red (IR) brane),
$\Lambda_{\rm UV} \gg \Lambda_{\rm IR}$. In the RS framework the gauge
hierarchy problem is thus solved by gravitational red-shifting.

There are numerous possibilities for building models of electroweak
symmetry breaking in AdS$_5$. The basic building blocks for the
construction of a viable theory include, among others, the choice of
the bulk gauge group, the zero-mode fermion localization, and the
dynamical mechanism for localizing the Higgs field on (or near) the IR
brane. While in the original RS proposal all SM fields were
constrained to reside on the IR boundary and the gauge group was taken
to be $SU(2)_L \times U(1)_Y$, it was soon realized that allowing
gauge \cite{Davoudiasl:1999tf, Pomarol:1999ad, Chang:1999nh} and
matter fields \cite{Grossman:1999ra,Gherghetta:2000qt} to spread in
the AdS$_5$ bulk not only avoids dangerous higher-dimensional
operators suppressed only by powers of $\Lambda_{\rm IR}$, but also
admits a natural explanation of the flavor structure of the SM
\cite{Huber:2000ie, Huber:2003tu} via geometrical sequestering
\cite{ArkaniHamed:1999dc}. This way of generating fermion hierarchies
also implies a certain amount of suppression of dangerous
flavor-changing neutral currents (FCNCs) \cite{Gherghetta:2000qt}, a
scheme referred to as the RS-GIM mechanism \cite{Agashe:2004ay,
  Agashe:2004cp}. Harmful contributions to the $T$ parameter can be
cured in an elegant way by extending the bulk hypercharge group to
$SU(2)_R\times U(1)_X$ and breaking it to $U(1)_Y$ on the UV brane
\cite{Agashe:2003zs}. An appropriate embedding of the down-type SM
quarks into the custodial RS model further furnishes the possibility
to reduce the tree-level corrections to the $Z b_L \bar b_L$ vertex
\cite{Agashe:2006at} and its flavor-changing counterparts
\cite{Blanke:2008zb}. As a result, all existing electroweak precision
and CP-conserving FCNC constraints are typically satisfied for the
mass of the lightest Kaluza-Klein (KK) gauge boson below a few
TeV. However, in spite of the RS-GIM mechanism, CP-violating effects
in the neutral kaon system \cite{Csaki:2008zd, Gedalia:2009ws} and
corrections to the neutron electric dipole moment
\cite{Agashe:2004ay, Agashe:2004cp} tend to be too large in models
with flavor anarchy, pushing the new-physics scale to at least 10 TeV
and thus beyond the reach of the CERN Large Hadron Collider
(LHC). Relaxing the latter bounds seems to require an additional
flavor alignment in warped models and has triggered a lot of
model-building activity.\footnote{A list of relevant references can be
  found in \cite{Bauer:2009cf}.}

The purpose of this article is to perform a thorough analysis of the
structure of the RS variant with extended gauge symmetry $SU(2)_L
\times SU(2)_R\times U(1)_X \times P_{LR}$ in the bulk, where $P_{LR}$
interchanges the two $SU(2)$ groups and is responsible for the protection 
of the $Z b_L \bar b_L$ vertex. While in the existing literature
on the custodial RS model the couplings of the bulk fields to the
Higgs sector are treated as a perturbation, we instead construct the
exact solutions to the bulk equations of motion (EOMs) subject to
appropriate boundary conditions (BCs). In that way we obtain exact
results for the profiles and masses of the various SM particles and
their KK excitations. This approach is not only more elegant but also
offers several advantages over the perturbative approach. In
particular, it facilitates the analytic calculation of all terms of
order $\Lambda^2_W/\Lambda_{\rm IR}^2$, including those arising from
the breaking of the $P_{LR}$ symmetry by the BCs and possibly the bulk
masses. The physical interpretation of the obtained results in terms
of (ir)reducible sources of symmetry breaking is thus evident in our
approach, while it remains somewhat hidden if the couplings of the
bulk fields to the Higgs sector are treated as a perturbation from the
very beginning. The exact approach also permits to include the mixing
of fermions between different generations in a completely general way,
making the dependence on the exact realization of the matter sector
explicit. In turn, it is straightforward to address questions about
the model-dependence of the resulting gauge- and Higgs-boson
interactions with the SM fermions. In summary, our work puts the
theory of custodial warped extra dimensions on a more sound basis,
both at the field theoretical and phenomenological level. In a
forthcoming paper we will apply the derived results to tree-level
flavor-violating $\Delta F = 1$ and $\Delta F = 2$ processes in the
quark sector.

This article is organized as follows. After recalling important
definitions and notations, we discuss in Section~\ref{sec:gauge} the
KK decomposition of the bulk gauge fields in the presence of the
brane-localized Higgs sector, working in a covariant $R_\xi$ gauge. We
also show how to compute sums over KK towers of gauge bosons in closed
form. The analogous discussion for bulk fermions is presented in
Section~\ref{sec:fermions}. Special attention is devoted to the
correct implementation of Yukawa couplings containing $Z_2$-odd
fermion profiles. In Sections~\ref{sec:custodialprotection} and
\ref{sec:higgscouplings} we present the main results of our work. We
first analyze the structure of gauge-boson interactions with SM
fermions and then study the couplings of the Higgs boson to matter. In
the first case, we give analytic formulas that expose, on one hand,
the prerequisites for achieving a custodial protection of the
left-handed $Z$-boson couplings and, on the other, which are the terms
that necessarily escape protection. In addition, we show explicitly
that no protection mechanism is present in the charged-current
sector. In the second case, the exact dependence on the realization of
the fermion sector of the Higgs-fermion couplings is worked out.  In
our article we concentrate on the leading contributions to the
observables of interest, ignoring possible effects of brane-localized
kinetic terms \cite{Davoudiasl:2002ua, Carena:2002dz,
  Carena:2003fx}. Although the UV dynamics is not specified, it is
natural to assume that these terms are loop suppressed, so that they
can be neglected to first order.  The most important phenomenological
implications of our findings are discussed in
Section~\ref{sec:pheno}. We begin by studying the constraints imposed
by the precision measurements of the bottom-quark pseudo observables,
including all tree-level corrections that avoid protection. We further
discuss the phenomenology of rare top decays in the extended RS model
and compare it to the one of the minimal formulation. Finally, we
explore the possible changes of the Higgs production cross section and
branching fractions at the LHC, including all leading-order quantum
corrections stemming from the extended electroweak gauge-boson and
fermion sectors. In a series of appendices we collect details on the
derivation of the IR BCs and Higgs-boson FCNCs in the presence of both
$Z_2$-even and -odd Yukawa couplings, our input values for the SM
parameters, and the explicit expressions for the form factors needed
to calculate the production cross section and the branching ratios of
the Higgs boson in the RS model.

\section{Preliminaries}
\label{sec:prelim}

We work with the non-factorizable RS geometry
\begin{equation}
  ds^2 = e^{-2\sigma(\phi)}\,\eta_{\mu\nu}\,dx^\mu dx^\nu - r^2
  d\phi^2 \,, \qquad \sigma(\phi) = kr|\phi| \,,
\end{equation}
where $x^\mu$ denote the coordinates on the 4D hyper-surfaces of
constant $\phi$ with metric $\eta_{\mu\nu} = \mbox{diag}
(1,-1,-1,-1)$. The fifth dimension is an $S^1/Z_2$ orbifold of size
$r$ labeled by $\phi \in [-\pi, \pi]$. The extra dimension has
orbifold fixed points at $\phi = 0$ (the UV brane) and $\phi = \pi$
(the IR brane). Since the ratio of the warp factor and the curvature,
$e^{\sigma (\phi)}/k$, corresponds to an inverse energy scale in the
4D theory, the gauge hierarchy problem can be tamed by an appropriate
choice of the product $k r \pi$. In order to address the hierarchy
between the electroweak scale, $\Lambda_W \approx M_W$, and the
fundamental Planck scale, $\Lambda_{\rm UV} \approx M_{ \rm Pl}$,
one has to choose
\begin{equation}
  L \equiv kr\pi \, \approx \,  
  \ln \left ( \frac{ M_{\rm Pl}}{M_W} \right ) 
  \approx \ln \left (10^{16} \right ) \approx 37 \,.
\end{equation}
Below we will sometimes refer to $L$ as the \quotes{volume} of
the extra dimension. The quantity $\epsilon \equiv e^{-k r \pi}$ also
sets the mass scale for the low-lying KK excitations of the SM fields
to be of order of the \quotes{KK scale}
\begin{equation}
  \Mkk\equiv k\epsilon \approx \Lambda_{\rm IR} = \ord(\mbox{few TeV}) \,.
\end{equation}
For instance, the masses of the first KK photon and gluon are
approximately $2.5 \hspace{0.25mm} \Mkk$.

It will often be convenient to introduce a coordinate $t \equiv
\epsilon\,e^{\sigma(\phi)}$, which equals $\epsilon$ on the UV brane
and 1 on the IR brane \cite{Grossman:1999ra}. Integrals over the
orbifold are then obtained using the following replacements
\begin{equation}
  \int_{-\pi}^\pi\!d\phi \to \frac{2\pi}{L}
  \int_\epsilon^1\!\frac{dt}{t} \,, \qquad
  \int_{-\pi}^\pi\!d\phi\,e^{\sigma(\phi)} \to \frac{2\pi}{L\epsilon}
  \int_\epsilon^1\!dt \,.
\end{equation}

We now have enough definitions and notations in place to start our
discussion.

\section{Bulk Gauge Fields in the Custodial RS Model}
\label{sec:gauge}

In this section we construct the KK decomposition in the gauge sector
and derive exact solutions for the bulk fields, including the effects
of an IR brane-localized Higgs sector. In all previous works on the RS
model with custodial protection, the couplings of the Higgs sector to
bulk fields were treated as a perturbation, expanding the theory in
powers of $v^2/\Mkk^2$. This leads to the necessity of diagonalizing
infinite-dimensional matrices which have to be truncated, including
only one (or a few) KK excitations. While such an approach should in
general lead to sensible results \cite{Goertz:2008vr}, it is
worthwhile to study the set-up within the basis of mass eigenstates,
thereby obtaining exact results \cite{Gherghetta:2000qt,
Casagrande:2008hr}. Indeed, we will see that the summation over the
entire KK tower receives additional contributions, which are lost
through truncation \cite{Bauer:2008xb}.  As the sum can be evaluated
in closed form, it is convenient to work with the complete sum and
afterwards expand the obtained expressions in powers of $v^2/\Mkk^2$.
Proceeding in this way one can clearly distinguish between leading and
subleading terms. Alternatively one could use five-dimensional (5D)
propagators \cite{Randall:2001gb}, which would be equivalent to our
method. An exhaustive treatment of the perturbative approach featuring
truncation after the first mode can be found in
\cite{Albrecht:2009xr}.

\subsection{Action of the 5D Theory}

We consider the RS model with custodial protection as proposed in
\cite{Agashe:2003zs}, with the bulk gauge symmetry $SU(2)_L\times
SU(2)_R\times U(1)_X\times P_{LR}$.  On the IR brane, the
symmetry-breaking pattern $SU(2)_L\times SU(2)_R\to SU(2)_V$ provides
a custodial symmetry, which protects the $T$ parameter. The additional
$P_{LR}$ symmetry prevents the left-handed $Z b\bar b$ coupling from
receiving excessively large corrections \cite{Agashe:2006at}.  On the
UV brane, the symmetry breaking $SU(2)_R\times U(1)_X\to U(1)_Y$
generates the SM gauge group.  The symmetry breaking down to
$U(1)_{\rm EM}$ is related to the interplay of UV and IR BCs and will
become clear later on. The 5D action of the gauge sector takes the
form
\begin{equation} \label{eq:Sgauge} 
  S_{\rm gauge} = \int \!d^4x\,r\int_{-\pi}^\pi\, d\phi\, \Big( {\cal
    L}_{\rm L,R,X} + {\cal L}_{\rm Higgs} + {\cal L}_{\rm GF}\Big) \,,
\end{equation}
with the gauge-kinetic terms
\begin{equation} \label{eq:Lgauge} 
  {\cal L}_{\rm L,R,X} = \frac{\sqrt{G}}{r}\,G^{KM} G^{LN} \left( -
    \frac14\,L_{KL}^a L_{MN}^a - \frac14\,R_{KL}^a R_{MN}^a -
    \frac14\,X_{KL} X_{MN} \right) ,
\end{equation}
where $G^{MN}$ denotes the 5D metric. As it is not needed for our
analysis, we ignore the Faddeev-Popov Lagrangian. The Higgs Lagrangian
\begin{equation} \label{eq:higgslag} 
  {\cal L}_{\rm Higgs} = \frac{\delta(|\phi|-\pi)}{r} \left( \frac12 \,
    {\rm Tr} \left|(D_\mu\Phi)\right|^2 - V(\Phi) \right)
\end{equation} 
is localized on the IR brane. A simple prescription of 
    how to deal with $\delta(|\phi|-\pi)$  has already been
    presented in \cite{Casagrande:2008hr} and we postpone a refined
  treatment to Section \ref{sec:FermKK}. The gauge-fixing Lagrangian,
${\cal L}_{\rm GF}$, will be given in the next 
    section. We choose the four-vector components of the gauge
fields to be even under the $Z_2$ parity, while the scalar fifth
components are odd, in order to arrive at a low-energy spectrum that
is consistent with observation. The Higgs bi-doublet, responsible for
breaking $SU(2)_L \times SU(2)_R$ to the diagonal subgroup $SU(2)_V$
on the IR brane, transforms as $\left(\bm{2},\bm{2}\right)_0$ and
explicitly reads\footnote{Notice that compared to
  \cite{Burdman:2008gm} the sign of the would-be Goldstone boson
  fields is switched.}
\begin{equation} \label{eq:Higgsbi}
  \Phi(x) = \frac{1}{\sqrt2} \left( \begin{array}{cc} v + h(x) -
      i\varphi^3(x) & -i\sqrt2\,\varphi^+(x) \\
      -i\sqrt2\,\varphi^-(x) & v + h(x) + i\varphi^3(x)
    \end{array} \right) ,
\end{equation}
where $\varphi^i$ are real scalar fields, $\varphi^\pm = (\varphi^1\mp
i\varphi^2)/\sqrt2$, and $v \approx 246 \, {\rm GeV}$ is the vacuum
expectation value (VEV) of the Higgs field. $SU(2)_L$
transformations act from the left on the bi-doublet, while the
$SU(2)_R$ transformations act from the right. The covariant derivative
in the Higgs sector reads
\begin{equation} \label{eq:D}
  D_\mu\Phi = \partial_\mu\Phi - i {g_L}_5\, L_\mu^a \hspace{0.25mm} 
  T_L^a\; \Phi + i {g_R}_5\, \Phi\, R_\mu^a \hspace{0.25mm} T_R^a \,,
\end{equation}
with $T^a_{L,R}=\sigma^a/2$. An explicit calculation leads to
\begin{align} \label{eq:cohigg}
    D_\mu\Phi & = \frac{1}{\sqrt2} \left( \begin{array}{c c}
        \partial_\mu \left(h - i\varphi^3 \right) -\displaystyle  i\, \frac v2
        \left({g_L}_5\, L_\mu^3 - {g_R}_5\, R_\mu^3 \right) & \phantom{i}
        -\partial_\mu i \sqrt 2 \varphi^+ - \displaystyle  i\,  \frac v2 \left({g_L}_5\,
          L_\mu^+ - {g_R}_5\, R_\mu^+ \right) \vspace{2mm} \\ 
        -\partial_\mu i \sqrt 2 \varphi^- - \displaystyle  i\,  
        \frac v2 \left({g_L}_5\,
          L_\mu^- - {g_R}_5\, R_\mu^- \right) & \phantom{i} 
        \partial_\mu \left(h +
          i\varphi^3 \right) + \displaystyle  i\, 
        \frac v2 \left({g_L}_5\, L_\mu^3 -
          {g_R}_5\, R_\mu^3 \right)
      \end{array} \right)  \hspace{4mm} \nonumber \\[-3mm] \\[-3mm] 
    & \phantom{xx} + \text{terms bi-linear in fields} \,, \nonumber 
\end{align}
where we have introduced
\begin{equation} \label{eq:LRmu}
  L_\mu^\pm = \frac{1}{\sqrt2} \left( L_\mu^1\mp i L_\mu^2 \right)
  ,\qquad R_\mu^\pm = \frac{1}{\sqrt2} \left( R_\mu^1\mp i R_\mu^2
  \right) .
\end{equation}
The structure of (\ref{eq:cohigg}) motivates us to define the new
fields \cite{Burdman:2008gm}
\begin{equation}
  \left( \begin{array}{c}
      \tilde A_M\\
      V_M
    \end{array} \right)= 
  \frac{1}{\sqrt{g_L^2 + g_R^2}} \,
  \left( \begin{array}{cr}
      g_L & - g_R \\
      g_R & g_L
    \end{array} \right)
  \left( \begin{array}{c}
      L_M\\
      R_M 
    \end{array} \right) ,
\end{equation}
which lead to a diagonal mass matrix. We have introduced the 4D gauge
couplings $g_a={g_a}_5/\sqrt{2 \pi r}$. The rotations are analogous to
the usual definitions of the $Z$ boson and photon fields in the SM,
which are themselves postponed to (\ref{eq:ZA}). Finally, the mass 
term adopts the form

\begin{equation} \label{eq:Lm} 
  {\cal L}_{\rm mass} = \frac{\delta(|\phi|-\pi)}{r} \,
  \frac{\left({g_L^2}_5+{g_R^2}_5\right) v^2}{8} \, \tilde A_\mu^a
  \tilde A^{\mu\, a} \equiv \frac{\delta(|\phi|-\pi)}{r} \, \frac 1 2 \,
  M_{\tilde A}^2 \, \tilde A_\mu^a \tilde A^{\mu\, a} \,,
\end{equation}
and reveals the breaking pattern
\begin{equation} \label{eq:IRbreaking}
  SU(2)_L \times SU(2)_R \xrightarrow{\rm IR} SU(2)_V \,,
\end{equation}
induced by the Higgs VEV $\left \langle \Phi\right \rangle= v/\sqrt{2}
\; \bm{1}$. Appropriate BCs break the extended electroweak gauge group
down to the SM gauge group on the UV boundary
\begin{equation} \label{eq:UVbreaking}
  SU(2)_R \times U(1)_X \xrightarrow{\rm UV} U(1)_Y \,.
\end{equation}
Explicitly, this is done by introducing the new fields
\begin{equation} \label{eq:Zp}
  \left( \begin{array}{c}
      Z_M^\prime\\
      B_M^Y 
    \end{array} \right)= 
  \frac{1}{\sqrt{g_R^2 + g_X^2}} \, 
  \left( \begin{array}{cr}
      g_R & - g_X \\
      g_X & g_R
    \end{array} \right)
  \left( \begin{array}{c}
      R_M^3\\
      X_M
    \end{array} \right) ,
\end{equation}
and giving Dirichlet BCs to $Z_\mu^\prime$ and $R_\mu^{1,2}$ on the UV
brane. The $U(1)_Y$ hypercharge coupling is related to the $SU(2)_R
\times U(1)_X$ couplings by
\begin{equation} \label{eq:gY}
  g_Y= \frac{g_R\,g_X}{\sqrt{g_R^2+g_X^2}} \,.
\end{equation}
The SM-like neutral electroweak gauge bosons are defined in the
standard way through
\begin{equation} \label{eq:ZA}
  \left( \begin{array}{c}
      Z_M\\
      A_M
    \end{array} \right)= 
  \frac{1}{\sqrt{g_L^2 + g_Y^2}} \,
  \left( \begin{array}{cr}
      g_L & - g_Y \\
      g_Y & g_L
    \end{array} \right)
  \left( \begin{array}{c}
      L_M^3\\
      B_M^Y
    \end{array} \right) .
\end{equation}
It follows that the definitions of the sine and cosine of the
weak-mixing angle,
\begin{equation} \label{eq:weakmixing}
  \sin\theta_w=\frac{g_Y}{\sqrt{g_L^2+g_Y^2}} \,, \qquad
  \cos\theta_w=\frac{g_L}{\sqrt{g_L^2+g_Y^2}} \,,
\end{equation}
agree with the ones in the SM. Furthermore, the fields $V_M^3$ and
$X_M$ can be rotated to the photon field $A_M$ and a state $Z_M^H$ via
\begin{equation}
  \left( \begin{array}{c}
      Z_M^H\\
      A_M
    \end{array} \right)= 
  \frac{1}{g_{LRX}^2} \,
  \left( \begin{array}{cc}
      g_L\,g_R & -g_X \sqrt{g_L^2+g_R^2} \\
      g_X \sqrt{g_L^2+g_R^2}& g_L\,g_R
    \end{array} \right)
  \left( \begin{array}{c}
      V_M^3\\
      X_M
    \end{array} \right) ,
\end{equation}
where 
\begin{equation} \label{eq:gLRX2}
  g_{LRX}^2= \sqrt{g_L^2\,g_R^2+g_L^2\,g_X^2+g_R^2\,g_X^2} \,,
\end{equation}
and we write $\tilde Z_M \equiv \tilde A_M^3$, as it is a linear
combination of $Z_M$ and $Z_M^\prime$, which is orthogonal to $Z_M^H$
as we will see below.

\begin{table}
  \begin{equation}
    \begin{tabular}{|c|c|}
      \hline
      $\partial_\phi L_\mu^\pm(x,0)=0$ & $L_5^\pm(x,0)=0$\\
      $R_\mu^\pm(x,0)=0$ & $R_5^\pm(x,0)=0$\\
      $\partial_\phi Z_\mu(x,0)=0$ & $Z_5(x,0)=0$\\
      $Z_\mu^\prime(x,0)=0$ & $Z_5^\prime(x,0)=0$\\
      $\partial_\phi A_\mu(x,0)=0$ & $A_5(x,0)=0$\\
      \hline
    \end{tabular}
    \quad
    \begin{tabular}{|c|c|}
      \hline
      $\partial_\phi \tilde A_\mu^\pm(x,\pi^-)=-\frac{r}{2 \epsilon ^2} 
      M_{\tilde A}^2 \tilde A_\mu^\pm(x,\pi)$
      & $\tilde A_5^\pm(x,\pi)=0$\\
      $\partial_\phi V_\mu^\pm(x,\pi)= 0$ & $V_5^\pm(x,\pi)=0$\\
      $\partial_\phi \tilde Z_\mu(x,\pi^-)=-\frac{r}{2 \epsilon ^2} 
      M_{\tilde A}^2 \tilde Z_\mu(x,\pi)$
      & $\tilde Z_5(x,\pi)=0$\\
      $\partial_\phi Z_\mu^H(x,\pi)=0$ & $Z_5^H(x,\pi)=0$\\
      $\partial_\phi A_\mu(x,\pi)=0$ & $A_5(x,\pi)=0$\\
      \hline
    \end{tabular}\nonumber
  \end{equation}
  \parbox{15.5cm}{\caption{\label{tab:BCs} UV (left) and IR (right) BCs.}}
\end{table}

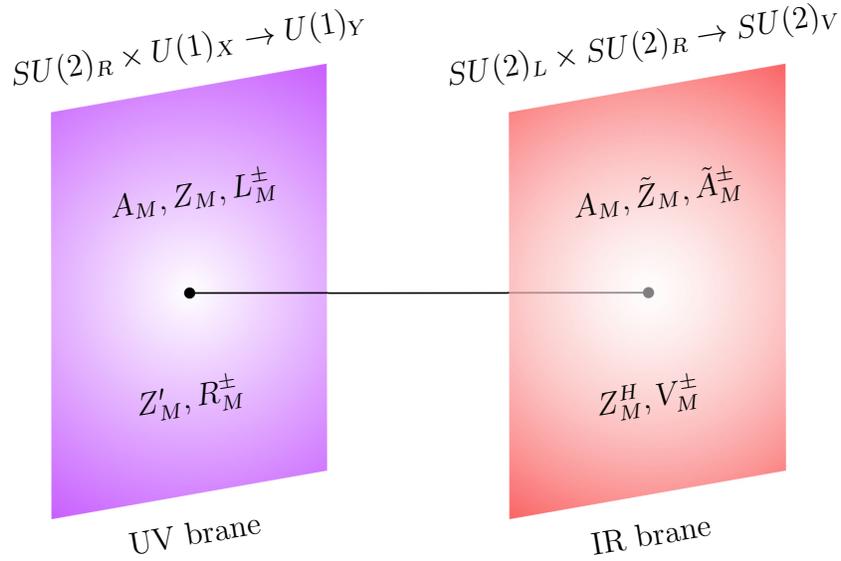
\begin{figure}[!t]
\begin{center}
\mbox{\includegraphics[width=12cm]{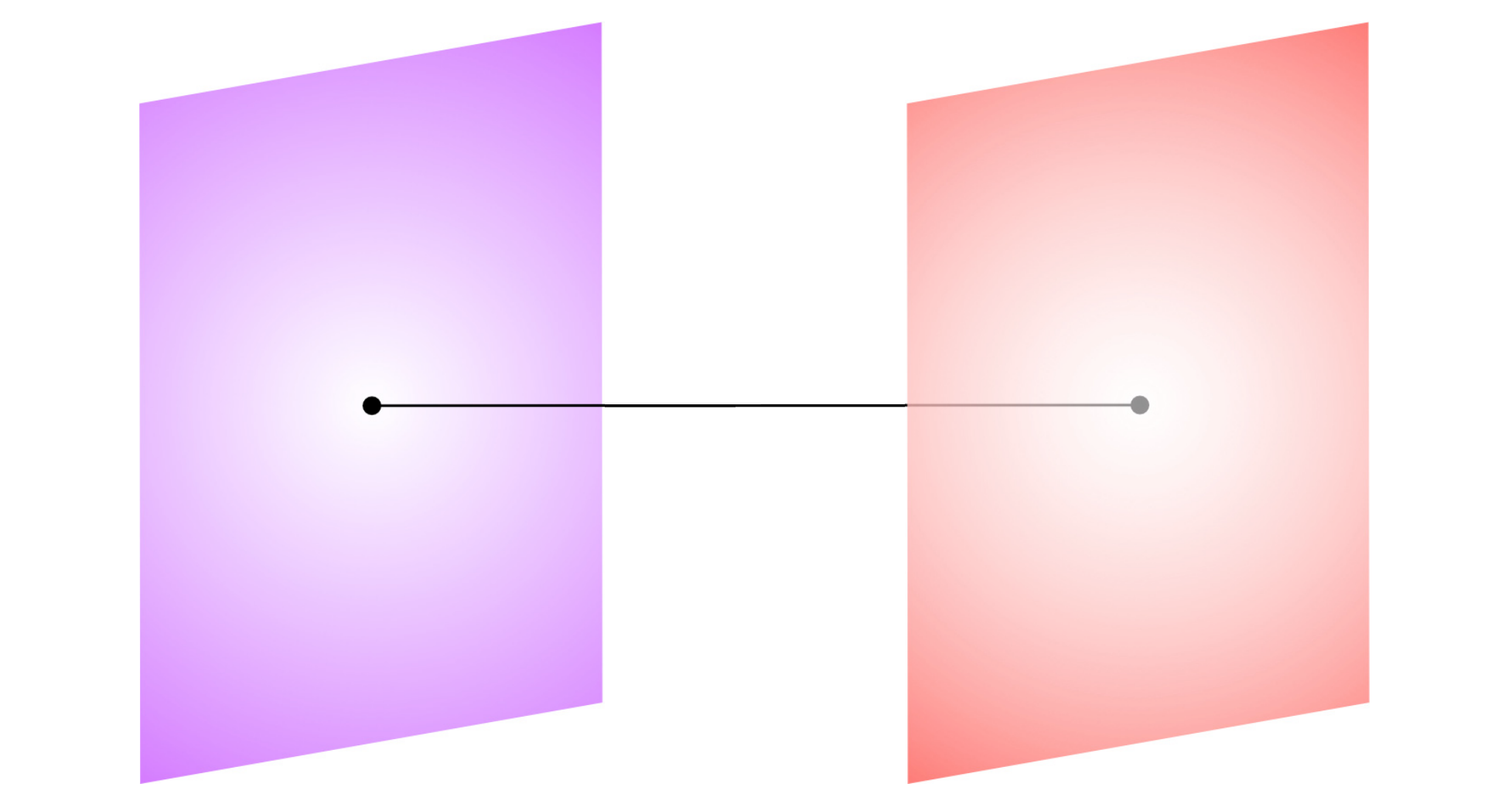}}
\begin{picture}(0,0)(0,0) 
\put(-285,-7.5){\rotatebox{9.25}{UV brane}}
\put(-330,170){\rotatebox{9.25}{$SU(2)_R \times U(1)_X \to U(1)_Y$}}
\put(-282.5,45){\rotatebox{9.25}{$Z^\prime_M, R_M^\pm$}}
\put(-292.5,120){\rotatebox{9.25}{$A_M, Z_M, L_M^\pm$}}
\put(-110,-7.5){\rotatebox{9.25}{IR brane}}
\put(-165,170){\rotatebox{9.25}{$SU(2)_L \times SU(2)_R \to SU(2)_V$}}
\put(-108.5,45){\rotatebox{9.25}{$Z^H_M, V_M^\pm$}}
\put(-117.5,120){\rotatebox{9.25}{$A_M, \tilde{Z}_M, \tilde{A}_M^\pm$}}
\end{picture}
\vspace{6mm}

\parbox{15.5cm}{\caption{ UV and IR basis, \ie, gauge fields
  with individual BCs on the corresponding branes. The fields in the
  first (second) row on the UV brane do (do not) possess a zero
  mode. The symmetry-breaking pattern on the UV and IR brane is also
  indicated. See text for details.}}
\end{center}
\label{fig:bases}
\end{figure}

In Table~\ref{tab:BCs} we summarize the BCs that we choose for the
fields in order to obtain the correct mass spectrum for the SM gauge
bosons. They are given in terms of fields with individual BCs at the
two different branes. In the following we will refer to these sets of
fields as the UV and IR basis, respectively. The situation is
summarized in Figure~\ref{fig:bases}, where we also recall the
symmetry-breaking patterns on the different branes. The BCs can easily
be transformed to another basis at the expense of obtaining
expressions that mix different fields. The photon $A_{\mu}$ has
individual and source-free Neumann BCs at both branes, and therefore
its zero mode remains massless. Note that there is just one mass
parameter $M_{\tilde A}$ entering the IR BCs, in contrast to the two
parameters $M_Z$ and $M_W$ appearing in the minimal model. In the
custodial model, the different masses for the lightest electroweak
gauge bosons are accomplished through the mixed UV BCs of the gauge
fields in the IR basis (see (\ref{eq:rots}) below).  The fact that
there is just one fundamental mass parameter is crucial for the
custodial protection of the $T$ parameter. We will elaborate on this
in Section \ref{sec:MWMZSTU}.

The action of the theory still contains mixing terms between gauge
fields and scalars, which can be removed by an appropriate
gauge-fixing Lagrangian. As the Higgs sector is localized on the IR
brane, it is natural to work in the IR basis for that purpose. For
this reason, we define the 5D theory in the IR basis. The concrete
form of the gauge fixing will be given below in (\ref{eq:Sgf}).

Before discussing the KK decomposition, we summarize the relations
between the UV (right) and IR basis (left). They read
\begin{equation} \label{eq:rots}
  \begin{split}
    \left( \begin{array}{c}
        \tilde Z_M\\
        Z_M^H
      \end{array} \right) & = 
    \left( \begin{array}{cr}
        \cos\theta_Z & -\sin\theta_Z \\
        \sin\theta_Z & \cos\theta_Z
      \end{array} \right)
    \left( \begin{array}{c}
        Z_M\\
        Z_M^\prime 
      \end{array} \right)
    \equiv {\bm R}_Z \left( \begin{array}{c}
        Z_M\\
        Z_M^\prime 
      \end{array} \right) , \\[1mm]
    \left( \begin{array}{c}
        \tilde A_M^\pm\\
        V_M^\pm 
      \end{array} \right) & = 
    \left( \begin{array}{cr}
        \cos\theta_W & -\sin\theta_W \\
        \sin\theta_W & \cos\theta_W
      \end{array} \right)
    \left( \begin{array}{c}
        L_M^\pm\\
        R_M^\pm 
      \end{array} \right)
    \equiv {\bm R}_W \left( \begin{array}{c}
        L_M^\pm\\
        R_M^\pm 
      \end{array} \right) ,
  \end{split}
\end{equation}
where 
\begin{align} \label{eq:thetas}
    \sin\theta_Z & =\frac{g_R^2}{\sqrt{(g_L^2+g_R^2)(g_R^2+g_X^2)}}\,,
    & \qquad \cos\theta_Z &
    =\frac{g_{LRX}^2}{\sqrt{(g_L^2+g_R^2)(g_R^2+g_X^2)}} \, , \nonumber \\[1mm]
    \sin\theta_W & =\frac{g_R}{\sqrt{g_L^2+g_R^2}} \,, & \qquad
    \cos\theta_W & =\frac{g_L}{\sqrt{g_L^2+g_R^2}} \,,
\end{align}
and $g_{LRX}^2$ has been defined in (\ref{eq:gLRX2}). In order to
shorten the notation we will hereafter employ the abbreviations $s_a
\equiv \sin \theta_a$ and $c_a \equiv \cos \theta_a$ for $a = w, Z,
W$.

\subsection{KK Decomposition}
\label{sec:KKdec}

We now perform the KK decomposition of the 5D fields. It is convenient
to work with profiles that obey definite Neumann ($+$) or Dirichlet
($-$) BCs at the UV brane. Therefore we include a rotation to the UV
basis, \ie, the basis in which the UV BCs decouple, in our
decomposition. Furthermore, as different UV fields get mixed by the IR
BCs, these fields should be expressed through the same 4D basis. We
consequently introduce the vectors $\vec Z_M=(\tilde Z_M, Z_M^H)^T$
and $\vec W_M^\pm=(\tilde A_M^\pm, V_M^\pm)^T$ and write
\begin{eqnarray} \label{eq:KKdec}
  \begin{aligned}
    A_\mu(x,\phi) &= \frac{1}{\sqrt r} \sum_n \chi_n^{(+)}(\phi)\,
    A_\mu^{(n)}(x) \,, &\; A_\phi(x,\phi) &= \frac{1}{\sqrt r} \sum_n
    \partial_\phi\chi_n^{(+)}(\phi)\,a_n^A\, \varphi_A^{(n)}(x)\,, \\
    \vec Z_\mu(x,\phi) &= \frac{{\bm R}_Z}{\sqrt r} \sum_n {\bm
      \chi}_n^+(\phi)\,\vec A_n^{\, Z}\, Z_\mu^{(n)}(x)\,, &\; \vec
    Z_\phi(x,\phi) &= \frac{{\bm R}_Z}{\sqrt r} \sum_n \partial_\phi
    {\bm \chi}_n^+(\phi)\,\vec A_n^{\, Z} a_n^Z\,\varphi_Z^{(n)}(x)\,,\\
    \vec W_\mu^\pm(x,\phi) &= \frac{{\bm R}_W}{\sqrt r} \sum_n {\bm
      \chi}_n^+(\phi)\,\vec A_n^{\, W}\, W_\mu^{\pm(n)}(x)\,, &\; \vec
    W_\phi^\pm(x,\phi) &= \frac{{\bm R}_W}{\sqrt r} \sum_n
    \partial_\phi {\bm \chi}_n^+(\phi)\,\vec A_n^{\, W}
    a_n^W\,\varphi_W^{\pm(n)}(x)\, , \hspace{1mm}
  \end{aligned}
\end{eqnarray}
where the sums run over $n = 0, \ldots, \infty$. Note that
$A_\mu^{(n)}(x)$ \etc \ are 4D mass eigenstates and the lightest modes
are identified with the SM gauge bosons. The matrices ${\bm R}_{Z,W}$
are defined in (\ref{eq:rots}) and we have introduced the diagonal
matrix

\begin{equation} \label{eq:chimatrix} 
  {\bm \chi}_n^+(\phi)=
  \left( \begin{array}{cc}
      \chi_n^{(+)}(\phi) & 0 \\
      0 & \chi_n^{(-)}(\phi)
    \end{array} \right) ,
\end{equation}
as well as two-component vectors $\vec A_n^a$, with $a=Z,W$,
representing the mixings between the different gauge fields and their
KK excitations.  These vectors are normalized according to
\begin{equation}
 {(\vec A_n^a)}^T \vec A_n^a =1 \,.
\end{equation}

Notice that the matrix ${\bm \chi}_n^+ (\phi)$ should in principle
also carry a superscript $a$, indicating the field to which it
belongs, but we will not show it, as the correct index should be
always clear from the context. The superscripts $(+)$ and $(-)$ label
the type of BC we impose on the profiles at the UV brane, \ie, they
indicate untwisted and twisted even functions\footnote{We use the term
twisted even functions for profiles with even $Z_2$-parity, which obey
Dirichlet BC on the UV brane and are thus not smooth at this orbifold
fix point. These fields are sometimes called odd, as they look like an
odd function if one just considers half of the orbifold.  Untwisted
even functions correspond to ordinary profiles with Neumann UV BCs.}
on the orbifold. Remember from Table \ref{tab:BCs} that both profiles
satisfy a Neumann BC at the IR boundary, which we do not indicate
explicitly by a superscript $(+)$ to avoid unnecessary clutter of
notation. Let us also introduce the shorthand notations
\begin{equation} \label{eq:ZWvecs}
  \vec{\chi}_n^{\, Z} (\phi) =\left( \begin{array}{c} \chi_n^Z(\phi)\\
      \chi_n^{Z^\prime}(\phi)
    \end{array} \right)={\bm \chi}_n^+(\phi)\,\vec A_n^{\, Z} \,,\qquad
  \vec{\chi}_n^{\, W} (\phi) =\left( \begin{array}{c} \chi_n^L(\phi)\\
      \chi_n^R(\phi)
    \end{array} \right)={\bm \chi}_n^+(\phi)\,\vec A_n^{\, W} \,,
\end{equation}
for the profiles of the UV fields. In analogy to the fermion
decomposition in \cite{Casagrande:2008hr}, the profiles ${\bm
\chi}_n^+(\phi)$ do not obey exact orthonormality conditions. This
fact is related to the decomposition of fields with Neumann and
Dirichlet BCs into the same 4D gauge-boson basis. The complete vectors
$\vec{\chi}_n^{\hspace{0.5mm} a} (\phi)$ with $a = Z, W$ are however
orthonormal on each other,
\begin{equation} \label{eq:ortho}
  \int_{-\pi}^\pi\!d\phi\,{\vec \chi}_m^{\, a \, T}(\phi)\,{\vec
    \chi}_n^{\, a} (\phi) = \delta_{mn} \,.
\end{equation}
Note also that the photon obeys a standard orthonormality
condition. We also expand the 4D Goldstone bosons in the basis of mass
eigenstates $\varphi_Z^{(n)}(x)$ and $\varphi_W^{ \pm(n)}(x)$ by
writing \cite{Casagrande:2008hr}
\begin{equation} \label{eq:phivecs}
  \vec \varphi^{\, 3}(x) = \sum_n \vec b_n^{\, Z}\,\varphi_Z^{(n)}(x)
  \,, \qquad \vec \varphi^{\, \pm}(x) = \sum_n \vec b_n^{\,
    W}\,\varphi_W^{\pm(n)}(x) \,.
\end{equation}
Employing the notation introduced in this section, the gauge-fixing
Lagrangian takes the form 
\begin{align} \label{eq:Sgf}
  \begin{split} {\cal L}_{\rm GF} &= - \frac{1}{2\xi}
    \left( \partial^\mu A_\mu - \xi \left[
        \frac{\partial_\phi\,e^{-2\sigma(\phi)}}{r^2} \, A_\phi
      \right] \right)^2\\
    &\quad\mbox{}- \frac{1}{2\xi} \left( \partial^\mu \vec Z_\mu - \xi
      \left[ \frac{\delta(|\phi|-\pi)}{r}\,M_{\tilde
          A}\,\vec\varphi^{\, 3} +
        \frac{\partial_\phi\,e^{-2\sigma(\phi)}}{r^2}\, \vec
        Z_\phi \right] \right)^2 \\
    &\quad\mbox{}- \frac{1}{\xi} \left( \partial^\mu \vec W_\mu^+ -
      \xi \left[ \frac{\delta(|\phi|-\pi)}{r}\,M_{\tilde
          A}\,\vec\varphi^{\, +} +
        \frac{\partial_\phi\,e^{-2\sigma(\phi)}}{r^2} \, \vec
        W_\phi^+ \right] \right)^T \\
    &\qquad\times \left( \partial^\mu \vec W_\mu^- - \xi \left[
        \frac{\delta(|\phi|-\pi)}{r}\,M_{\tilde A}\,\vec\varphi^{\, -}
        + \frac{\partial_\phi\,e^{-2\sigma(\phi)}}{r^2} \, \vec
        W_\phi^- \right] \right).
  \end{split}
\end{align}

Inserting the decomposition (\ref{eq:KKdec}) into the action and
defining the projectors ${\bm P}_{(+)}={\rm diag}(1,0)$ and ${\bm
P}_{(-)}={\rm diag}(0,1)$, we derive the EOMs
\cite{Casagrande:2008hr, Davoudiasl:1999tf, Pomarol:1999ad}
\begin{equation} \label{eq:gaugeeom} 
  -\frac{1}{r^2}\,\partial_\phi\,e^{-2\sigma(\phi)}\,
  \partial_\phi {\bm R}_a\,{\bm \chi}_n^+(\phi) \vec A_n^{\, a} =
  (m_n^a)^2\, {\bm R}_a\,{\bm \chi}_n^+(\phi) \vec A_n^{\, a} -
  \frac{\delta(|\phi|-\pi)}{r}\,M_a^2\, {\bm P}_{(+)} {\bm R}_a\,{\bm
    \chi}_n^+(\phi) \vec A_n^{\, a} \,,
\end{equation} 
where $a = Z, W, A$ with $M_Z=M_W=M_{\tilde A}$ and $M_A=0$, as well
as ${\bm R}_A={\bm 1}$ and $\vec A_n^{\, A}=(1,0)^T$. In order to
avoid boundary terms due to integration by parts, we move the
$\delta$-distribution by an infinitesimal amount into the bulk
\cite{Casagrande:2008hr}. We will indicate this limiting procedure by
a superscript in the argument of the profiles, \eg, by writing ${\bm
\chi}_n^+(\pi^-)$. The appropriate IR BCs for the profiles can be
obtained by integrating the EOMs (\ref{eq:gaugeeom}) over an
infinitesimal interval around $|\phi|=\pi$. At the 5D level they have
already been given in Table \ref{tab:BCs}. However, since the profiles
of the scalar components are taken to be proportional to the
$\phi$-derivative of the vector profiles, they develop discontinuities
at the IR brane \cite{Casagrande:2008hr}. We arrive at
\begin{equation} \label{eq:IRBC2} 
  \frac{m_n^a}{\Mkk} \, {\bm R}_a\, {\bm \chi}_n^-(\pi^-) \vec A_n^{\,
    a} =- X^2\hspace{0.25mm} L\hspace{0.25mm} {\bm P}_{(+)} {\bm R}_a\,
  {\bm \chi}_n^+(\pi) \vec A_n^{\, a} \,,
\end{equation} 
where
\begin{equation} \label{eq:chiminus} 
  {\bm \chi}_n^- (\phi) \equiv \frac{1}{m_n^a r }\,
  e^{-\sigma(\phi)}\partial_\phi \, {\bm \chi}_n^+ (\phi) \,, \qquad X^2
  \equiv \frac{ (g_L^2+g_R^2) \, v^2}{4 \Mkk^2} \,.
\end{equation} 
Notice that for the photon the right-hand side in (\ref{eq:IRBC2}) is
equal to zero.

After applying the EOMs and the orthonormality condition
(\ref{eq:ortho}), we observe that the 4D action takes the desired
canonical form, if
\begin{equation} \label{eq:coeffs} 
  a_n^a = - \frac{1}{m_n^a} \,, \qquad \vec b_n^{\, a} =
  \frac{M_a}{\sqrt r\, m_n^a} {\bm P}_{(+)} {\bm R}_a\,{\bm
    \chi}_n^+(\pi^-)\vec A_n^{\, a} \,.
\end{equation} 
The spectrum of the theory is determined by the IR BCs
(\ref{eq:IRBC2}). The eigenvalues $x_n^a \equiv m_n^a/M_{\rm KK}$ are
thus solutions of
\begin{equation} \label{eq:IRBC3}
  {\rm det}\left[x_n^a \, {\bm \chi}_n^-(\pi^-) + L \hspace{0.25mm} X^2
    \hspace{0.25mm} {\bm D}_a {\bm \chi}_n^+(\pi)\right]=0\,,
\end{equation} 
with 
\begin{equation} \label{eq:Dmatrix} 
  {\bm D}_a = {\bm R}_a^{-1} {\bm P}_{(+)} {\bm R}_a =
  \left( \begin{array}{cc}
      c_a^2 & -s_a c_a \\
      -s_a c_a & s_a^2
    \end{array} \right) .
\end{equation}
Once the eigenvalues are known, the eigenvectors $\vec A_n^{\, a}$ are
determined by (\ref{eq:IRBC2}).

\subsection{Bulk Profiles}

We now derive expressions for the profiles $\chi_n^{(\pm)} (\phi)$.
In order to get the EOMs for the UV basis we multiply
(\ref{eq:gaugeeom}) by ${\bm R}_a^T$ from the left. Introducing the
coordinate $t=\epsilon \, e^{\sigma(\phi)}$, we then write the
solutions as \cite{Casagrande:2008hr}
\begin{equation} \label{eq:sol} 
  \chi_n^{(+)}(t) = N_n^{(+)}\sqrt{\frac{L}{\pi}}\,t\,c_n^{(+)+}(t) \,,
  \qquad \chi_n^{(-)}(t) = N_n^{(-)}\sqrt{\frac{L}{\pi}}\,
  t\,c_n^{(-)+}(t)\,,
\end{equation}
with 
\begin{equation}
  \begin{split}
    c_n^{(+)+}(t) &= Y_0(x_n \epsilon) J_1(x_n t)
    -J_0(x_n \epsilon) \,Y_1(x_n t) \,, \\[2mm]
    c_n^{(-)+}(t) &= Y_1(x_n \epsilon) J_1(x_n t)
    -J_1(x_n \epsilon) \,Y_1(x_n t) \,, \\[1mm]
    c_n^{(+)-}(t) &= \frac{1}{x_n t}\, \frac{d}{dt} \Big(
    t\,c_n^{(+)+}(t) \Big) =Y_0(x_n \epsilon) \,J_0(x_n t)
    -J_0(x_n \epsilon) \,Y_0(x_n t) \,, \\
    c_n^{(-)-}(t) &= \frac{1}{x_n t}\, \frac{d}{dt} \Big(
    t\,c_n^{(-)+}(t) \Big) =Y_1(x_n \epsilon) \,J_0(x_n t) -J_1(x_n
    \epsilon) \,Y_0(x_n t) \,.
  \end{split}
\end{equation}
The masses of the KK states normalized to the KK scale, $x_n$, are
determined by the IR BCs as explained above. From the latter
expressions, it is obvious that the profiles fulfill the UV BCs, since
$c_n^{(+)-}(\epsilon) = c_n^{(-)+}(\epsilon) =0$. The normalization
constants $N_n^{(\pm)}$ are determined from the orthonormality
condition (\ref{eq:ortho}). With respect to the formula given in
\cite{Casagrande:2008hr}, they contain additional terms
due to the different UV BCs. We obtain
\begin{equation} \label{eq:Nngauge}
  \begin{split}
    \big ( N_n^{(\pm)} \big )^{-2} = &\left[ c_n^{(\pm)+}(1) \right]^2
    + \left[ c_n^{(\pm)-}(1^-) \right]^2 - \frac{2}{x_n} \Big
    (\,c_n^{(\pm)+}(1)\,c_n^{(\pm)-}(1^-)-
    \epsilon \,c_n^{(\pm)+}(\epsilon)\,c_n^{(\pm)-}(\epsilon^+) \Big)\\
    &- \epsilon^2 \left( \left[ c_n^{(\pm)+}(\epsilon) \right]^2 +
      \left[ c_n^{(\pm)-}(\epsilon^+) \right]^2\right) .
  \end{split}
\end{equation} 
Note that, depending on the type of the UV BCs, some of the terms
in (\ref{eq:Nngauge}) vanish identically.

\subsection{Zero-Mode Masses and Oblique Corrections}
\label{sec:MWMZSTU}

It will turn out to be useful to have simple analytical expressions
for the masses and profiles of the lightest modes. Expanding
(\ref{eq:IRBC3}) in powers of $v^2/\Mkk^2$ and inserting the
definitions of the mixing angles (\ref{eq:thetas}), which connect the
UV and IR bases, we arrive at analytic expressions for the masses of
the $W$ and $Z$ bosons. They read
\begin{eqnarray} \label{eq:mwmz}
  \begin{split}
    m_W^2 & = \frac{g_L^2 v^2}{4} \left[ 1 - \frac{g_L^2 v^2}{8\Mkk^2}
      \left( L - 1 + \frac{1}{2L}\right) - \frac{g_R^2 v^2}{8\Mkk^2}
      \,L
      + \ord\left( \frac{v^4}{\Mkk^4} \right) \right] , \\
    m_Z^2 & = \frac{(g_L^2+g_Y^2) \, v^2}{4} \left[ 1 -
      \frac{(g_L^2+g_Y^2) \, v^2}{8\Mkk^2} \left( L - 1 +
        \frac{1}{2L}\right) - \frac{(g_R^2-g_Y^2) \, v^2}{8\Mkk^2} \,
      L + \ord\left( \frac{v^4}{\Mkk^4} \right) \right] , \quad 
  \end{split}
\end{eqnarray}
where the last terms inside the square brackets are new compared to
the minimal model studied in \cite{Casagrande:2008hr}.

Interestingly, the latter terms are responsible for the custodial
protection of the Peskin-Takeuchi \cite{Peskin:1990zt, Peskin:1991sw}
parameter $T$, which is sensitive to the difference between the
corrections to the $W$- and $Z$-boson vacuum-polarization functions
and thus measures isospin violation. The set of oblique corrections,
which are defined as
\begin{equation} \label{eq:STUdef}
  \begin{split}
    S & = \frac{16 \hspace{0.5mm} \pi s^2_w c^2_w}{e^2} \left[ \,
      \Pi_{ZZ}^{\hspace{0.25mm} \prime}(0)+ \frac{s^2_w-c^2_w}{s_w
        c_w} \, \Pi_{ZA}^{\hspace{0.25mm}
        \prime}(0)-\Pi_{AA}^{\hspace{0.25mm}
        \prime}(0) \, \right] ,\\
    T & = \frac{4 \hspace{0.5mm} \pi}{e^2 c^2_w m_Z^2} \, \Big
    [\Pi_{WW}(0)-c^2_w \, \Pi_{ZZ}(0) -2 \, s_w c_w \, \Pi_{ZA}(0)-
    s^2_w \, \Pi_{AA}(0) \Big]\,,\\
    U & = \frac{16 \hspace{0.5mm} \pi s^2_w}{e^2} \, \Big
    [\Pi_{WW}^{\hspace{0.25mm} \prime}(0)-c^2_w \,
    \Pi_{ZZ}^{\hspace{0.25mm} \prime}(0) -2 \, s_w c_w \,
    \Pi_{ZA}^{\hspace{0.25mm} \prime}(0) -s^2_w \,
    \Pi_{AA}^{\hspace{0.25mm} \prime}(0) \Big ] \,,
  \end{split}
\end{equation} 
can be computed in an effective Lagrangian approach
\cite{Csaki:2002gy}. Gauge invariance guarantees that $\Pi_{AA}(0)=0$
to all orders in perturbation theory, and one further has
$\Pi_{ZA}(0)=\Pi_{ZA}^{\hspace{0.25mm} \prime}(0)=0$ as long as one
works at tree level.

The non-zero tree-level correlators $\Pi_{aa} (0)$ with $a = W, Z$ are
calculated from the corrections to the zero-mode masses
(\ref{eq:mwmz}) and profiles (\ref{eq:expprof}), where the latter also
give rise to non-zero derivatives $\Pi_{aa}^{\hspace{0.25mm} \prime}
(0)$ of the correlators at zero momentum. We find
\begin{equation} \label{eq:cor}
  \begin{split}
    \Pi_{WW}(0)&=-\frac{g_L^2 v^4}{32 \Mkk^2}
    \left[g_L^2\left(L-\frac 1{2L}\right)+g_R^2 L\right] ,\\
    \Pi^{\hspace{0.25mm} \prime}_{WW}(0)&=\frac{g_L^2v^2}{8 \Mkk^2}
    \left(1-\frac 1 L\right) ,\\
    \Pi_{ZZ}(0)&=-\frac{(g_L^2+g_Y^2) \, v^4}{32
      \Mkk^2}\left[\left(g_L^2+g_Y^2\right)\left(L-\frac
        1{2L}\right)+\left(g_R^2-g_Y^2\right) L\right] ,\\
    \Pi^{\hspace{0.25mm} \prime}_{ZZ}(0)&=\frac{(g_L^2+g_Y^2) \,
      v^2}{8 \Mkk^2} \left(1-\frac 1 L\right) .
  \end{split}
\end{equation}
Inserting these expressions into (\ref{eq:STUdef}) yields 
\begin{equation} \label{eq:STUres}
  S=\frac{2\pi v^2}{\Mkk^2}\left(1-\frac 1L\right) , \qquad
  T=-\frac{\pi v^2}{4 \, c^2_w \Mkk^2}\, \frac1L\,, \qquad U=0 \,,
\end{equation}
in agreement with \cite{Agashe:2003zs, Delgado:2007ne}.  In contrast
to the minimal model \cite{Carena:2002dz, Casagrande:2008hr,
  Delgado:2007ne} there is no $L$-enhanced term in the $T$
parameter. It has been cancelled by the additional corrections
appearing in the contributions to the mass formula (\ref{eq:mwmz}),
which introduces extra terms in the correlators $\Pi_{WW}(0)$ and
$\Pi_{ZZ}(0)$. A related discussion including estimates of loop
effects on the $T$ parameter has been given in \cite{Carena:2006bn,
  Carena:2007ua}. The one-loop corrections to the $S$ parameter in the
custodial RS model arising from Higgs loops have been calculated in
\cite{Burdman:2008gm} and found to be logarithmically UV
divergent. This could result in a large and positive $S$ parameter,
which is rather problematic \cite{Casagrande:2008hr} in view of the
consistency of the global fit of the oblique electroweak precision
observables.

The zero-mode profiles, which were used for the above derivations, read
\begin{equation} \label{eq:expprof}
  \begin{split}
    \chi_0^{(+)}(t) & = \frac{1}{\sqrt{2\pi}} \left[ \, 1 +
      \frac{x_{a}^2}{4} \left( 1 - \frac{1}{L} + t^2 \, \big( 1 - 2L -
        2\ln t\big) \right)
      + \ord\left( x_{a}^4\right) \right] ,\\[1mm]
    \chi_0^{(-)}(t) & = \sqrt{\frac{L}{2\pi}}\; t^2 \left[ \, -2 +
      \frac{x_{a}^2}{4} \left(t^2-\frac 2 3\right) + \ord\left( x_a^4
      \right) \right] ,
  \end{split}
\end{equation} 
for $a = W, Z$. Here $x_a^2 \equiv (m_0^a)^2/M_{\rm KK}^2$ denotes the
relevant zero-mode solution in (\ref{eq:mwmz}). The profiles
$\chi_0^{(+)}(t)$ with Neumann IR BC are identical to those appearing
in the minimal model, while the profiles $\chi_0^{(-)} (t)$ satisfying
Dirichlet IR BC scale like $\sqrt{L}$, reflecting the localization of
KK modes close to the IR boundary.  Notice that (\ref{eq:expprof})
contains, besides the $t$-independent terms that are included in
(\ref{eq:STUres}), also $t$-dependent contributions that will in
general lead to non-universal vertex corrections. While these
corrections modify the interactions of the SM fermions with the $W$
and $Z$ bosons, they turn out to be negligibly small for light
fermions localized near the UV brane. This is the case for the first
two generations of SM fermions, and it helps to avoid excessive
contributions to FCNCs. In such a case the oblique corrections are
adequately parametrized by the $S$, $T$, $U$ parameters as given in
(\ref{eq:STUres}).

Finally, we can also expand $\vec A_0^{\,a}$. Including corrections up
to $v^2/\Mkk^2$, we find
\begin{equation} \label{eq:vecA0a}
  \vec A_0^{\, a} = \left( \begin{array}{c}
      1\\
      \displaystyle -s_a c_a\, \frac{X^2}{4}\, \sqrt L
    \end{array} \right),
\end{equation}
where the second component parametrizes the admixture of $\chi_0^{(-)}
(t)$ in the zero mode. As we will see below in Section
\ref{sec:custodialprotection}, the results (\ref{eq:expprof})
  and (\ref{eq:vecA0a}) play a crucial role in the custodial protection
mechanism of the $Z b_L \bar b_L$ vertex and its flavor-changing
counterparts.

\subsection{Summing over KK Modes}
\label{sec:kksum}

In this section we will show how to compute the following sum over
gauge-boson profiles $\vec{\chi}_n^{\, a}(t)$ weighted by inverse
powers of the normalized KK mode masses $\left (x_n^a \right )^2$,
\begin{equation} \label{eq:KKsum1}
  \bm{\Sigma}_a (t, t^\prime) \equiv \sum_n \, \frac{ \vec{\chi}_n^{\,
      a} (t) \, \vec{\chi}_n^{\, a \, T} (t^\prime)}{\left ( x_n^a
    \right)^2} \,,
\end{equation}
which arises when one attempts to calculate the tree-level exchange of a
SM electroweak gauge boson accompanied by its KK excitations in the
limit of zero (or negligibly small) momentum transfer. 
The infinite sum in (\ref{eq:KKsum1}) can be calculated by employing
the methods developed in \cite{Casagrande:2008hr, Hirn:2007bb}. We
first integrate the EOMs (\ref{eq:gaugeeom}) twice, accounting for the
BCs on both the UV and IR brane. After switching to $t$ coordinates
this yields
\begin{equation}
  \frac{\vec{\chi}_n^{\, a} (t)}{(x_n^a)^2} \, = \, \vec{\cal I}_n^{\,
    a} (t) - \left (t^2 - \epsilon^2 \right ) {\bm X}_a \, \vec{\cal
    I}_n^{\, a} (1) + \left [ \, \bm{1} - \left (t^2 - \epsilon^2 \right )
    {\bm X}_a \, \right ] \! {\bm P}_{(+)} \, \frac{\vec{\chi}_n^{\, a}
    (\epsilon)}{\big ( x_n^a \big)^2} \,,
\end{equation}
where we have defined
\begin{equation}
  {\vec{\cal I}}_n^{\, a} (t) \equiv \int_{\epsilon}^t \! dt^\prime \,
  t^\prime \int_{t^\prime}^{1^-} \! \frac{dt^{\prime \prime}}{t^{\prime
      \prime}} \, \vec{\chi}_n^{\, a} (t^{\prime\prime}) \,, \qquad {\bm
    X}_a \equiv \tilde{X}^2 \, {\bm D}_a \equiv \frac{L X^2}{2 + L X^2 \,
    (1-\epsilon^2)} \, {\bm D}_a \,.
\end{equation}
Using the completeness relation
\begin{equation} 
  \sum_n \frac{1}{t} \, {\vec \chi}_n^{\, a} (t) \, {\vec \chi}_n^{\,
    a\, T} (t^\prime) = \frac{L}{2 \pi} \, \delta (t - t^\prime) \, {\bm
    1}
\end{equation}
for the gauge-boson profiles, it is then easy to prove that
\begin{equation}
  \sum_n {\vec{\cal I}}_n^{\, a} (t) \, \vec{\chi}_n^{\, a \, T}
  (\phi^\prime) = \frac{L}{4 \pi} \left (t_<^2 - \epsilon^2 \right )
  {\bm 1} \,,
\end{equation}
where $t_< \equiv {\rm min} (t, t^\prime)$. With these results at hand
it is now a matter of simple algebraic manipulations to arrive at
\begin{equation}
  \begin{split}
    \bm{\Sigma}_a (t, t^\prime) & = \frac{L}{4 \pi} \, \Big [ \left
      (t_<^2 - \epsilon^2 \right ) {\bm 1} + \left (t^2 - \epsilon^2
    \right ) \left
      (t^{\prime \, 2} - \epsilon^2 \right ) {\bm X}_a \Big ] \\[1mm]
    & \phantom{xx} + \left [ \,\bm{1} - \left (t^2 - \epsilon^2 \right
      ) {\bm X}_a \, \right ] {\bm P}_{(+)} \, {\bm \Sigma}_a
    (\epsilon, \epsilon) \, {\bm P}_{(+)} \left [ \, \bm{1} - \left
        (t^{\prime \, 2} - \epsilon^2 \right ) {\bm X}_a \, \right]^T ,
  \end{split}
\end{equation}
which is exact to all orders in $v^2/{M_{\rm KK}^2}$.

With the help of the orthonormality relation (\ref{eq:ortho}), the
remaining sum over gauge profiles evaluated at the UV brane can be
brought into the form 
\begin{equation}
  \begin{split}
    {\bm P}_{(+)} \, {\bm \Sigma}_a (\epsilon, \epsilon) \, {\bm
      P}_{(+)} = \frac{L}{2 \pi x_a^2} \, \big ( \vec{\chi}_0^{\, a}
    (\epsilon) \big )_1 \, & \bigg [ \, \int_{\epsilon}^1 \frac{dt}{t} \,
    \Big [ \left ( 1 - c_a^2 \hspace{0.25mm} \tilde X^2 \left ( t^{2} -
        \epsilon^2 \right ) \right ) \big ( \vec{\chi}_0^{\, a} (t) \big )_1
    \\ & \phantom{x} + s_a c_a \hspace{0.25mm} \tilde X^2 \left ( t^{2} -
      \epsilon^2 \right ) \big ( \vec{\chi}_0^{\, a} (t) \big )_2 \Big ] \,
    \bigg ]^{-1} {\bm P}_{(+)}\,,
  \end{split}
\end{equation}
where $\big(\vec{\chi}_0^{\,a}(t)\big)_i$ denotes the $i^{\rm th}$
component of the corresponding zero-mode vector. This formula has the
advantage that it can be easily expanded in powers of $v^2/\Mkk^2$
using (\ref{eq:expprof}) and (\ref{eq:vecA0a}). Retaining the first
two terms in the expansion leads to
\begin{equation}
  {\bm P}_{(+)} \, {\bm \Sigma}_a (\epsilon, \epsilon) \, {\bm P}_{(+)}
  = \left ( \frac{1}{2 \pi x_a^2} + \frac{1}{4 \pi} \left [ \, 1 -
      \frac{1}{2 L} - \epsilon^2 \left ( L - \frac{1}{2 L} \right ) \right ]
    + {\cal O} (x_a^2) \right ) {\bm P}_{(+)}\,.
\end{equation}
Keeping in mind that $X^2 = x_a^2/c_a^2 + O(x_a^4)$ and dropping
phenomenologically irrelevant terms of second order in $\epsilon
\approx 10^{-16}$, we finally arrive at
\begin{equation} \label{eq:Sigmafinal}
  \bm{\Sigma}_a (t, t^\prime) = \frac{L}{4 \pi} \, \Big [ \, t_<^2 \,
  {\bm 1} - {\bm P}_a \, t^2 - {\bm P}_a^{\, T} \, t^{\prime \, 2} \,
  \Big ] + \left [ \frac{1}{2 \pi x_a^2} + \frac{1}{4 \pi} \left ( 1 -
      \frac{1}{2 L} \right ) \right ] {\bm P}_{(+)} + {\cal O} (x_a^2) \,,
\end{equation}
where
\begin{equation} \label{eq:Pa}
  {\bm P}_a= \left( \begin{array}{cc}
      1~ & 0\\
      -\frac{s_a}{c_a}~ & 0
    \end{array} \right) .
\end{equation}

Having at hand an analytic expression for the zero-mode contribution
to (\ref{eq:KKsum1}) alone,
\begin{equation}
  \bm{\Pi}_a (t, t^\prime) \equiv \frac{ \vec{\chi}_0^{\, a} (t) \,
    \vec{\chi}_0^{\, a \, T} (t^\prime)}{ x_a^2 } \,,
\end{equation}
will also prove useful later in our discussion. Employing the results
(\ref{eq:expprof}) and (\ref{eq:vecA0a}), a straightforward
calculation leads to
\begin{eqnarray} \label{eq:Pifinal}
  \begin{split}
    \bm{\Pi}_a (t, t^\prime) & = -\frac{L}{4 \pi} \, \Big [ {\bm P}_a
    \, t^2
    + {\bm P}_a^{\hspace{0.25mm} T} \, t^{\prime \, 2} \, \Big ] \\[1mm]
    & \phantom{xx} + \left [ \, \frac{1}{2 \pi x_a^2} + \frac{1}{4
        \pi} \left ( 1 - \frac{1}{L} + t^2 \left ( \frac{1}{2} - \ln t
        \right ) + t^{\prime \, 2} \left ( \frac{1}{2} - \ln t^\prime
        \right ) \right ) \right ] {\bm P}_{(+)} + {\cal O} (x_a^2)
    \,. \hspace{4mm}
  \end{split}
\end{eqnarray}
Comparing (\ref{eq:Sigmafinal}) to (\ref{eq:Pifinal}) we see that all
$L$-enhanced terms in $\bm{\Sigma}_a (t, t^\prime)$ besides the one
proportional to the non-factorizable term $t_<^2$ arise from the
zero-mode contribution $\bm{\Pi}_a (t, t^\prime)$. Factorizable
contributions due to the ground-state $W$ and $Z$ bosons are thus
enhanced by the logarithm of the warp factor with respect to the
contributions from the tower of KK excitations
\cite{Casagrande:2008hr}. We also recall that the term $t_<^2$
reflects the full 5D structure of the RS model, which is lost when one
considers only a few low-lying KK modes \cite{Bauer:2008xb}.

Our analytic results for $\bm{\Sigma}_a (t,t^\prime)$ and $\bm{\Pi}_a
(t, t^\prime)$ will turn out to be phenomenologically quite important,
as they allow for a clear understanding of the cancellation of certain
terms in $\Delta F = 1$ and $\Delta F = 2$ FCNC interactions. In
particular, the exact form of the matrix ${\bm P}_a$ and its interplay
with the terms proportional to the $2 \times 2$ unit matrix ${\bm 1}$
are key ingredients for the custodial protection of the
flavor-conserving $Z b_L \bar b_L$ coupling as well as of the
flavor-violating $Z d_L^{\hspace{0.25mm} i} \bar d^{\hspace{0.25mm}
j}_L$ vertices. We will for the moment no further dwell on this issue,
but will return to it in detail in
Section~\ref{sec:custodialprotection}.  Before moving on, let us
remark that the KK sums involving photon and gluon excitations do not
depend on whether the electroweak gauge group is minimal or extended,
so that the results for the corresponding sums (excluding the zero
modes) can be taken over from \cite{Casagrande:2008hr}.

\section{Bulk Fermions}
\label{sec:fermions}

We will now present the explicit realization of the quark sector in
the model under consideration. Then we will turn to the KK
decomposition and derive the bulk profiles for the corresponding
fields. As we want to have a custodial protection of the $Z b_L \bar
b_L$ vertex \cite{Agashe:2006at}, we impose a discrete $P_{LR}$
symmetry that interchanges the two $SU(2)$ groups. As a consequence,
the left-handed bottom quark has to be part of a $SU(2)_L \times
SU(2)_R$ bi-doublet with isospin quantum numbers $T_L^3 =-T_R^3 =
-1/2$ (see Section \ref{sec:custodialprotection}). This fixes the
quantum numbers of the other fields uniquely and implies the following
multiplet structure for the quark fields with even $Z_2$ parity:
\begin{equation} \label{eq:multiplets}
  \begin{split}
    & \quad Q_L \equiv \left(\begin{array}{cc}
        {u_L^{(+)}}_{\frac 23} & {\lambda_L^{(-)}}_{\frac 53}\\
        {d_L^{\hspace{0.25mm} (+)}}_{-\frac 13} & {u_L^{\prime \,
            (-)}}_{\frac 23}
      \end{array}\right)_{\frac 23}\, ,\qquad \qquad 
    u_R^c \equiv \left ( {u_R^{c\, (+)}}_{\frac 23}
    \right )_{\frac 23} \, ,\\[1mm]
    {\cal T}_R & \equiv {\cal T}_{1R}\oplus{\cal
      T}_{2R}\equiv\left(\begin{array}{c}
        {\Lambda_R^{\prime\, (-)}}_{\frac 53}\\
        {U_R^{\prime\, (-)}}_{\frac 23}\\
        {D_R^{\prime\, (-)}}_{-\frac 13}
      \end{array}\right)_{\frac 23}
    \oplus\left(\begin{array}{ccc} {D_R^{(+)}}_{-\frac 13}\
        {U_R^{(-)}}_{\frac 23}\ {\Lambda_R^{(-)}}_{\frac 53}
      \end{array}\right)_{\frac 23}\,.
  \end{split}
\end{equation}
Here the superscripts $(+)$ and $(-)$ of the chiral fields specify the
type of BC on the UV boundary, and as before we have not explicitly
shown the BCs at the IR brane, which are understood to be of Neumann
type in all cases. The choice of the parities is motivated by the
constraint to arrive at a low-energy spectrum of the theory that is
consistent with observations. The subscripts correspond to the
$U(1)_{\rm EM}$ and $U(1)_X$ charges, respectively, which are
connected through the relations $Y=-T_R^3+Q_X$ and $Q=T_L^3+Y$. For
completeness and future reference, we summarize the quantum numbers of
the quark fields in Table~\ref{tab:charges}. The right-handed
down-type quarks have to be embedded in a $SU(2)_R$ triplet in order
to arrive at an $U(1)_X$-invariant Yukawa coupling. Note that we have
chosen the same $SU(2)_L \times SU(2)_R$ representations for all three
generations, which is necessary if one wants to consistently
incorporate quark mixing in the fully anarchic approach to flavor in
warped extra dimensions. The chosen representations also play a
crucial role in the suppression of flavor-changing left-handed
$Z$-boson couplings \cite{Blanke:2008zb}. Altogether they feature 15
different quark fields in the up-type and nine in the down-type
sector. Due to the BCs, there will be three light modes in each sector
to be identified with the SM quarks. These are accompanied by KK
towers which consist of groups of 15 and nine modes of similar masses
in the up- and down-type quark sector, respectively.  Moreover one
also faces a KK tower of exotic fermion fields of electric charge
$5/3$, which exhibits nine excitations with small mass splitting in
each level. In addition to (\ref{eq:multiplets}) we have a second set
of multiplets, belonging to the components of opposite chirality. The
corresponding states have opposite BCs. In particular, they all obey
Dirichlet BCs at the IR brane. Remember that the $SU(2)_L$
transformations act vertically, while the $SU(2)_R$ transformations
act horizontally on the multiplets.

\begin{table}
\begin{center}
\begin{tabular}{|c|c|c|c|c|c|}
\hline
& $Q$ & $Q_X$ & $Y$ & $T_L^3$ & $T_R^3$ \\
\hline
$u_L^{(+)}$ & $\phantom{-} 2/3$ & $2/3$ & $1/6$ & $\phantom{-} 1/2$ 
& $\phantom{-} 1/2$ \\
$d_L^{\hspace{0.25mm} (+)}$ & $-1/3$ & $2/3$ & $1/6$ & 
$-1/2$ & $\phantom{-} 1/2$ \\
$\lambda_L^{(-)}$ & $\phantom{-} 5/3$ & $2/3$ & $7/6$ & 
$\phantom{-} 1/2$ & $-1/2$ \\
$u_L^{\prime \, (-)}$ & $\phantom{-} 2/3$ & $2/3$ & $7/6$ & 
$-1/2$ & $-1/2$ \\ \hline
\end{tabular} \quad 
\begin{tabular}{|c|c|c|c|c|c|} 
\hline
& $Q$ & $Q_X$ & $Y$ & $T_L^3$ & $T_R^3$ \\
\hline
$u_R^{c \, {(+)}}$ & $\phantom{-} 2/3$ & $2/3$ & $2/3$ & 
$\phantom{-} 0$ & $0$ \\
$\Lambda_R^{\prime \, (-)}$ & $\phantom{-} 5/3$ & $2/3$ & $2/3$ & 
$\phantom{-} 1$ & $ 0$ \\
$U_R^{\prime \, {(-)}}$ & $\phantom{-} 2/3$ & $2/3$ & $2/3$ & 
$\phantom{-} 0$ & $0$ \\
$D_R^{\prime \, {(-)}}$ & $-1/3$ & $2/3$ & $2/3$ & 
$-1$ & $ 0$ \\ \hline
\end{tabular}
\end{center}
\begin{center}
\begin{tabular}{|c|c|c|c|c|c|} 
\hline
& $Q$ & $Q_X$ & $Y$ & $T_L^3$ & $T_R^3$ \\
\hline
$D_R^{(+)}$ & $-1/3$ & $2/3$ & $-1/3$ & 
$0$ 
& $\phantom{-} 1$ \\
$U_R^{(-)}$ & $\phantom{-} 2/3$ & $2/3$ & $\phantom{-} 2/3$ & $0$ & 
$\phantom{-} 0$ \\
$\Lambda_R^{(-)}$ & $\phantom{-} 5/3$ & $2/3$ & $\phantom{-} 5/3$ & 
$0$ & $-1$ \\ \hline
\end{tabular}
\end{center}
\parbox{15.5cm}{\caption{\label{tab:charges} Charge assignments of the different quark fields in the 
  extended RS model.}}
\end{table}

\subsection{Fermionic Action and Yukawa Couplings}

The structure of the 5D action of the quark fields has already been
given in \cite{Casagrande:2008hr}. It is straightforward to
generalize the action to the custodial model \cite{Albrecht:2009xr}. 
The only non-trivial part are the Yukawa couplings, where the possible
gauge-invariant terms take the form
\begin{eqnarray} \label{Sfermyuk}
  \begin{split} 
    S_{\rm Yukawa} & = -\int d^4x\,r \int_{-\pi}^\pi\!d\phi\;
    \delta(|\phi|-\pi)\; \frac{e^{-3\sigma(\phi)}}{r} \, \Bigg [ \,
    \big( Y_u^{\rm (5D)}\big)_{ij} \, \Big \{ \! \left(\bar
      Q_L^i\right)_{a\alpha}\,u_R^{c \, j} + \left(\bar
      Q_R^i\right)_{a\alpha}\,u_L^{c \, j}
    \Big \} \, \Phi_{a \alpha} \hspace{4mm} \\
    & \phantom{xx} \hspace{1cm} + \frac{\big( Y_d^{\rm
        (5D)}\big)_{ij}}{\sqrt{2}}\, \bigg\{ \Big [ \left(\bar
      Q_L^i\right)_{a\alpha} \left({\cal T}_{1R}^j\right)^c +
    \left(\bar Q_R^i\right)_{a\alpha} \left({\cal
        T}_{1L}^j\right)^c\Big ] \left(\sigma^c\right)_{ab} \Phi_{b
      \alpha} \\
    & \phantom{xx} \hspace{3cm}+\Big [ \left(\bar
      Q_L^i\right)_{a\alpha} \left({\cal T}_{2R}^j\right)^\gamma +
    \left(\bar Q_R^i\right)_{a\alpha} \left({\cal
        T}_{2L}^j\right)^\gamma \Big ]
    \left(\sigma^\gamma\right)_{\alpha\beta} \Phi_{a \beta} \bigg\}
    +{\rm h.c.} \, \Bigg]\,.
  \end{split}
\end{eqnarray}
Here $\Phi$ is the Higgs bi-doublet introduced in (\ref{eq:Higgsbi}),
and repeated indices are understood to be summed over. Notice that
(\ref{Sfermyuk}) contains operators like $(Y_u^{\rm (5D)})_{ij}
\left(\bar Q_R^i\right)_{a\alpha} \hspace{0.25mm} u_L^{c \,j} \,
\Phi_{a\alpha}$ not included in \cite{Casagrande:2008hr}. 
   Furthermore, we will choose the same Yukawa matrix for the
    coupling of both chirality structures, \ie, $LR$ and $RL$.
  The action (\ref{Sfermyuk}) can thus be regarded
  as the limit of a set-up with a bulk Higgs approaching the
  IR brane. The generalization of our results to the case of
  different Yukawa matrices, which would in general be allowed for a
  perfectly brane-localized Higgs, is straightforward. Since
  the 5D Lorentz group is irreducible, a splitting of different
  chiralities is not possible in the case of a bulk Higgs.

In (\ref{Sfermyuk}) the Latin (Greek) letters from the beginning of
the alphabet refer to $SU(2)$ indices, while superscripts $i,j$
label the quark generations. Moreover, the components of the
triplets in the expression above refer to the representations
\begin{equation} 
  {\cal T}_{1R}=\left(\begin{array}{c} \frac{1}{\sqrt
        2}\left({D_R^{\prime\, (-)}}_{-\frac 13}+ {\Lambda_R^{\prime\,
            (-)}}_{\frac 53}\right)\\
      \frac{i}{\sqrt 2}\left({D_R^{\prime\, (-)}}_{-\frac 13}-
        {\Lambda_R^{\prime\, (-)}}_{\frac 53}\right)\\
      {U_R^{\prime\, (-)}}_{\frac 23} \end{array}\right) , \qquad {\cal
    T}_{2R}=\left(\begin{array}{c} \frac{1}{\sqrt
        2}\left({D_R^{(+)}}_{-\frac 13}+ {\Lambda_R^{(-)}}_{\frac 53}\right)\\
      \frac{i}{\sqrt 2}\left({-D_R^{(+)}}_{-\frac 13}+
        {\Lambda_R^{(-)}}_{\frac 53}\right)\\
    {U_R^{(-)}}_{\frac 23} \end{array}\right)^T \!,
\end{equation}
which ensures that one ends up in the desired mass basis, and
$\sigma^{c, \gamma}$ are the Pauli matrices. After electroweak
symmetry breaking, the Yukawa couplings (\ref{Sfermyuk}) give rise to
mass terms which mix different 5D fields with the same $U(1)_{\rm EM}$
charge. 

In analogy to the KK decomposition for the gauge bosons in
(\ref{eq:KKdec}), we will later work in the basis of 4D
mass eigenstates for the quark fields. Therefore it is convenient to
introduce the vectors
\begin{equation} \label{eq:fermvec}
  \vec U \equiv
  \left(\begin{array}{c}
      u\\
      u^\prime
    \end{array}\right) ,\quad 
  \vec u \equiv \left(\begin{array}{c}
      u^c\\
      U^\prime\\
      U
    \end{array}\right) ,\quad 
  \vec D \equiv d\,,\quad 
  \vec d \equiv \left(\begin{array}{c}
      D\\
      D^\prime
    \end{array}\right) ,\quad 
  \vec \Lambda \equiv \lambda\,,\quad \vec \lambda \equiv
  \left(\begin{array}{c}
      \Lambda^\prime\\
      \Lambda
    \end{array}\right) ,
\end{equation}
which leads to a one-to-one correspondence between the analysis of
\cite{Casagrande:2008hr} and the one presented here. The action can
now be expressed in the simple form
\begin{eqnarray} \label{Sferm2}
  \begin{split}
    S_{\rm ferm,2} & = \int d^4x\,r\int_{-\pi}^\pi\!d\phi\;\Bigg\{
    \sum_{q=U,u,D,d,\Lambda,\lambda} \Bigg( e^{-3\sigma(\phi)}\,
    \bar{\vec q}\,i\delslash\,\vec q -
    e^{-4\sigma(\phi)}\,\sgn(\phi)\,
    \bar{\vec q}\,\bm{M}_{\vec q}\,\vec q\\
    &\quad\mbox{}- \frac{1}{2r} \, \bigg[ \, \bar{\vec
      q}_L\,e^{-2\sigma(\phi)} \,
    {\stackrel{\leftrightarrow}{\partial}}_\phi \,
    e^{-2\sigma(\phi)}\,\vec q_R + \mbox{h.c.}\,
    \bigg] \Bigg)\\
    & \quad\mbox{}- \delta(|\phi|-\pi)\,e^{-3\sigma(\phi)}\,
    \frac{v}{\sqrt2 r} \, \Big [ \, \bar{\vec U}_L\,\bm{Y}_{\vec
      u}^{\rm (5D)}\,\vec u_R + \bar{\vec D}_L\,\bm{Y}_{\vec d}^{\rm
      (5D)}\,\vec d_R + \bar{\vec \Lambda}_L\,\bm{Y}_{\vec
      \lambda}^{\rm (5D)}\,\vec \lambda_R \\
    &\hspace{4.75cm} + \bar{\vec U}_R\,\bm{Y}_{\vec u}^{\rm
      (5D)}\,\vec u_L + \bar{\vec D}_R\,\bm{Y}_{\vec d}^{\rm
      (5D)}\,\vec d_L + \bar{\vec \Lambda}_R\,\bm{Y}_{\vec
      \lambda}^{\rm (5D)}\,\vec \lambda_L + \mbox{h.c.} \, \Big ]
    \Bigg\} \,, \hspace{4mm}
  \end{split}
\end{eqnarray}
with the Yukawa matrices
\begin{equation} \label{eq:yukawam}
  \bm{Y}_{\vec u}^{\rm (5D)} \equiv
  \left(\begin{array}{ccc} \bm{Y}_u^{\rm (5D)}& \frac{1}{\sqrt
        2}\bm{Y}_d^{\rm (5D)}&
      \, \frac{1}{\sqrt 2}\bm{Y}_d^{\rm (5D)}\\
      \bm{Y}_u^{\rm (5D)}& - \frac{1}{\sqrt 2}\bm{Y}_d^{\rm (5D)}& -
      \, \frac{1}{\sqrt 2}\bm{Y}_d^{\rm (5D)}
    \end{array}\right) , \qquad
  \bm{Y}_{\vec d}^{\rm (5D)} \equiv \bm{Y}_{\vec \lambda}^{\rm (5D)}
  \equiv \left(\begin{array}{cc} 
      \hspace{0.5mm} \bm{Y}_d^{\rm (5D)}&
      \hspace{0.5mm} \bm{Y}_d^{\rm (5D)}
    \end{array}\right) .
\end{equation}
The symmetric derivative
$\stackrel{\leftrightarrow}{\partial}_\phi\,\equiv\,
\stackrel{\rightarrow}{\partial}_\phi
-\stackrel{\leftarrow}{\partial}_\phi$ ensures hermicity of the action
in the presence of boundary terms. In the case of the SM with three
generations each entry of (\ref {eq:yukawam}) corresponds to a
$3\times 3$ matrix.  We define dimensionless 4D Yukawa couplings via
\begin{equation}
  \bm{Y}_q^{\rm (5D)}\equiv \frac {2 \bm{ Y}_q}{k}\,,
\end{equation}
where $k$ denotes the $\rm{AdS}_5$ curvature  and $q = u,d$. 
The generalized bulk mass matrices ${\bm M}_{\vec q}$ take the
form
\begin{equation}
  \begin{array}{ccc}
    \bm{M}_{\vec U} \equiv \left(\begin{array}{cc}
        \bm{M}_Q&0\\
        0&\bm{M}_Q
      \end{array}\right) ,&
    \bm{M}_{\vec D} \equiv \bm{M}_Q\, , &
    \bm{M}_{\vec \Lambda} \equiv \bm{M}_Q\, ,\\[8mm]
    \bm{M}_{\vec u} \equiv \left(\begin{array}{ccc}
        \bm{M}_{u^c}&0&0\\
        0&\bm{M}_{{\cal T}_1}&0\\
        0&0&\bm{M}_{{\cal T}_2}
      \end{array}\right) ,&\quad 
    \bm{M}_{\vec d} \equiv \left(\begin{array}{cc}
        \bm{M}_{{\cal T}_2}&0\\
        0&\bm{M}_{{\cal T}_1}
      \end{array}\right) , &\quad 
    \bm{M}_{\vec \lambda} \equiv \left(\begin{array}{cc}
        \bm{M}_{{\cal T}_1}&0\\
        0&\bm{M}_{{\cal T}_2}
      \end{array}\right) ,
  \end{array}
\end{equation}
where $\bm{M}_A$ are the $3 \times 3$ bulk mass matrices of the
corresponding multiplets $A = Q,u^c,{\cal T}_1,{\cal T}_2$.

\subsection{KK Decomposition}\label{sec:FermKK}

As motivated above, we decompose different 5D spinors that get mixed
by the IR BCs (\ie, by the Yukawa couplings) into the same basis of 4D
spinors\footnote{Here we have already made use of the fact that $\vec
Q_{L,R} (x,\phi)$ ($\vec q_{L,R} (x,\phi)$) can be expanded in terms
of the same vector $\vec a_n^{\hspace{0.5mm} Q}$ ($\vec
a_n^{\hspace{0.25mm} q}$). With this choice, the profiles ${\bm C}^Q_n
(\phi)$ and ${\bm S}^Q_n (\phi)$ (${\bm C}^q_n (\phi)$ and ${\bm
S}^q_n (\phi)$) are normalized in the same way.}
\begin{equation}\label{eq:KKdecferm}
  \begin{split}
    \vec Q_L(x,\phi) &= \frac{e^{2\sigma(\phi)}}{\sqrt r} \sum_n
    \bm{C}_n^{Q}(\phi)\,\vec a_n^{\hspace{0.5mm} Q}\,q_L^{(n)}(x) \,,
    \qquad \vec Q_R(x,\phi) = \frac{e^{2\sigma(\phi)}}{\sqrt r} \sum_n
    \bm{S}_n^{Q}(\phi)\, \vec a_n^{\hspace{0.5mm} Q}\,q_R^{(n)}(x) \,,
    \hspace{4mm} \\
    \vec q_L(x,\phi) &= \frac{e^{2\sigma(\phi)}}{\sqrt r} \sum_n
    \bm{S}_n^{\hspace{0.25mm} q}(\phi)\, \vec a_n^{\hspace{0.25mm}
      q}\,q_L^{(n)}(x) \,, \hspace{1.2cm} \vec q_R(x,\phi) =
    \frac{e^{2\sigma(\phi)}}{\sqrt r} \sum_n \bm{C}_n^{\hspace{0.25mm}
      q}(\phi)\,\vec a_n^{\hspace{0.5mm} q}\,q_R^{(n)}(x) \,,
  \end{split}
\end{equation}
where $Q=U,D,\Lambda$ and $q=u,d,\lambda$. Furthermore
\begin{equation}
  \begin{aligned}
    \bm{C}_n^{\hspace{0.25mm} U} &\equiv{\rm diag} \big
    (\bm{C}_n^{\hspace{0.25mm} Q(+)}, \bm{C}_n^{\hspace{0.25mm} Q(-)}
    \big )\, , & \qquad \bm{C}_n^{\hspace{0.25mm} u} &\equiv{\rm diag}
    \big (\bm{C}_n^{\hspace{0.25mm} u^c(+)}, \bm{C}_n^{\hspace{0.25mm}
      {\cal T}_1(-)},
    \bm{C}_n^{\hspace{0.25mm} {\cal T}_2(-)} \big )\, ,\\[1mm]
    \bm{S}_n^{\hspace{0.25mm} U} &\equiv{\rm diag} \big
    (\bm{S}_n^{\hspace{0.25mm} Q(+)}, \bm{S}_n^{\hspace{0.25mm} Q(-)}
    \big )\, , & \qquad \bm{S}_n^{\hspace{0.25mm} u} &\equiv{\rm diag}
    \big (\bm{S}_n^{\hspace{0.25mm}u^c(+)},
    \bm{S}_n^{\hspace{0.25mm}{\cal T}_1(-)},
    \bm{S}_n^{\hspace{0.25mm} {\cal T}_2(-)} \big )\, ,\, \\[4mm]
    \bm{C}_n^{\hspace{0.25mm} D} &\equiv \bm{C}_n^{\hspace{0.25mm}
      Q(+)}\, ,& \bm{C}_n^{\hspace{0.25mm} d} &\equiv{\rm diag} \big
    (\bm{C}_n^{\hspace{0.25mm} {\cal T}_2(+)},
    \bm{C}_n^{{\cal T}_1(-)} \big )\, ,\\[1mm]
    \bm{S}_n^{\hspace{0.25mm} D}&\equiv \bm{S}_n^{\hspace{0.25mm}
      Q(+)}\, ,& \bm{S}_n^{\hspace{0.25mm} d} &\equiv{\rm diag}\big
    (\bm{S}_n^{\hspace{0.25mm} {\cal T}_2 (+)},
    \bm{S}_n^{\hspace{0.25mm} {\cal T}_1 (-)} \big )\, , \, \\[4mm]
    \bm{C}_n^{\hspace{0.25mm} \Lambda}&\equiv
    \bm{C}_n^{\hspace{0.25mm} Q(-)}\, ,& \bm{C}_n^{\hspace{0.25mm}
      \lambda}&\equiv{\rm diag} \big (\bm{C}_n^{\hspace{0.25mm} {\cal
        T}_1(-)},
    \bm{C}_n^{\hspace{0.25mm} {\cal T}_2(-)})\, ,\\[1mm]
    \bm{S}_n^{\hspace{0.25mm} \Lambda}&\equiv
    \bm{S}_n^{\hspace{0.25mm} Q(-)}\, ,& \bm{S}_n^{\hspace{0.25mm}
      \lambda}&\equiv{\rm diag} \big (\bm{S}_n^{\hspace{0.25mm} {\cal
        T}_1(-)}, \bm{S}_n^{\hspace{0.25mm} {\cal T}_2(-)} \big )\,,
  \end{aligned}
\end{equation} 
and
\begin{eqnarray} \label{eq:avecs}
  \begin{array}{cccccc}
    \vec a_n^{\hspace{0.35mm} U} \equiv 
    \left(\begin{array}{c} a_n^{\hspace{0.25mm} u}\\
        a_n^{\hspace{0.25mm} u^\prime}\end{array}\right) , & 
    ~~ \vec a_n^{\hspace{0.5mm} u} \equiv 
    \left(\begin{array}{c} a_n^{\hspace{0.25mm} u^c}\\ 
        a_n^{\hspace{0.25mm} U^\prime}\\ 
        a_n^{\hspace{0.25mm} U}\end{array}\right) , & ~~ 
    \vec a_n^{\hspace{0.25mm} D} \equiv a_n^{\hspace{0.25mm} d}\, , & 
    ~~ \vec a_n^{\hspace{0.35mm} d} \equiv 
    \left(\begin{array}{c}a_n^{\hspace{0.25mm} D}\\
        a_n^{\hspace{0.25mm} D^\prime}\end{array}\right) , &
    ~~ \vec a_n^{\hspace{0.25mm} \Lambda} \equiv 
    a_n^{\hspace{0.25mm} \lambda}\, , & ~~ 
    \vec a_n^{\hspace{0.5mm} \lambda} 
    \equiv \left(\begin{array}{c}a_n^{\hspace{0.25mm} \Lambda^\prime}\\
        a_n^{\hspace{0.25mm} \Lambda}\end{array}\right) , \hspace{6mm}
  \end{array}
\end{eqnarray}
where in order to simplify the notation we have dropped the argument
$\phi$ in the profile functions. Each component of the vector of
spinors on the left-hand side of (\ref{eq:KKdecferm}) contains three
entries belonging to the three quark generations. The $3 \times 3$
matrices $\bm{C}_n^{A(\pm)}(\phi)$ ($\bm{S}_n^{A(\pm)}(\phi)$) with
$A=Q,u^c,{\cal T}_1,{\cal T}_2$ correspond to even (odd) profiles on
the orbifold, and the superscript $(\pm)$ indicates the type of BC on
the UV brane. With some abuse of notation, the superscripts of the odd
profiles refer to the UV BC of the associated even profiles. The
quarks present already in the minimal RS model hence all carry a $(+)$
superscript. Labels for the IR BCs are again omitted to simplify the
notation. The flavor structure is encoded in the three-component
vectors $a_n^{A}$ with $A=u, u^\prime, u^c, U^\prime, U, d, D^\prime,
D, \lambda, \Lambda^\prime, \Lambda$, which are then combined into
larger flavor vectors $\vec{a}_n^{\, Q,q}$ with $Q$ and $q$ defined as
above. Finally, $q_{L}^{(n)}(x)$ and $q_{R}^{(n)}(x)$ are 4D spinors,
and the index $n$ labels the different mass eigenstates with masses
$m_n$, \ie, $m_1=m_u,\ m_2=m_c, \ m_3 = m_t$, \etc \ in the case of
up-type quarks, and similarly for down-type quarks.

By virtue of the vector notation (\ref{eq:fermvec}), we have reached
complete analogy to the decomposition of bulk quark fields in the
minimal model \cite{Casagrande:2008hr}. The further analysis in this
section thus follows almost entirely the corresponding part in that
article. An exception is the inclusion of the Yukawa couplings
involving $Z_2$-odd fermion profiles, which, as pointed out in
\cite{Azatov:2009na}, have not been considered in our previous
work. In the following we will close this gap, confirming the ${\cal
O} (v^2/M_{\rm KK}^2)$ results for a brane-localized Higgs presented
in \cite{Azatov:2009na} and extending them to all orders in the ratio
of the weak over the KK scale. A detailed discussion of the impact of
the inclusion of the $Z_2$-odd fermion couplings on the flavor
misalignment between the SM fermion masses and the Yukawa couplings is
deferred to Section~\ref{sec:higgscouplings}.

The 5D variational principle requires all the variations of the action
(\ref{Sferm2}) to vanish for arbitrary infinitesimal changes of the
fermionic fields. After KK decomposition (\ref{eq:KKdecferm}), this
leads to the following EOMs
\begin{align} \label{eq:EOM}
    \left( - \frac{1}{r}\,\partial_\phi - \bm{M}_{\vec Q}\,\sgn(\phi)
    \right) \bm{S}_n^Q(\phi)\, \vec a_n^{\hspace{0.25mm} Q} &= -
    m_n\,e^{\sigma(\phi)}\,\bm{C}_n^Q(\phi)\, \vec
    a_n^{\hspace{0.25mm} Q} \nonumber \\ & \phantom{=} + \delta (|\phi|-\pi) \,
    e^{\sigma(\phi)} \, \frac{\sqrt{2} \, v}{kr} \, {\bm Y}_{\vec q}
    \hspace{1mm} \bm{C}_n^{\hspace{0.25mm} q}(\phi) \, \vec
    a_n^{\hspace{0.5mm} q} \,,  \nonumber \\[1mm]
    \left( \frac{1}{r}\,\partial_\phi - \bm{M}_{\vec q}\,\sgn(\phi)
    \right) \bm{S}_n^{\hspace{0.25mm} q}(\phi)\, \vec
    a_n^{\hspace{0.5mm} q} &= -m_n\,e^{\sigma(\phi)}\,
    \bm{C}_n^{\hspace{0.25mm} q}(\phi)\, \vec a_n^{\hspace{0.5mm} q}  \nonumber  \\
    & \phantom{=} + \delta (|\phi|-\pi) \, e^{\sigma(\phi)} \,
    \frac{\sqrt{2} \, v}{kr} \, {\bm Y}_{\vec q}^\dagger \hspace{1mm}
    \bm{C}_n^{Q}(\phi) \, \vec a_n^{\hspace{0.25mm} Q} \,,  \nonumber  \\[1mm] \nonumber \\
    \left( \frac{1}{r}\,\partial_\phi - \bm{M}_{\vec Q}\,\sgn(\phi)
    \right) \bm{C}_n^Q(\phi)\,\vec a_n^{\hspace{0.25mm} Q} &= -
    m_n\,e^{\sigma(\phi)}\,\bm{S}_n^Q(\phi)\,\vec a_n^{\hspace{0.5mm}
      Q}  \nonumber \\
    & \phantom{=} + \delta (|\phi|-\pi) \, e^{\sigma(\phi)} \,
    \frac{\sqrt{2} \, v}{kr} \, {\bm Y}_{\vec q} \hspace{1mm}
    \bm{S}_n^{\hspace{0.25mm} q}(\phi) \, \vec a_n^{\hspace{0.5mm} q}
    \,,  \nonumber  \\[1mm]
    \left( - \frac{1}{r}\,\partial_\phi - \bm{M}_{\vec q}\,\sgn(\phi)
    \right) \bm{C}_n^{\hspace{0.25mm} q}(\phi)\,\vec
    a_n^{\hspace{0.5mm} q} &= -
    m_n\,e^{\sigma(\phi)}\,\bm{S}_n^{\hspace{0.25mm}
      q}(\phi)\,\vec a_n^{\hspace{0.5mm} q}  \nonumber  \\
    & \phantom{=} + \delta (|\phi|-\pi) \, e^{\sigma(\phi)} \,
    \frac{\sqrt{2} \, v}{kr} \, {\bm Y}_{\vec q}^\dagger \hspace{1mm}
    \bm{S}_n^{Q}(\phi) \, \vec a_n^{\hspace{0.25mm} Q} \,.
\end{align}

Within the bulk, \ie, for $\phi \neq 0, \pm \pi$, where no
brane-localized terms are present, the general solutions
\cite{Gherghetta:2000qt, Grossman:1999ra} to the above equations can
be written as linear combinations of Bessel functions (see
Section~\ref{sec:fermionprofiles}). The presence of the source terms
on the IR brane dictates the boundary behavior of the fields and
causes both the $Z_2$-even and -odd profiles to become discontinuous
at the IR brane with $\bm{C}_n^{Q,q} (\pm \pi) \neq \bm{C}_n^{Q,q}
(\pm \pi^-)$ and $\bm{S}_n^{Q,q} (\pm \pi) = 0$ but
$\bm{S}_n^{Q,q}(\pm \pi^-) \neq 0$ \cite{Bagger:2001qi}. Finding the
correct IR BCs requires a proper regularization of the
$\delta$-functions appearing in (\ref{eq:EOM}). In the
following, we will view the $\delta$-function as the limit of a
sequence of regularized functions $\delta^\eta$ with support on the
interval $x \in [-\eta, 0]$. This limit is understood in the weak
sense so that
\begin{equation} \label{eq:deltadis}
  \lim_{\eta \to 0^+} \int_{-\infty}^{+\infty} dx \; \delta^{\eta} (x) 
  \hspace{0.25mm} f(x) = f(0) \,,
\end{equation}
for all test functions $f(x)$, \ie, smooth functions having compact
support.

In an infinitesimal interval around $|\phi| = \pi$, the source terms
in (\ref{eq:EOM}) are formally singular and as a result the behavior
of the profiles ${\bm C}^{Q, q}_n (\phi)$ and ${\bm S}^{Q,q}_n (\phi)$
becomes independent of the mass terms entering the EOMs. We regularize
the $\delta$-functions, switch to $t$ coordinates and integrate these
equations from $t$ to $1$, taking into account that the odd fermion
profiles vanish identically on the IR brane, \ie, ${\bm S}^{Q,q}_n (1)
= 0$. In this way we find
\begin{equation} \label{eq:BCsintegrated}
  \begin{split}
    \bm{S}_n^Q(t)\, \vec a_n^Q & = \frac{v}{\sqrt{2} M_{\rm KK}} \,
    {\bm Y}_{\vec{q}} \int_t^1 dt^\prime \, \big [ \delta^\eta
    (t^\prime - 1 ) \, \bm{C}_n^{\hspace{0.25mm} q}(t^\prime) \big ]
    \vec a_n^{\hspace{0.25mm} q} \,, \\[1mm]
    \bm{S}_n^{\hspace{0.25mm} q}(t)\, \vec a_n^{\hspace{0.25mm} q} & =
    -\frac{v}{\sqrt{2} M_{\rm KK}} \, {\bm Y}_{\vec{q}}^\dagger
    \int_t^1 dt^\prime \, \big [ \delta^\eta (t^\prime - 1 ) \,
    \bm{C}_n^{Q}(t^\prime) \big ] \vec a_n^{Q} \,, \\[1mm]
    \bm{C}_n^Q(t)\, \vec a_n^Q & = \bm{C}_n^Q(1)\, \vec a_n^Q
    -\frac{v}{\sqrt{2} M_{\rm KK}} \, {\bm Y}_{\vec{q}} \int_t^1
    dt^\prime \, \big [ \delta^\eta (t^\prime - 1 ) \,
    \bm{S}_n^{\hspace{0.25mm} q}(t^\prime) \big ] \vec
    a_n^{\hspace{0.25mm} q} \,, \\[1mm]
    \bm{C}_n^{\hspace{0.25mm} q}(t)\, \vec a_n^{\hspace{0.25mm} q} & =
    \bm{C}_n^{\hspace{0.25mm} q}(1)\, \vec a_n^{\hspace{0.25mm} q}
    +\frac{v}{\sqrt{2} M_{\rm KK}} \, {\bm Y}_{\vec{q}}^\dagger
    \int_t^1 dt^\prime \, \big [ \delta^\eta (t^\prime - 1 ) \,
    \bm{S}_n^{Q}(t^\prime) \big ] \vec a_n^{Q} \,.
  \end{split}
\end{equation}

In order to solve (\ref{eq:BCsintegrated}), we first introduce the
regularized Heaviside function
\begin{equation}
  \bar \theta^\eta (x) \equiv 1 - \int_{-\infty}^x dy \; \delta^\eta (y) \,,
\end{equation}
which obeys 
\begin{equation}
  \bar \theta^\eta (0) = 0 \,, \qquad 
   \bar \theta^\eta (-\eta) = 1 \,, \qquad 
  \partial_x \hspace{0.25mm} \bar \theta^\eta (x) = - \delta^\eta (x) \,.
\end{equation}
Using the latter properties it is readily shown that
\begin{equation} \label{eq:delthe}
  \int_t^1 dt^\prime \, \delta^\eta (t^\prime - 1) \, \big [ \bar 
  \theta^\eta (t^\prime - 1) \big ]^n = \frac{1}{n+1} \, \big [ \bar 
  \theta^\eta (t - 1) \big ]^{n+1} \,.
\end{equation}
It follows that for any arbitrary invertible matrix ${\bm A}$ one has
\begin{equation} \label{eq:magic1}
  \begin{split}
    \int_t^1 dt^\prime \, \delta^\eta (t^\prime - 1) \, \sinh
    \hspace{0.25mm} \big ( \bar \theta^\eta (t^\prime - 1)
    \hspace{0.25mm} {\bm A} \big ) & = \left [ \cosh \hspace{0.25mm}
      \big ( \bar \theta^\eta (t - 1) \hspace{0.25mm} {\bm A} \big ) -
      {\bm 1} \right ] {\bm A}^{-1} \,, \\
    \int_t^1 dt^\prime \, \delta^\eta (t^\prime - 1) \, \cosh
    \hspace{0.25mm} \big ( \bar \theta^\eta (t^\prime - 1)
    \hspace{0.25mm} {\bm A} \big ) & =\sinh \hspace{0.25mm} \big (
    \bar \theta^\eta (t - 1) \hspace{0.25mm} {\bm A} \big ) \, {\bm
      A}^{-1} \,,
  \end{split}
\end{equation} 
where the hyperbolic sine and cosine are defined via their power
expansions. 

The relations (\ref{eq:magic1}) now allow to determine the solutions
to (\ref{eq:BCsintegrated}). We find the following expressions
\begin{equation} \label{eq:BCssolution}
  \begin{split}
    \bm{S}_n^Q(t)\, \vec a_n^Q & = {\bm Y}_{\vec{q}} \left (
      \sqrt{{\bm Y}_{\vec{q}}^\dagger \hspace{0.25mm} {\bm
          Y}_{\vec{q}}} \right )^{-1} \, \sinh \left (
      \frac{v}{\sqrt{2} M_{\rm KK}} \, \bar \theta^\eta (t - 1) \,
      \sqrt{{\bm Y}_{\vec{q}}^\dagger \hspace{0.25mm} {\bm
          Y}_{\vec{q}}}\right ) \bm{C}_n^{\hspace{0.25mm} q}(1) \,
    \vec a_n^{\hspace{0.25mm} q} \,, \\[1mm]
    \bm{S}_n^{\hspace{0.25mm} q}(t)\, \vec a_n^{\hspace{0.25mm} q} & =
    - {\bm Y}_{\vec{q}}^\dagger \left ( \sqrt{{\bm Y}_{\vec{q}}
        \hspace{0.25mm} {\bm Y}_{\vec{q}}^\dagger} \right )^{-1} \,
    \sinh \left ( \frac{v}{\sqrt{2} M_{\rm KK}} \, \bar \theta^\eta (t
      - 1) \, \sqrt{{\bm Y}_{\vec{q}} \hspace{0.25mm} {\bm
          Y}_{\vec{q}}^\dagger} \right ) \bm{C}_n^{Q}(1) \, \vec
    a_n^{Q} \,, \\[1mm]
    \bm{C}_n^Q(t)\, \vec a_n^Q & = \cosh \left ( \frac{v}{\sqrt{2}
        M_{\rm KK}} \, \bar \theta^\eta (t - 1) \, \sqrt{{\bm
          Y}_{\vec{q}} \hspace{0.25mm} {\bm Y}_{\vec{q}}^\dagger}
    \right ) \bm{C}_n^Q(1)\, \vec a_n^Q
    \,, \\[1mm]
    \bm{C}_n^{\hspace{0.25mm} q}(t)\, \vec a_n^{\hspace{0.25mm} q} & =
    \cosh \left ( \frac{v}{\sqrt{2} M_{\rm KK}} \, \bar \theta^\eta (t
      - 1) \, \sqrt{{\bm Y}_{\vec{q}}^\dagger \hspace{0.25mm} {\bm
          Y}_{\vec{q}}}\right ) \bm{C}_n^{\hspace{0.25mm} q}(1)\, \vec
    a_n^{\hspace{0.25mm} q} \,.
  \end{split}
\end{equation}

 Since the $t$-integration has already been performed in
(\ref{eq:BCssolution}), one can now safely take the limit $\eta \to
0^+$ and trade the on-brane values $\bm{C}^{Q,q}_n (1)$ for the bulk
values $\bm{C}^{Q,q}_n (1^-)$ obtained from the solutions to
(\ref{eq:EOM}) by a limiting procedure. Introducing the rescaled
Yukawa matrices\footnote{Generalizing this result to the case where
$Z_2$-even and -odd fermion fields couple differently to the
brane-localized Higgs sector only requires to perform the replacements
$\bm{Y}_{\vec q} \to \bm{Y}_{\vec q}^{C}$ and $\bm{Y}_{\vec q}^\dagger
\to \bm{Y}_{\vec q}^{S \, \dagger}$, where the superscript in
$\bm{Y}_{\vec q}^{C,S}$ denotes the fields the Higgs couples to. The
same replacement rules also apply in the case of (\ref{eq:gtil1}) to
(\ref{eq:bmh}).}
\begin{equation} \label{eq:Yukresc}
  \bm{\tilde Y}_{\vec q} \equiv \bm{f} \left (
    \frac{v}{\sqrt{2} M_{\rm KK}} \, \sqrt{{\bm Y}_{\vec{q}}
      \hspace{0.25mm} {\bm Y}_{\vec{q}}^\dagger} \right ) \bm{
    Y}_{\vec q} \,, \qquad \bm{f} (\bm{A}) =  
    \tanh \left (\bm{A} \right ) \bm{A}^{-1} \,,
\end{equation}
which coincide at leading order with the original ones, \ie,
$\bm{\tilde Y}_{\vec q} = \bm{Y}_{\vec q} + {\cal O} (v^2/M_{\rm
KK}^2)$, it is then easy to show that the sought IR BCs are manifestly
regularization independent\footnote{It seems instructive to rederive
(\ref{eq:bcIRrescaled}) using a rectangular function to regularize the
$\delta$-functions appearing in (\ref{eq:EOM}). The explicit
calculation is presented in Appendix \ref{app:higgsstuff}. There it is
also shown how to obtain (\ref{eq:gtil1}) to (\ref{eq:bmh}) using the
latter regularization.}  and can be written in $\phi$ coordinates as
\begin{equation} \label{eq:bcIRrescaled}
  \begin{split}
    \bm{S}_n^Q(\pi^-)\, \vec a_n^Q &= \frac{v}{\sqrt2\Mkk}\,\bm{\tilde
      Y}_{\vec q}\;\bm{C}_n^{\hspace{0.25mm}
      q}(\pi^-)\, \vec a_n^{\hspace{0.5mm} q} \,, \\
    - \bm{S}_n^{\hspace{0.25mm} q}(\pi^-)\, \vec a_n^{\hspace{0.5mm}
      q} &= \frac{v}{\sqrt2\Mkk}\,\bm{\tilde Y}_{\vec q}^{
      \dagger}\;\bm{C}_n^Q(\pi^-)\, \vec a_n^Q \,.
  \end{split}
\end{equation}
They hence take precisely the same form as in
\cite{Casagrande:2008hr}, with the original Yukawa couplings replaced
by the rescaled ones as defined in (\ref{eq:Yukresc}).  Since in
practice the Yukawa matrices (together with the quark profiles) are
chosen such that the zero-mode masses as well as the quark mixing
angles match the ones determined by experiment, such a rescaling has
no observable effect on the mass spectrum and the mixing
pattern. However, as we will explain in detail in Section
\ref{sec:higgscouplings}, the inclusion of the Yukawa coupling
involving $Z_2$-odd fermions alters the form of the tree-level
interactions of the Higgs-boson with fermions with respect to the
results derived in \cite{Casagrande:2008hr}.

The $\delta$-distributions in the 5D action (\ref{Sferm2}) forces one
to impose generalized orthonormality conditions for the individual
profiles \cite{Casagrande:2008hr},
\begin{equation} \label{eq:orthonorm}
  \begin{split}
    \int_{-\pi}^\pi\!d\phi\;e^{\sigma(\phi)}\,
    \bm{C}_m^{Q,q}(\phi)\,\bm{C}_n^{Q,q}(\phi) &= \delta_{mn}\,\bm{1}
    + \bm{\Delta C}_{mn}^{Q,q} \,, \\
    \int_{-\pi}^\pi\!d\phi\;e^{\sigma(\phi)}\,
    \bm{S}_m^{Q,q}(\phi)\,\bm{S}_n^{Q,q}(\phi) &= \delta_{mn}\,\bm{1}
    + \bm{\Delta S}_{mn}^{Q,q} \,,
  \end{split}
\end{equation}
where $\bm{1}$ is a unit matrix of dimension $3 \times 3$, $6 \times
6$, or $9 \times 9$, depending on the value of the indices $Q$ or
$q$. We then find that the 4D action reduces to the desired canonical
form if and only if, in addition to the BCs, the relation
\begin{equation} \label{eq:CS}
  \vec a_m^{Q,q\, \dagger} \, \big ( \delta_{mn} \bm 1 + \bm{\Delta
    C}_{mn}^{Q,q} \big ) \; \vec a_n^{Q,q} + \vec a_m^{\hspace{0.5mm}
    q,Q\, \dagger}\,\big (\delta_{mn} \bm 1 + \bm{\Delta
    S}_{mn}^{\hspace{0.25mm} q,Q} \big ) \; \vec a_n^{\hspace{0.5mm} q,Q}
  = \delta_{mn}
\end{equation}
is fulfilled. It is also straightforward to show that the profiles
satisfy
\begin{equation}\label{eq:Deltamn} 
  m_m\,\bm{\Delta C}_{mn}^{Q,q} - m_n\,\bm{\Delta S}_{mn}^{Q,q} =
  \pm\frac{2}{r}\,\bm{C}_n^{Q,q}(\pi^-)\, \bm{S}_m^{Q,q}(\pi^-) \,.
\end{equation}
Since an overall normalization can always be reshuffled between the
profiles ${\bm{ C}}^{Q,q}_n (\phi)$ and ${\bm{S}}^{Q,q}_n (\phi)$ and
the eigenvectors $\vec a_n^{\hspace{0.5mm} Q,q}$, the sum $\bm{\Delta
C}_{nn}^{Q,q} + \bm{\Delta S}_{nn}^{Q,q}$ can be chosen freely without
changing the physical result. The option $\bm{\Delta C}_{nn}^{Q,q} +
\bm{\Delta S}_{nn}^{Q,q}=0$ turns out to be particular convenient and
thus will be adopted hereafter. With this choice (\ref{eq:CS}) splits
into
\begin{equation} \label{abrel} 
  \vec a_n^{Q\, \dagger}\,\vec a_n^Q + \vec a_n^{\hspace{0.5mm} q\,
    \dagger}\,\vec a_n^{\hspace{0.5mm} q} = 1 \,,
\end{equation} 
and 
\begin{equation} \label{eq:magicCS} 
  \vec a_m^{Q,q\, \dagger}\,\bm{\Delta C}_{mn}^{Q,q}\; \vec a_n^{Q,q} +
  \vec a_m^{\hspace{0.5mm} q,Q\, \dagger}\,\bm{\Delta
    S}_{mn}^{\hspace{0.25mm} q,Q}\; \vec a_n^{\hspace{0.5mm} q,Q} = 0\,.
\end{equation}

Finally, the mass eigenvalues $m_n$ follow from the solutions to the
equation
\begin{equation} \label{fermeigenvals} 
  \det\left( \bm{1} + \frac{v^2}{2\Mkk^2} \, \bm{\tilde Y}_{\vec q}\,
    \bm{C}_n^{\hspace{0.25mm} q}(\pi^-) \left [ \bm{S}_n^{\hspace{0.25mm}
        q}(\pi^-) \right]^{-1} \bm{ \tilde Y}_{\vec
      q}^\dagger\,\bm{C}_n^Q(\pi^-) \left [ \bm{S}_n^Q(\pi^-) \right]^{-1}
  \right) = 0 \,.
\end{equation} 
Once they are known, the eigenvectors $\vec a_n^{Q,q}$ can be
determined from (\ref{eq:bcIRrescaled}). Note that, while it is always
possible to work with real and diagonal profiles $\bm{C}_n^{Q,q}
(\phi)$ and $\bm{S}_n^{Q,q} (\phi)$, the vectors $\vec a_n^{Q,q}$ are,
in general, complex-valued objects.

\subsection{Bulk Profiles}
\label{sec:fermionprofiles}

The explicit form of the solutions $\big (C_n^{A(+)} (\phi) \big )_i$
and $ \big (S_n^{A(+)} (\phi) \big )_i$ associated with bulk mass
parameters $M_{A_i}$ was obtained in \cite{Gherghetta:2000qt,
Grossman:1999ra}. The functions $\big (C_n^{A(-)} (\phi) \big )_i$ and
$\big (S_n^{A(-)} (\phi) \big )_i$ can be derived in a similar fashion
by requiring a Dirichlet condition for the even mode, $\big(C_n^{A(-)}
(0) \big )_i = 0$, to account for the additional twist of the
non-SM-like fermions at the UV boundary. The treatment is therefore
analogous to the odd modes of the SM-like fermions, for which
$\big(S_n^{A(+)} (0) \big )_i = 0$. We will drop the label $A$ and the
index $i$ for the purposes of most of the discussion, since they
should be clear from the context. In terms of $t=\epsilon\,e^{\sigma
(\phi)}$, one finds in the bulk (\ie, for $t \in ]\epsilon, 1[$)
\begin{equation} \label{fermprofiles}
  \begin{split}
    C_n^{(\pm)}(\phi) &= {\cal N}_n^{(\pm)}(c)\,\sqrt{\frac{L\epsilon
        t}{\pi}}\, f_n^{(\pm)+}(t,c) \,, \\
    S_n^{(\pm)}(\phi) &= \pm {\cal N}_n^{(\pm)}(c)\,\sgn(\phi)\,
    \sqrt{\frac{L\epsilon t}{\pi}}\,f_n^{(\pm)-}(t,c) \,,
  \end{split}
\end{equation} 
where the overall ``$+$'' sign entering the $Z_2$-odd profiles holds
if $c = c_Q \equiv +M_{Q}/k$ refers to the bi-doublet, while the
``$-$'' sign applies in the case of $c = c_A \equiv -M_A/k$, where $A
= u^c,{\cal T}_1,{\cal T}_2$. The functions $f^{(\pm)\pm}_n (t,c)$ are
given by
\begin{equation} \label{fplmi}
  \begin{split}
    f_n^{(+)\pm}(t,c) & =
    J_{-\frac12-c}(x_n\epsilon)\,J_{\mp\frac12+c}(x_n t) \pm
    J_{+\frac12+c}(x_n\epsilon)\,J_{\pm\frac12-c}(x_n t) \,, \\
    f_n^{(-)\pm}(t,c) & =
    J_{+\frac12-c}(x_n\epsilon)\,J_{\mp\frac12+c}(x_n t) \mp
    J_{-\frac12+c}(x_n\epsilon)\,J_{\pm\frac12-c}(x_n t) \,.
  \end{split}
\end{equation} 
They satisfy the equalities 
\begin{equation} \label{eq:frel}
  f_n^{(+)+}(t,c)=f_n^{(-)-}(t,-c)\,, \qquad 
  f_n^{(+)-}(t,c)=-f_n^{(-)+}(t,-c)\,.
\end{equation}
The orthonormality relations (\ref{eq:orthonorm}) imply the
normalization conditions
\begin{equation} 
  2\int_\epsilon^1\!dt\,t \left[ f_n^{(a)\pm}(t,c)
  \right]^2 = \frac{1}{\big [ {\cal N}_n^{(a)}(c) \big ]^2} \pm
  \frac{f_n^{(a)+}(1,c)\,f_n^{(a)-}(1^-,c)}{x_n} \,, 
\end{equation} 
where $a=\pm$. From these relations we derive
\begin{equation} \label{fermnorm}
  \begin{split}
    \big[{\cal N}_n^{(a)}(c)\big]^{-2} & = \left[ f_n^{(a)+}(1,c)
    \right]^2 + \left[ f_n^{(a)-}(1^-,c) \right]^2 \\
    & \phantom{xx} -
    \frac{2c}{x_n}\,f_n^{(a)+}(1,c)\,f_n^{(a)-}(1^-,c) - \epsilon^2
    \left ( \left[ f_n^{(a)+}(\epsilon,c) \right]^2 + \left[
        f_n^{(a)-}(\epsilon^+,c) \right]^2 \right ) ,
  \end{split}
\end{equation}
which extends the result obtained in \cite{Casagrande:2008hr} to the
case of $Z_2$-odd profiles with non-zero value at the UV boundary. For
the special cases where $c+1/2$ is an integer, the profiles must be
obtained from the above relations by a limiting procedure.

For the SM fermions, it is a very good approximation to expand the
profiles in the limit $x_n\ll 1$, since even the top-quark mass is
much lighter than the KK scale. We will hereafter refer to such an
expansion as the \quotes{zero-mode approximation} (ZMA).  Using
results from \cite{Casagrande:2008hr} in combination with
(\ref{eq:frel}), we obtain
\begin{equation} \label{eq:profileexp}
\begin{split}  
  C_n^{(+)}(\phi) \approx \sqrt{\frac{L\epsilon}{\pi}}\, F(c) \, t^{c}
  \, , \qquad S_n^{(+)}(\phi) \approx
  \pm\sgn(\phi)\,\sqrt{\frac{L\epsilon}{\pi}}\; x_n F(c)\;
  \frac{t^{1+c} - \epsilon^{1+2c}\,t^{-c}} {1+2c} \,,
  \quad \\[1mm]
  C_n^{(-)}(\phi) \approx -\sqrt{\frac{L\epsilon}{\pi}}\; x_n F(-c)\;
  \frac{t^{1-c} - \epsilon^{1-2c}\,t^{c}} {1-2c} \, , \qquad
  S_n^{(-)}(\phi) \approx \pm\sgn(\phi) \sqrt{\frac{L\epsilon}{\pi}}\,
  F(-c) \, t^{-c} \,, 
\end{split}
\end{equation}
where we have introduced the zero-mode profile
\cite{Gherghetta:2000qt, Grossman:1999ra}
\begin{equation} \label{Fdef} 
  F(c) \equiv \sgn \mkern+1mu \big[ \mkern-3mu \cos(\pi c) \big]\,
  \sqrt{\frac{1+2c}{1-\epsilon^{1+2c}}} \,.
\end{equation} 
The sign factor in (\ref{Fdef}) is chosen such that the signs in
(\ref{eq:profileexp}) agree with those derived from the exact profiles
(\ref{fermprofiles}). Notice that the profiles $C_n^{(+)} (\phi)$ and
$S_n^{(-)} (\phi)$ are of ${\cal O}(1)$, while $C_n^{(-)} (\phi)$ and
$S_n^{(+)} (\phi)$ are of ${\cal O} (v/M_{\rm KK})$. As we will
explain in detail in the next section, this feature will allow to
partially shield the $Z b_L \bar b_L$ and $Z d_L^{\hspace{0.25mm} i}
\bar d^{\hspace{0.25mm} j}_L$ vertices from corrections due to mixing
of zero-mode quarks with their KK excitations.

The quantity $F(c)$ depends strongly on the value of $c$. One obtains
\begin{equation}
  F(c) \approx \begin{cases} -\sqrt{-1-2c}\,\, \epsilon^{-c-\frac12} \,,
    & -3/2<c<-1/2 \,, \\[4mm] \sqrt{1+2c} \,, & -1/2<c<1/2 \,, \end{cases}
\end{equation}
which implies that for UV-localized fermions the corresponding
zero-mode profile is exponentially small, while it is of ${\cal O}
(1)$ for IR-localized fields.

\section{Gauge Interactions with Fermions}
\label{sec:custodialprotection}

In this section we reexamine how a protection of the left-handed
down-type couplings of the $Z$ boson can be achieved by choosing an
appropriate embedding of the fermions into the enlarged gauge group in
the bulk.  We also derive the four-fermion charged-current
interactions and show explicitly that a custodial protection is not at
work in this case. To start, we give the covariant derivative in the
UV basis,
\begin{align}
  \begin{split}
    D_\mu=\partial_\mu & - i \, \frac{{g_L}_5}{\sqrt 2}
    \left(L_\mu^+\hspace{0.25mm} T_L^++L_\mu^-\hspace{0.25mm}
      T_L^-\right) + i \, \frac{{g_R}_5}{\sqrt 2} \left(R_\mu^+
      \hspace{0.25mm}
      T_R^++R_\mu^- \hspace{0.25mm} T_R^-\right) \\
    & - i \, {g_Z}_5\, Q_Z \hspace{0.25mm} Z_\mu - i \, {g_{Z'}}_5 \,
    Q_{Z^{\, \prime}} \hspace{0.25mm} Z_\mu^\prime - i \, e_5
    \hspace{0.25mm} Q A_\mu \, .
  \end{split}
\end{align}
The $Z$-boson couplings to fermionic currents are defined by
\begin{equation} \label{eq:gQZZp}
  \begin{aligned}
    & g_Z=\sqrt{g_L^2+g_Y^2}\,, && \qquad g_{Z^{\, \prime}} =
    \sqrt{g_R^2+g_X^2}\,, \\
    & Q_Z =T_L^3 - \frac{g_Y^2}{g_Z^2}Q\,, && \qquad Q_{Z^{\,
        \prime}}= -T_R^3 - \frac{g_X^2}{g_{Z^{\, \prime}}^2}Y\,.
  \end{aligned}
\end{equation}
At this point we have to specify the form of the $SU(2)_{L,R}$
generators when acting on different fermion representations. Acting on
fermion bi-doublets, the generators $T_{L,R}^i$ with $i=1,2,3$ are
given by the Pauli matrices in the standard convention times a factor
of $1/2$. As usual we define $T_{L,R}^\pm=T_{L,R}^1 \pm i \,
T_{L,R}^2$. If, on the other hand, the generators act on $SU(2)_{
L,R}$ triplets, then one has
\begin{equation}
  T_{L,R}^+=\begin{pmatrix} \,0\, & \sqrt 2 & 0 \\
    \,0\, & 0 & \sqrt 2 \\
    \,0\, & 0 & 0
  \end{pmatrix} ,\qquad
  T_{L,R}^-=\begin{pmatrix} 0 & 0\, & 0\, \\
    \sqrt 2 & 0\, & 0\, \\
    0 & \sqrt 2\, & 0\,
  \end{pmatrix} ,\qquad
  T_{L,R}^3=\begin{pmatrix} \,1\; & 0 & 0 \\
    \,0\; & 0 & 0 \\
    \,0\; & 0 & -1
  \end{pmatrix} .
\end{equation}
Note that the current operators involve a trace with respect to the
fundamental gauge indices, and that again the $T_R^i$ act from the
right. It will furthermore turn out to be useful to introduce
\begin{equation}
  \vec{g}_Z = \begin{pmatrix} g_Z \hspace{0.25mm} Q_Z \\
    g_{Z^\prime} \hspace{0.25mm} Q_{Z^{\, \prime}} \end{pmatrix} ,
\end{equation}
as well as the charged-current vectors
\begin{equation} \label{eq:currents}
  \begin{split}
    \vec J_{W_Q}^{\, \mu\pm}&=\frac 1{\sqrt 2}\, \big(g_L\,
    \text{Tr}\left[\bar Q\,\gamma^\mu\,T^\pm\,Q\right],
    g_R\, \text{Tr}\left[\bar Q\,\gamma^\mu\,Q\, T^\pm\right]\big)\,,\\
    \vec J_{W_{\cal T}}^{\, \mu\pm}& = \frac 1{\sqrt 2}\, \big( g_L
    \bar {\cal T}_1\,\gamma^\mu\,T^\pm\,{\cal T}_1\, ,\, g_R\,
    \text{Tr}\left[\bar {\cal T}_2\,\gamma^\mu\,{\cal T}_2
      \,T^\pm\right] \big)\,,
  \end{split}
\end{equation}
which act on $\left(L_\mu^\pm\ , R_\mu^\pm \right)^T$ from the left.
These latter formulas are needed to calculate the corrections to the
quark-mixing matrices, which we postpone until
Section~\ref{sec:4Fint}.

\subsection{Custodial Protection: Gauge-Boson Contributions}

Using (\ref{eq:expprof}) and (\ref{eq:vecA0a}), we find that the
coupling of the $Z$ boson to a quark current is proportional to
\begin{eqnarray} \label{eq:Zqq}
  \big( \vec{g}_Z^{\, q} \big)^{\mkern-3mu T} \vec{\chi}^{\, Z}_0 (\phi)
  = \frac{g_Z \hspace{0.25mm} Q_Z^{\, q}}{\sqrt{2 \pi}} \left \{ 1 +
    \frac{m_Z^2}{4 M_{\rm KK}^2} \left [ \, 1 - \frac{1}{L} - 2 L \, t^2
      \omega_Z^q + 2 \, t^2 \!\left( \frac12 - \ln t \right) \right ] \right
  \} + \ord\left(\frac{m_Z^4}{\Mkk^4}\right) , \hspace{6mm}
\end{eqnarray}
with 
\begin{equation} \label{eq:omegaZ_definition}
\omega_Z^q = 1 - \frac{s_Z}{c_Z} \, \frac{g_{Z^{\, \prime}}
\hspace{0.25mm} Q^q_{Z^{\, \prime}} }{ g_Z \hspace{0.25mm} Q_Z^q
}\,.
\end{equation} 
This is an important result, as it allows us to understand the
custodial protection of the $Z b_L\bar b_L$ vertex. Note that the term
in (\ref{eq:Zqq}) that is enhanced by the volume factor $L$ gets
modified by a prefactor $\omega_Z^q$, \ie, a combination of the
fundamental charges and couplings. While $\omega_Z^q = 1$ for all
quarks in the minimal RS model, it is possible to arrange for
\begin{equation} \label{eq:custodialQ}
  \omega_Z^{b_L} = 0 \quad \Longleftrightarrow \quad g_Z \hspace{0.25mm}
  Q_Z^{b_L} = \frac{s_Z}{c_Z} \, g_{Z^{\, \prime}} \hspace{0.5mm}
  Q^{b_L}_{Z^{\, \prime}} \;,
\end{equation}
by virtue of the extension of the gauge group in the bulk. It is
interesting to observe that only the leading term in $L$ can be
protected, while no such mechanism is available for the subleading
terms in $L$, since they arise from the fact that the fields
$\chi_0^{(\pm)} (t)$ obey different BCs. The latter effects hence
represent an irreducible source of $P_{LR}$ symmetry breaking.
Numerically, the corrections to the $Z b_L \bar b_L$ vertex arising
from the gauge sector are thus suppressed by a factor of $L\approx 37$
in the $SU(2)_L \times SU(2)_R \times P_{LR}$ custodial model relative
to the minimal RS model.

Formula (\ref{eq:omegaZ_definition}) can be recast into the form
\begin{equation} \label{eq:omegaZ}
  \omega_Z^q = \frac{c_w^2}{2 g_L^2} \, \frac{ \, (g_L^2 + g_R^2) \,
    (T_L^{3 \, q} + T_R^{3 \, q}) + (g_L^2 - g_R^2) \, (T_L^{3 \, q} -
    T_R^{3 \, q}) \, } { T_L^{3 \, q} - s_w^2 \hspace{0.25mm} Q_q } \,,
\end{equation}
which allows one to read off that the choices 
\begin{equation} \label{PC}
  T_L^{3 \, q} = T_R^{ 3 \, q} = 0 \,, \qquad (P_C \ \text{symmetry})
\end{equation} 
and 
\begin{equation} \label{PLR}
  g_L = g_R\,, \qquad T_L^{3 \, q} = - T_R^{ 3 \, q} \,, \qquad
  (P_{LR} \ \text{symmetry})
\end{equation}
are suitable to protect the $Z$-boson vertices from receiving
$L$-enhanced corrections. Since the representation
(\ref{eq:multiplets}) features $T_L^{3\, d_L} = -T_R^{3\, d_L} = -1/2$
and $T_L^{3\, u_R} = T_R^{3\, u_R} = 0$, it is then immediately clear
that the $Z d_L^{\hspace{0.25mm} i} \bar d_L^{\hspace{0.25mm} j}$ and
$Z u_R^{\hspace{0.25mm} i} \bar u_R^{\hspace{0.25mm} j}$ vertices are
protected to leading order in $L$ by the $P_{LR}$ and $P_C$
symmetries, respectively. On the other hand, the $Z
d_R^{\hspace{0.25mm} i} \bar d_R^{\hspace{0.25mm} j}$ and $Z
u_L^{\hspace{0.25mm} i} \bar u_L^{\hspace{0.25mm} j}$ vertices do
receive $L$-enhanced corrections, since the corresponding quantum
numbers are $T_L^{3\, d_R} = 0$, $T_R^{3\, d_R} = 1$ and $T_L^{3\,
u_L} = T_R^{3\, u_L} = 1/2$. We also add that devising the quark
sector as in (\ref{eq:multiplets}) implies $\omega_Z^{b_R} > 0$, so
that the shift in the right-handed $Z$-boson coupling to bottom quarks
arising from the gauge-boson sector is predicted to be strictly
negative. This suggests that the well-known tension in the global fit
to the $Z \to b \bar b$ pseudo observables cannot be softened in the
model under considerations. We will come back to this point in Section
\ref{sec:Zbbnumerics}.

\subsection{Fermion Couplings to the $\bm Z$ Boson}

We turn to the phenomenologically relevant expressions for weak gauge
interactions of quark currents in the custodial model and identify
the RS contribution at relative order $v^2/\Mkk^2$. The $Z$-boson
couplings to left- and right-handed quarks can be read off the
Lagrangian
\begin{equation} \label{eq:Zff}
  {\cal L}_{\rm 4D} \, \ni \, \frac{g_L}{c_w} \left[ 1 +
    \frac{m_Z^2}{4\Mkk^2} \left( 1 - \frac{1}{L} \right) \right] \;
  \sum_{q,m,n} \Big[ \big( g_L^{\, q} \big)_{m n} \left ( \bar
    q_{L}^{\hspace{0.25mm} m} \gamma_\mu q_{L}^n \right ) + \big( g_R^{\,
    q} \big)_{mn} \left ( \bar q_{R}^{\hspace{0.25mm} m} \gamma_\mu
    q_{R}^n \right ) \Big] Z^\mu \,,
\end{equation}
where the prefactor accounts for a universal correction due to the
$t$-independent terms in (\ref{eq:Zqq}).  The left- and right-handed
couplings $\bm{g}_{L,R}^q$ are infinite-dimensional matrices in the
space of quark modes, and can be parametrized as
\begin{eqnarray}\label{gLR}
  \begin{split}
    \bm{g}_L^q &= \left(T_L^{3 \, q_L} - s_w^2 Q_q\right) \left[
      \bm{1} - \frac{m_Z^2}{2\Mkk^2} \left ( \omega_Z^{q_L} L
        \hspace{0.5mm} \bm{\Delta}_Q -\bm{\Delta}'_Q \right )\right]
    -\bm{\delta}_Q + \frac{m_Z^2}{2\Mkk^2} \,\left (
      \frac{c_w^2}{g_L^2} \, L \hspace{0.5mm} \bm{\varepsilon}_Q -
      \bm{\varepsilon}'_Q \right ) ,
    \hspace{6mm} \\
    \bm{g}_R^q &= -s_w^2 Q_q \left[ \bm{1} - \frac{m_Z^2}{2\Mkk^2}
      \left ( \omega_Z^{q_R} L \hspace{0.5mm} \bm{\Delta}_q -
        \bm{\Delta}'_q \right ) \right ] +\bm{\delta}_q
    -\frac{m_Z^2}{2\Mkk^2} \left ( \frac{c_w^2}{g_L^2} \, L
      \hspace{0.5mm} \bm{\varepsilon}_q - \bm{\varepsilon}'_q \right),
  \end{split}
\end{eqnarray}
where the charges appearing in the expressions are understood to be
the ones of the zero-mode fermions. In $\bm{g}_L^q$ they read $T_L^{3
  \, u_L} (= T_L^{3 \, u}) = T_R^{3 \, u_L} (= T_R^{3 \, u}) = T_R^{3
  \, d_L} (= T_R^{3 \, d}) = 1/2$ and $T_L^{3 \, d_L} (= T_L^{3 \, d})
= ~-~1/2$, whereas for $\bm{g}_R^q$ one has $T_L^{3 \, u_R} (= T_L^{3
  \, u^c}) = T_R^{3 \, u_R} (= T_R^{3 \,u^c}) = T_L^{3 \, d_R} (=
T_L^{3 \, D}) = 0$ and $T_R^{3 \, d_R} (= T_R^{3 \, D}) = 1$. The
quoted numerical values correspond to the choice
(\ref{eq:multiplets}). We do not consider the sector of $\lambda$ and
$\Lambda^{(\prime)}$ quarks at this point, as these fields do not
possess zero modes. The $L$-enhanced term proportional to
$\omega_Z^{q}$ vanishes for the assignments (\ref{PC}) and
(\ref{PLR}), making the custodial protection explicit. Following
\cite{Casagrande:2008hr}, we have split the corrections to the
$Z$-boson couplings into leading contributions in the ZMA, denoted by
$\bm{\Delta}^{(\prime)}_{Q,q}$, and subleading ones, parametrized by
$\bm{\varepsilon}^{(\prime)}_{Q,q}$. The elements of the leading-order
matrices ${\bm \Delta}^{(\prime)}_{Q,q}$ are defined as
\begin{equation}\label{overlapintsLO}
  \begin{split} 
    \left( \Delta_Q \right)_{mn} &= \frac{2\pi}{L\epsilon}
    \int_\epsilon^1\!dt\,t^2 \, \Big [ \vec a_m^{\hspace{0.25mm} Q
      \dagger}\,\bm{C}_m^{Q}(t)\, \bm{C}_n^{Q}(t)\,\vec
    a_n^{\hspace{0.25mm} Q} + \vec a_m^{\hspace{0.25mm}
      q\dagger}\,\bm{S}_m^{q}(t)\,
    \bm{S}_n^{q}(t)\,\vec a_n^{\hspace{0.25mm}q} \Big] , \\
    \left( \Delta_q \right)_{mn} &= \frac{2\pi}{L\epsilon}
    \int_\epsilon^1\!dt\,t^2 \, \Big[ \vec
    a_m^{\hspace{0.25mm}q\dagger}\,\bm{C}_m^{q}(t)\,
    \bm{C}_n^{q}(t)\,\vec a_n^{\hspace{0.25mm}q} + \vec
    a_m^{\hspace{0.25mm} Q\dagger}\,\bm{S}_m^{Q}(t)\,
    \bm{S}_n^{Q}(t)\,\vec a_n^{\hspace{0.25mm}Q} \Big] , \\
    \left( \Delta'_Q \right)_{mn} &= \frac{2\pi}{L\epsilon}
    \int_\epsilon^1\!dt\,t^2 \left( \frac12 - \ln t \right) \Big [
    \vec a_m^{\hspace{0.25mm}Q\dagger}\,\bm{C}_m^{Q}(t)\,
    \bm{C}_n^{Q}(t)\,\vec a_n^{\hspace{0.25mm}Q} + \vec
    a_m^{\hspace{0.25mm}q\dagger}\,\bm{S}_m^{q}(t)\,
    \bm{S}_n^{q}(t)\,\vec a_n^{\hspace{0.25mm}q} \Big] , \\
    \left( \Delta'_q \right)_{mn} &= \frac{2\pi}{L\epsilon}
    \int_\epsilon^1\!dt\,t^2 \left( \frac12 - \ln t \right) \Big[ \vec
    a_m^{\hspace{0.25mm} q\dagger}\,\bm{C}_m^{q}(t)\,
    \bm{C}_n^{q}(t)\,\vec a_n^{\hspace{0.25mm} q} + \vec
    a_m^{\hspace{0.25mm} Q\dagger}\,\bm{S}_m^{Q}(t)\,
    \bm{S}_n^{Q}(t)\,\vec a_n^{\hspace{0.25mm} Q} \Big] ,
  \end{split}
\end{equation} 
while the elements of the matrices ${\bm\varepsilon}^{(\prime)}_{Q,q}$
describing subleading effects take the form
\begin{eqnarray}\label{overlapintsNLO}
  \begin{split}
    \left( \varepsilon_Q \right)_{mn} &= \frac{2\pi}{L\epsilon}
    \int_\epsilon^1\!dt\,t^2 \, \Big[ \vec a_m^{\hspace{0.25mm} Q
      \dagger}\,\bm{C}_m^{Q}(t) \, \Big \{g_L^2 \hspace{0.5mm}
    \Big(T_L^{3 \, q_L}{{\bm 1}-{\bm T}_L^{3 \, Q}}\Big)+
    g_R^2\hspace{0.5mm} \Big(T_R^{3 \, q_L}{{\bm 1}-{\bm T}_R^{3 \,
        Q}}\Big)\Big\}\,
    \bm{C}_n^{Q}(t)\, \vec a_n^{\hspace{0.25mm} Q}\\
    &\qquad \qquad \qquad + \vec
    a_m^{\hspace{0.25mm}q\dagger}\,\bm{S}_m^{q}(t)\, \Big \{g_L^2
    \hspace{0.5mm} \Big (T_L^{3 \, q_L}{\bm 1}-{\bm T}_L^{3 \, q}\Big
    )+ g_R^2 \hspace{0.5mm} \Big (T_R^{3 \, q_L}{\bm 1}-{\bm T}_R^{3\,
      q}\Big
    )\Big \} \, \bm{S}_n^{q}(t)\, \vec a_n^{\hspace{0.25mm}q} \Big] , \\
    \left( \varepsilon_q \right)_{mn} &= \frac{2\pi}{L\epsilon}
    \int_\epsilon^1\!dt\,t^2 \, \Big[ \vec a_m^{\hspace{0.25mm} q
      \dagger}\,\bm{C}_m^{q}(t) \, \Big \{g_L^2 \hspace{0.5mm} {\bm
      T}_L^{3 \, q}- g_R^2\hspace{0.5mm} \Big(T_R^{3 \, q_R}{{\bm
        1}-{\bm T}_R^{3 \,
        q}}\Big)\Big\}\, \bm{C}_n^{q}(t)\, \vec a_n^{\hspace{0.25mm} q}\\
    &\qquad \qquad \qquad + \vec
    a_m^{\hspace{0.25mm}Q\dagger}\,\bm{S}_m^{Q}(t)\, \Big \{g_L^2
    \hspace{0.5mm} {\bm T}_L^{3 \, Q}- g_R^2 \hspace{0.5mm} \Big
    (T_R^{3 \, q_R}{\bm 1}-{\bm T}_R^{3\, Q}\Big )\Big \} \,
    \bm{S}_n^{Q}(t)\,
    \vec a_n^{\hspace{0.25mm}Q} \Big] , \hspace{6mm} \\
    \left( \varepsilon'_Q \right)_{mn} &= \frac{2\pi}{L\epsilon}
    \int_\epsilon^1\!dt\,t^2 \left( \frac12 - \ln t \right) \Big[ \vec
    a_m^{\hspace{0.25mm} Q\dagger}\,\bm{C}_m^{Q}(t)\, \Big(T_L^{3 \,
      q_L}{\bm 1}-{\bm T}_L^{3 \, Q}\Big) \bm{C}_n^{Q}(t)\, \vec
    a_n^{\hspace{0.25mm} Q}\\
    &\hspace{46.5mm} + \vec a_m^{\hspace{0.25mm}
      q\dagger}\,\bm{S}_m^{q}(t)\, \Big(T_L^{3\, q_L}{\bm 1}-{\bm
      T}_L^{3\,
      q}\Big)\bm{S}_n^{q}(t)\, \vec a_n^{\hspace{0.25mm}q} \Big] , \\
    \left( \varepsilon'_q \right)_{mn} &= \frac{2\pi}{L\epsilon}
    \int_\epsilon^1\!dt\,t^2 \left( \frac12 - \ln t \right) \Big[ \vec
    a_m^{\hspace{0.25mm}q\dagger}\,\bm{C}_m^{q}(t)\, {\bm T}_L^{3\,
      q}\, \bm{C}_n^{q}(t)\,\vec a_n^{\hspace{0.25mm} q} + \vec
    a_m^{\hspace{0.25mm} Q\dagger}\,\bm{S}_m^{Q}(t)\, {\bm
      T}_L^{3\,Q}\, \bm{S}_n^{Q}(t)\,\vec a_n^{\hspace{0.25mm} Q}
    \Big] \,. \hspace{11mm}
  \end{split}
\end{eqnarray} 
Finally, the elements of the matrices $\bm{\delta}_{Q,q}$, which arise
because of the non-orthonormality of the quark profiles and describe
mixings between the different multiplets, read
\begin{align} \label{eq:delta2}
    \left( \delta_Q \right)_{mn} &= \frac{2\pi}{L\epsilon}
    \int_\epsilon^1\!dt \, \Big[ \vec a_m^{\hspace{0.25mm} Q
      \dagger}\,\bm{C}_m^{Q}(t)\, \Big(T_L^{3 \, q_L}{\bm 1}-{\bm
      T}_L^{3\,
      Q}\Big) \, \bm{C}_n^{Q}(t)\,\vec a_n^{\hspace{0.25mm} Q} \nonumber \\
    & \hspace{2.1cm} + \vec a_m^{\hspace{0.25mm}
      q\dagger}\,\bm{S}_m^{q}(t) \, \, \Big (T_L^{3 \, q_L}{\bm
      1}-{\bm T}_L^{3\, q} \Big) \, \bm{S}_n^{q}(t)\,\vec
    a_n^{\hspace{0.25mm} q}
    \Big ] , \nonumber \\
    \left( \delta_q \right)_{mn} &= \frac{2\pi}{L\epsilon}
    \int_\epsilon^1\!dt \, \Big [ \vec a_m^{\hspace{0.25mm}
      q\dagger}\,\bm{C}_m^{q}(t)\, {\bm T}_L^{3\,q}\,
    \bm{C}_n^{q}(t)\,\vec a_n^{\hspace{0.25mm} q} + \vec
    a_m^{\hspace{0.25mm} Q\dagger}\,\bm{S}_m^{Q}(t)\, {\bm T}_L^{3\,
      Q}\, \bm{S}_n^{\hspace{0.25mm} Q}(t)\,\vec a_n^{\hspace{0.25mm}
      Q} \Big] .
\end{align} 
In the expressions above we have used the charge matrices ${\bm
T}_{L,R}^{3 \, Q,q}$, defined as
\begin{equation}
  \begin{split}
    {\bm T}_{L,R}^{3\, U} & =\left(\begin{array}{cc}
        T_{L,R}^{3 \, u}&0\\
        0&T_{L,R}^{3 \, u^\prime}
      \end{array}\right) , \qquad 
    {\bm T}_{L,R}^{3 \, u}=\left(\begin{array}{ccc}
        T_{L,R}^{3 \, u^c}&0&0\\
        0&T_{L,R}^{3 \, U^\prime}&0\\
        0&0&T_{L,R}^{3 \, U}
      \end{array}\right) , \\
    {\bm T}_{L,R}^{3 \, D} & =T_{L,R}^{3 \, d}\, , \qquad \qquad
    \qquad \phantom{i} {\bm T}_{L,R}^{3 \, d}=\left(\begin{array}{cc}
        T_{L,R}^{3 \, D}&0\\
        0&T_{L,R}^{3 \, D^\prime}
      \end{array}\right) .
  \end{split}
\end{equation}
One can easily check that for our choice (\ref{eq:multiplets}) the
quantities $\bm{\varepsilon}^{(\prime)}_{Q,q}$ are indeed suppressed
by $v^2/\Mkk^2$ with respect to the matrices
$\bm{\Delta}^{(\prime)}_{Q,q}$. Note that with the chosen embedding
(\ref{eq:multiplets}), the matrices ${\bm T}_{L,R}^{3 \, u}$ vanish
identically.

\subsection{Custodial Protection: Fermionic Contributions}

Finally, we want to have a look at the custodial protection of the $Z
b_L \bar b_L$ vertex from effects arising from quark mixings,
parametrized by $\bm{\delta}_{Q,q}$. These objects scale in general as
$v^2/\Mkk^2$, but as they come with an $\ord(1)$ coefficient in
(\ref{gLR}) they are parametrically of the same order as the matrices
$\bm{\Delta}^{(\prime)}_{Q,q}$. In the case of the left-handed
down-type quark sector, one has
\begin{equation} \label{deltaD}
  \begin{split}
    \left( \delta_D \right)_{mn} &= \frac{2\pi}{L\epsilon}
    \int_\epsilon^1\!dt \, \Big[a_m^{ D\dagger}\,\bm{S}_m^{{\cal
        T}_2(+)}(t)\,
    \Big(T_L^{3\hspace{0.25mm}d_L}-T_L^{3\hspace{0.25mm}D}\Big)
    \bm{S}_n^{{\cal T}_2(+)}(t)
    \,a_n^{ D}\\
    &\hspace{22mm}+a_m^{D^\prime\dagger}\,\bm{S}_m^{{\cal
        T}_1(-)}(t)\, \Big(T_L^{3 \, d_L}-T_L^{3 \, D^\prime}\Big)
    \bm{S}_n^{{\cal T}_1(-)}(t)
    \,a_n^{D^\prime} \Big]\\
    &=- \frac 1 2 \, \frac{2\pi}{L\epsilon} \int_\epsilon^1\!dt \,
    \Big [a_m^{D\dagger}\,\bm{S}_m^{{\cal T}_2(+)}(t) \,
    \bm{S}_n^{{\cal T}_2(+)}(t)\,a_n^{D}
    -a_m^{D^\prime\dagger}\,\bm{S}_m^{{\cal T}_1(-)}(t)\,
    \bm{S}_n^{{\cal T}_1(-)}(t)\,a_n^{D^\prime} \Big] \,, \hspace{6mm}
  \end{split}
\end{equation} 
where in the second step we have inserted the quantum numbers
corresponding to our choice (\ref{eq:multiplets}) of multiplets. 

The relative sign between the two terms in the second line of
(\ref{deltaD}) suggests that also for the corrections due to quark
mixing a custodial protection mechanism could be at work. To see if
this is indeed the case, let us derive the ZMA expression for ${\bm
\delta}_D$. Using the approximate expressions (\ref{eq:profileexp}),
the system of equations (\ref{eq:bcIRrescaled}) can be brought into
the form
\begin{equation} \label{IRBCZMA}
  \frac{\sqrt 2\, m_n}{v} \, {\hat a}_n^d = {\bm Y}_d^{\rm eff} \, {\hat
    a}_n^D \,, \qquad \frac{\sqrt 2\, m_n}{v} \, {\hat a}_n^D = \left (
    {\bm Y}_d^{\rm eff} \right )^\dagger {\hat a}_n^d \,,
\end{equation}
and 
\begin{equation} \label{IRBC3}
  \hat a_n^{D^\prime} = \frac{m_n}{M_{\rm KK}} \, {\rm diag} \; \Big (
  F^{-1}(c_{{\cal T}_{2 i}}) \hspace{0.5mm} F^{-1}(-c_{{\cal T}_{1 i}})
  \Big ) \, \hat a_n^{D} \,,
\end{equation}
where the diagonal matrix contains the entries shown in brackets. We
have furthermore defined the effective Yukawa couplings
\begin{equation}\label{eq:effY}
  ({\bm Y}_d^{\rm eff} )_{ij} \equiv F(c_{Q_i}) \left (Y_d \right )_{ij}
  F(c_{{\cal T}_{2 j}}) \,,
\end{equation}
and the rescaled vectors $\hat a_n^A \equiv \sqrt{2} \, a_n^A$ with $A
=d,D,D^\prime$, which obey the normalization conditions
\begin{equation}
  \hat a_n^{D \hspace{0.5mm} \dagger} \, \hat a_n^{D} = 1 \, , \qquad
  \hat a_n^{d \hspace{0.5mm} \dagger} \, \hat a_n^{d} + \hat
  a_n^{D^\prime \hspace{0.5mm} \dagger} \, \hat a_n^{D^\prime} = 1 \,.
\end{equation}
Moreover, we obtain from (\ref{IRBCZMA}) the equalities
\begin{equation}\label{eq:EVeq1}
  \Big ( m_n^2\,\bm{1} - \frac{v^2}{2} \, \bm{Y}_d^{\rm eff} \left
    (\bm{Y}_d^{\rm eff} \right )^\dagger \Big ) \, \hat a_n^{d} = 0 \,,
  \qquad \Big ( m_n^2\,\bm{1} - \frac{v^2}{2} \, \left (\bm{Y}_d^{\rm
      eff} \right )^\dagger \bm{Y}_d^{\rm eff} \Big) \, \hat a_n^{D} = 0 \,,
\end{equation}
and the mass eigenvalues are the solutions to the equation 
\begin{equation}\label{eq:EVeq2}
  \det\left( m_n^2\,\bm{1} - \frac{v^2}{2} \, \bm{Y}_d^{\rm eff} \left
      (\bm{Y}_d^{\rm eff} \right )^\dagger \right) = 0 \,,
\end{equation}
which implies that to leading order in $v/M_{\rm KK}$ the values $m_n$
are unaffected by the presence of the $D^\prime$ quarks embedded in 
the multiplet ${\cal T}_1$. 
Notice that in the ZMA, but not in general, the vectors $\hat a_n^{d}$ and
$\hat a_n^{D}$ belonging to different $n$ are orthogonal on each
other.

The eigenvectors $\hat{a}_n^{d}$ and $\hat{a}_n^{D}$ with $n=1,2,3$ of
the matrices ${\bm Y}_d^{\rm eff} \left( {\bm Y}_d^{\rm eff}
\right)^\dagger$ and $\left( {\bm Y}_d^{\rm eff} \right)^\dagger {\bm
Y}_d^{\rm eff}$ form the columns of the unitary matrices ${\bm U}_d$
and ${\bm W}_d$ appearing in the singular-value decomposition
\begin{equation} \label{eq:singular}
  \bm{Y}_d^{\rm eff} = \bm{U}_d\,\bm{\lambda}_d\,\bm{W}_d^\dagger \,,
\end{equation}
where
\begin{equation} \label{eq:lambdad}
  \bm{\lambda}_d = \frac{\sqrt 2}{v}\, \mbox{diag} \, (m_d,m_s,m_b)
\end{equation}
is a diagonal matrix containing the masses of the SM down-type quarks
in units of $v/\sqrt2$. Similar relations hold in the up-type quark
sector and will be given explicitly in Section
\ref{sec:higgscouplings}. It follows that in the ZMA the relations
between the original 5D fields and the SM mass eigenstates involve the
matrices $\bm{U}_{u,d}$ and $\bm{W}_{u,d}$. In particular, the
Cabibbo-Kobayashi-Maskawa (CKM) matrix is given by $\bm{V}_{\rm CKM} =
\bm{U}_u^\dagger\,\bm{U}_d$.

With these results at hand, it is a matter of simple algebra to find
the expression for ${\bm \delta}_D$\ in the ZMA. Working to first
order in $v^2/M_{\rm KK}^2$, and using (\ref{eq:profileexp}) and
(\ref{IRBC3}) we arrive at
\begin{eqnarray} \label{deltaDZMA}
  {\bm \delta}_D = -\frac{1}{2} \, {\bm x}_d \, {\bm W}_d^\dagger \, \,
  {\rm diag } \left [ \frac{1}{1 - 2 \hspace{0.25mm}c_{{{\cal T}}_{2
          i}}} \left ( \frac{1}{F^2(c_{{{\cal T}}_{2 i}})} \left [ 1 - \frac{1 -
          2 \hspace{0.25mm} c_{{{\cal T}}_{2 i}}}{F^2(-c_{{{\cal T}}_{1 i}})}
      \right ] - 1 + \frac{F^2(c_{{{\cal T}}_{2 i}})}{3 + 2 \hspace{0.25mm}
        c_{{{\cal T}}_{2 i}}} \right ) \right ] {\bm W}_d \, {\bm x}_d
  \hspace{0.5mm} , \hspace{6mm}
\end{eqnarray}
where ${\bm x}_d \equiv {\rm diag} (m_d, m_s, m_b)/M_{\rm
  KK}$. Compared to the ZMA result in the minimal RS model
\cite{Casagrande:2008hr}, this relation contains an additional term
involving the zero-mode profile $F(-c_{{{\cal T}}_{1 i}})$. It stems
from the admixture of the ${\cal T}_1$ multiplet in the zero mode,
which is parametrized by the value of $\hat a^{D^\prime}_n$. Notice
that although this admixture is suppressed by $v/M_{\rm KK}$, the fact
that the profile ${\bm S}^{{\cal T}_1(-)}_n (t)$ is enhanced with
respect to ${\bm S}^{{\cal T}_2(+)}_n (t)$ by the reciprocal factor
promotes the second term in the last line of (\ref{deltaD}) to a
leading contribution.

The relations (\ref{IRBCZMA}) and (\ref{IRBC3}) are valid to leading
order in $v/M_{\rm KK}$. Beyond that order the first relation in
(\ref{IRBCZMA}) receives corrections from the profiles ${\bm C}^{{\cal
    T}_1(-)}_n (1^-)$ which scale like $x_n/F (-c_{{\cal T}_{1 i}})$
as can be seen from (\ref{eq:profileexp}). Thus in order to avoid
exponentially enhanced terms of the form $v/M_{\rm KK} \,
\epsilon^{1-2 \hspace{0.25mm} c_{{{\cal T}}_{1 i}}}$ in the mass
eigenvalues $m_n$, which, barring accidental cancellations, would make
it impossible to reproduce the observed zero-mode down-type quark
masses, one has to require that all the bulk mass parameters belonging
to the multiplet ${\cal T}_1$ obey the relation $c_{{{\cal T}}_{1 i}}
< 1/2$. In this case the profiles ${\bm C}^{{\cal T}_1(-)}_n (t)$ are
IR localized and one has to an excellent accuracy
\begin{eqnarray} \label{deltaD2}
  {\bm \delta}_D = -\frac{1}{2} \, {\bm x}_d \, {\bm W}_d^\dagger \, \,
  {\rm diag } \left [ \frac{1}{1 - 2 \hspace{0.25mm}c_{{{\cal T}}_{2
          i}}} \left ( \frac{1}{F^2(c_{{{\cal T}}_{2 i}})} \left [ 1 - \frac{1 -
          2 \hspace{0.25mm} c_{{{\cal T}}_{2 i}}}{1 - 2 \hspace{0.25mm}
          c_{{{\cal T}}_{1 i}}} \right ] - 1 + \frac{F^2(c_{{{\cal T}}_{2
            i}})}{3 + 2 \hspace{0.25mm} c_{{{\cal T}}_{2 i}}} \right ) \right ] \,
  {\bm W}_d \, {\bm x}_d \hspace{0.5mm} . \hspace{6mm}
\end{eqnarray}
This result implies that the leading term in ${\bm \delta}_D$, \ie,
the contribution inversely proportional to $F^2(c_{{{\cal T}}_{2
i}})$, is absent if the bulk mass parameters $c_{{{\cal T}}_{1
i}}$ satisfy
\begin{equation} \label{eq:cbcb}
  c_{{{\cal T}}_{1 i}} = c_{{{\cal T}}_{2 i}} \,.
\end{equation}

Unlike in the case of the gauge-boson corrections (\ref{eq:Zqq}), the
conditions (\ref{PC}) and (\ref{PLR}) alone are thus not sufficient to
entirely shield the $Z b_L \bar b_L$ vertex from the leading
corrections due to quark mixing. However, since for $c_{{{{\cal T}}_{2
i}}} \approx -1/2$ and $c_{{{{\cal T}}_{1 i}}} \lesssim 0$ the
first term in (\ref{deltaD2}) is smaller in magnitude than 1, a
partial protection is in place for a large range of bulk
parameters. In consequence, effects due to quark mixing entering the
$Z$-boson couplings are generically suppressed in the custodial RS
model relative to the minimal scenario as long as the $Z_2$-odd quark
fields are not too far localized in the UV. The subleading terms in
${\bm \delta}_D$ are independent of $c_{{{\cal T}}_{1 i}}$ and
therefore not protected even if $c_{{{\cal T}}_{1 i}} = c_{{{\cal
T}}_{2 i}}$. They embody irreducible sources of symmetry
breaking, originating from the different BCs of the $Z_2$-even and
-odd quark fields.

Notice that (\ref{eq:cbcb}) can be enforced by requiring that the
action (\ref{Sferm2}) be invariant under the exchange of the
$D^\prime$ and $D$ quark fields,
\begin{equation} \label{eq:extendedPLR}
P_{LR} (D^\prime) = D \,, 
\qquad (\text{extended} \ P_{LR} \ \text{symmetry})
\end{equation}
which extends the $P_{LR}$ symmetry to the part of the quark sector
that mixes with the left-handed down-type zero modes. Notice that
while (\ref{eq:cbcb}) can always be accommodated, the extended
symmetry will be necessarily broken by the different BCs of $D^\prime$
and $D$.  The symmetry (\ref{eq:extendedPLR}) can also be broken
softly by choosing bulk masses for $D^\prime$ that differ from those
of $D$, which is a phenomenological viable option as long as
$c_{{{\cal T}}_{1 i}} < 1/2$, because it does not affect the SM
down-type quark masses in an appreciable way.  The protection
mechanism discussed here has also been studied in \cite{Buras:2009ka}
employing a perturbative approach. Our analysis based on the exact
solution of the EOMs (\ref{eq:EOM}) including the BCs
(\ref{eq:bcIRrescaled}) goes beyond the latter work in the sense that
it makes explicit the dependence of ${\bm \delta}_D$ on the bulk
masses ${\bm M}_{{\cal T}_{1,2}}$. It therefore allows for a
transparent understanding of the custodial protection mechanism in two
respects. First, it makes clear which are the requirements that need
to be satisfied to achieve a protection and, second, which are the
terms in ${\bm \delta}_D$ that inevitably escape protection.  Compared
to the perturbative approach, the exact solution thus has again the
salient advantage that the protection of the $Z d_L^i \bar d_L^j$
vertices from effects due to quark mixing can be clearly deciphered.

\subsection{Four-Fermion Charged-Current Interactions}
\label{sec:4Fint}

Hereafter we derive the effective four-fermion interactions induced by
the exchange of charged weak $W$ bosons and their KK excitations. In
this case, of course, flavor-changing effects are unsuppressed already
in the SM. Restricting ourselves to the phenomenologically most
relevant case with leptons in the final state, and including
corrections up to ${\cal O} (v^2/M_{\rm KK}^2)$, we obtain
\begin{equation} \label{eq:HeffW}
  {\cal H}_{\rm eff}^{(W)} = 2 \sqrt{2} \, G_F \, \sum_{l} \, \Big \{
  \big [ \, \bar{u}_L \gamma^\mu \bm{V}_{\!L\,} d_L + \bar{u}_R
  \gamma^\mu \bm{V}_{\!R\,} d_R \, \big ] \, (\bar l_L \gamma_\mu \nu_{l
    \, L}) + {\rm h.c.} \Big \} \,,
\end{equation} 
where the elements of the mixing matrices $\bm{V}_{\!  \!  L,R}$ are
computed from (\ref{eq:currents}) and take the form
\begin{equation} \label{eq:VLVR}
  \begin{split}
    \bm{V}_{\! L} = \, \bm{\Delta}^{\! +\, Q} + \sqrt 2\,
    \bm{\varepsilon}^{+\, q} -\frac{m_W^2}{2 \Mkk^2} \, L
    \hspace{0.25mm} \left( \bm{\bar\Delta}^{\! +\, Q} + \sqrt 2\,
      \bm{\bar\varepsilon}^{\, +\, q} \right) , \\
    \bm{V}_{\! R} = \, \sqrt 2\, \bm{\Delta}^{\! +\, q} +
    \bm{\varepsilon}^{+\, Q} - \frac{m_W^2}{2 \Mkk^2} \, L
    \hspace{0.25mm} \left(\sqrt 2\, \bm{\bar\Delta}^{\! +\, q} +
      \bm{\bar\varepsilon}^{\, +\, Q} \right) ,
  \end{split}
\end{equation} 
with 
\begin{equation} \label{eq:Deltaepsilonplus}
  \begin{split}
    \Delta_{mn}^{+\, Q,q} & = \frac{2 \pi}{L \epsilon} \,
    \int_\epsilon^1 \! dt \; \vec a_m^{\hspace{0.5mm} U,u \dagger} \,
    \bm{C}_m^{U,u} (t) \, \bm{\Omega}^{Q,q} \, \bm{C}_n^{D,d}(t) \,
    \vec a_n^{D,d} \,, \\
    \epsilon_{mn}^{+\, Q,q} & = \frac{2 \pi}{L \epsilon} \,
    \int_\epsilon^1 \! dt \; \vec a_m^{\hspace{0.5mm} U,u \dagger} \,
    \bm{S}_m^{U,u} (t) \, \bm{\Omega}^{Q,q} \,
    \bm{S}_n^{D,d}(t) \, \vec a_n^{D,d} \,,\\
    \bar\Delta_{mn}^{+\, Q,q} & = \frac{2 \pi}{L \epsilon} \,
    \int_\epsilon^1 \! dt \; t^2 \, \vec a_m^{\hspace{0.5mm} U,u
      \dagger} \, \bm{C}_m^{U,u} (t) \, \bm{\bar\Omega}^{Q,q} \,
    \bm{C}_n^{D,d}(t) \,
    \vec a_n^{D,d} \,, \\
    \bar\epsilon_{mn}^{+\, Q,q} & = \frac{2 \pi}{L \epsilon} \,
    \int_\epsilon^1 \! dt \; t^2 \, \vec a_m^{\hspace{0.5mm} U,u
      \dagger} \, \bm{S}_m^{U,u} (t) \, \bm{\bar\Omega}^{Q,q} \,
    \bm{S}_n^{D,d}(t) \, \vec a_n^{D,d} \,,
  \end{split}
\end{equation} 
and
\begin{equation} 
   \bm{\Omega}^Q=\begin{pmatrix}
   \bm{1}\\[1mm]
   \bm{0}
 \end{pmatrix} , \qquad \bm{\Omega}^q=
 \begin{pmatrix}
   \bm{0}~ &  \;\bm{0}\; \\
   \bm{0}~ &  \;\bm{1}\;\\
   \bm{0}~ &\, \;\bm{0}\;
  \end{pmatrix} , \qquad 
 \bm{\bar\Omega}^Q=\begin{pmatrix}
   \bm{1}\\[1mm]
   - \displaystyle \frac{g_R^2}{g_L^2} \, \bm{1}
 \end{pmatrix} , \qquad \bm{\bar\Omega}^q=
 \begin{pmatrix}
   \bm{0}~ &  \;\bm{0}\; \\
   \bm{0}~ &  \;\bm{1}\;\\[1mm]
   - \displaystyle \frac{g_R^2}{g_L^2}\, \bm{1}~ &\, \;\bm{0}\;
  \end{pmatrix} .
\end{equation}
Notice that the definition of $\bm{V}_{\! L,R}$ includes the exchange
of the entire tower of $W$ bosons and their KK excitations and
therefore differs from the definition of the CKM matrix employed in
\cite{Casagrande:2008hr, Buras:2009ka}, which is based on the $W
u^i_L d^j_L$ and $W u^i_R d^j_R$ vertices. Because an extraction of
CKM elements generically involves the normalization of the
semileptonic amplitude to the Fermi constant, we have included in
$G_F$ in (\ref{eq:HeffW}) a universal factor $(1 + m_W^2/(2
\Mkk^2) ( 1 - 1/(2 \hspace{0.5mm} L) ))$ which describes the finite
correction to muon decay in the RS model \cite{Casagrande:2008hr} and
is independent of the gauge group. Proceeding in this way renders the
individual factors in the combination $G_F \hspace{0.05mm} \bm{V}_{\!
L,R}$ physically observable \cite{Bauer:2009cf}. We note finally that
in order to arrive at the effective Hamiltonian (\ref{eq:HeffW}), we
have made the simplified assumption that the left- and right-handed 5D
leptonic fields all have the same bulk mass parameter, and that they
are localized sufficiently close to the UV brane so as not to violate
the constraints imposed by the electroweak precision tests. By
construction, the interactions of the SM leptons with the $W$ boson
and its KK excitations are therefore flavor universal and numerically
insignificant. 

From the formulas (\ref{eq:VLVR}) it is evident that no custodial
protection mechanism is at work in the charged-current sector. This is
due to the embedding of the up-type quarks (\ref{eq:multiplets}) and
has already been pointed out in \cite{Agashe:2006at}.  
  The leading contribution to $(V_{L})_{mn}$ stems from $\Delta^{+ \,
    Q}_{mn}$, which is unitary to very good approximation. Corrections
  of order $v^2/\Mkk^2$ arise from the non-universality of KK gauge
  bosons encoded in $\bar\Delta_{mn}^{+\, Q}$ as well as the admixture
  from $U^\prime$ and $D^\prime$ quarks described by
  $\epsilon_{mn}^{+\, q}$. Contributions arising from the admixture of
  $U$, $D$, and $u^\prime$ quarks are of order $v^4/\Mkk^4$ and will
  be neglected in the following. The full expression for $\bm{V}_{\!
    L}$ obtained by employing the ZMA reads
\begin{equation}
\begin{aligned}
  \bm{V}_{\! L} &= \bm{U}_u^\dagger \, \Bigg [ \bm{1} - \frac{m_W^2}{2
    \Mkk^2} \, L \; {\rm diag}\bigg(
  \frac{F^2(c_{Q_i})}{3+2c_{Q_i}} \bigg) \\
  &\hspace{1.2cm}\mbox{} + \frac{v^2}{2 \Mkk^2} \; {\rm diag}\, \big (
  F(c_{Q_i}) \big ) \, \bm{Y}_{\! d} \, {\rm diag}\, \big
  (F^{-2}(-c_{{\cal T}_{1i}} ) \big ) \, \bm{Y}_{\! d}^{\dagger} \;
  {\rm diag}\, \big ( F(c_{Q_i}) \big ) \Bigg ] \, \bm{U}_d \,,
\end{aligned}
\end{equation}
which is obviously not unitary. As far as $(V_{R})_{mn}$ is concerned,
the dominant contribution is given by $\epsilon_{mn}^{+\, Q}$, which is
suppressed both by $v^2/M_{\rm KK}^2$ and a chiral factor $m_m^u
m_n^d/v^2$. The chiral suppression present in each of the
terms contributing to ${\bm V}_{\! R}$ reflects the mere fact that
they all originate from quark mixing. In consequence, right-handed
charged-current interactions are too small to give rise to any
observable effect \cite{Casagrande:2008hr}.

As a measure of unitarity violation, we consider the deviation from
unity of the sum of the squares of the matrix elements in the first
row of ${\bm V}_{\!  L}$,
\begin{equation}
  \Delta_1^{\rm non} = 1 - (|V_{ud}|^2 + |V_{us}|^2 + |V_{ub}|^2) = \left (\bm{1}
    - \bm{V}_{\! L}\bm{V}_{\! L}^\dagger \right )_{11} \,.
\end{equation}
After expanding the mixing matrices ${\bm U}_{u,d}$ in powers of the
Cabibbo angle $\lambda$ (using the warped-space Froggatt-Nielsen
formulas given in \cite{Casagrande:2008hr}) and normalizing the result
to the typical value of the bulk mass parameter $c_{Q_1} \approx
-0.63$, we obtain
\begin{equation} \label{eq:delta1non_approximate}
  \begin{aligned}
    \Delta_1^{\rm non} &\approx 2 \cdot 10^{-6} \left(
      \frac{F(c_{Q_1})}{F(-0.63)} \right)^2 \left( \frac{\Mkk}{\rm TeV}
    \right)^{-2} \\
    &\quad \times \left[ \, \left| \, {\rm diag}\left(
          \sqrt{\frac{2}{3+2c_{Q_i}}}\; \right) \vec{u} \; \right|^2 -
      \frac{1}{4} \left| {\rm diag}\left( \sqrt{\frac{2}{1-2c_{{\cal
                  T}_{1i}}}}\; \right) \hspace{0.5mm} \bm{Y}_{\!d}^T
        \vec{u} \hspace{0.5mm} \right|^2 \right] ,
  \end{aligned}
\end{equation} 
where the vector $\vec{u}$ is given by
\begin{equation}
  \vec{u} = \big(1, -(M_u)_{21}/(M_u)_{11}, (M_u)_{31}/(M_u)_{11}\big) \,.
\end{equation}
Here $\bm{M}_u$ denotes the matrix of minors of $\bm{Y}_{\!u}$. The
first contribution in the square brackets in
\eqref{eq:delta1non_approximate} stems from the exchange of the whole
tower of $W$ bosons and is also present in the minimal RS model.  It
gives a strictly positive contribution to $\Delta_1^{\rm non}$, which
is typically well below the current experimental uncertainty of 
$6.5 \cdot 10^{-4}$ \cite{Bauer:2009cf}. However, the effects due to 
the admixture of $U^\prime$ and $D^\prime$ quarks contribute to
\eqref{eq:delta1non_approximate} with opposite sign and can in
principle lead to negative values of $\Delta_1^{\rm non}$. This is not
possible in the minimal RS model. A detailed discussion of the
breakdown of the unitarity of the quark mixing matrix in the framework
of the RS model with custodial protection has been presented in
\cite{Buras:2009ka}. Unfortunately, in that paper the CKM
matrix is defined via the $W u^i_{L} d^j_{L}$ vertex and not the
effective four-fermion interactions induced by the exchange of the $W$
boson and its KK excitations. This difference prevents us from a
  straightforward comparison of the results in \cite{Buras:2009ka}
  with ours.

\section{Fermion Couplings to the Higgs Boson}
\label{sec:higgscouplings}

The mixing of fermion zero-modes with their KK excitations leads to
flavor-changing Higgs couplings in scenarios with warped extra
dimensions \cite{Agashe:2006wa}. Analytic expressions for the
relevant interactions have been presented first within the minimal RS
model in \cite{Casagrande:2008hr}, which however did not include the
Yukawa couplings that involve $Z_2$-odd fermion profiles. This
omission has been noticed in \cite{Azatov:2009na}, where it has been
pointed out that the latter terms provide the dominant corrections to
tree-level Higgs FCNCs in the case of light quark flavors. We confirm
the ${\cal O} (v^2/M_{\rm KK}^2)$ result for a brane-localized Higgs
sector obtained in the latter work and generalize it to our exact
treatment of KK profiles and thus to all orders in $v/M_{\rm KK}$ for
both the minimal as well as the custodial RS model. 

Working in unitary gauge, we first identify the relevant terms in the
4D Lagrangian describing the couplings of the Higgs boson to
quarks. They read
\begin{equation}\label{eq:hff}
  {\cal L}_{\rm 4D} \ni - \sum_{q,m,n}\,(g_h^q)_{m n}\, h\,\bar
  q_L^{\hspace{0.5mm} m}\,q_R^{\hspace{0.25mm} n} + {\rm h.c.} \,,
\end{equation}
and the couplings $(g_h^q)_{mn}$ are given by 
\begin{eqnarray}\label{eq:ghn1n2}
  (g_h^q)_{mn} = \frac{\sqrt2\hspace{0.25mm} \pi}{L} \!
  \int_{-\pi}^{\pi} \! d\phi \, \delta (|\phi| - \pi) \,
  e^{\sigma(\phi)} \! \left[ \vec a_{m}^{Q\hspace{0.25mm} \dagger}\,\bm{C}_{m}^{Q}
    (\phi)  \bm{Y}_{\vec q} \hspace{1mm} \bm{C}_{n}^q(\phi)\,\vec
    a_{n}^{\hspace{0.25mm} q} +\vec a_{m}^{\hspace{0.25mm} q
      \hspace{0.5mm} \dagger}\,\bm{S}_{m}^{q} (\phi) \bm{Y}_{\vec q}^\dagger
    \hspace{1mm} \bm{S}_{n}^Q(\phi)\,\vec a_{n}^Q\right] \! . \hspace{7mm}
\end{eqnarray} 
To simplify this expression, we follow \cite{Azatov:2009na} and
observe that the EOMs (\ref{eq:EOM}) imply
\begin{eqnarray} \label{eq:EOMscomb}
  \begin{split}
    & m_n \, e^{\sigma (\phi)}\, \vec a_m^{Q\, \dagger}\,
    \bm{C}_m^Q(\phi) \hspace{0.25mm} \bm{C}_n^Q(\phi) \,\vec a_n^Q -
    \vec a_m^{Q\, \dagger}\, \bm{S}_m^Q(\phi) \hspace{0.25mm}
    \bm{S}_n^Q(\phi)\,\vec a_n^Q \, e^{\sigma (\phi)}\, m_m
    -\frac{1}{r}\, \partial_\phi \, \vec a_m^{Q \, \dagger}\,
    \bm{C}_m^Q(\phi) \hspace{0.25mm}
    \bm{S}_n^Q(\phi) \,\vec a_n^Q  \hspace{6mm} \\
    & \; - \frac{\sqrt{2} \pi v}{L}\, \delta(|\phi|-\pi) \,
    e^{\sigma(\phi)} \left [ \vec a_m^{Q \, \dagger}\,
      \bm{C}_m^Q(\phi) \, \bm{Y}_{\vec q} \, \bm{C}_n^{\hspace{0.25mm}
        q}(\phi)\,\vec a_n^{\hspace{0.25mm} q} - \vec a_m^{q
        \hspace{0.25mm} \, \dagger}\, \bm{S}_m^{\hspace{0.25mm}
        q}(\phi) \, \bm{Y}_{\vec q}^{\dagger} \, \bm{S}_n^Q(\phi)
      \,\vec a_n^Q \right ] = 0\,.
  \end{split}
\end{eqnarray}
After integrating this relation over the whole orbifold, the total
derivative in (\ref{eq:EOMscomb}) does not contribute since the
$Z_2$-odd profiles obey ${\bm S}_n^{Q,q} (0) = {\bm S}_n^{Q,q} (\pm
\pi) = 0$. One thus finds, after making use of the canonical
normalization of the kinetic terms, as encoded in (\ref{eq:orthonorm})
and (\ref{eq:CS}), the following expression
\begin{eqnarray} \label{eq:followingexpression}
  \begin{split}
    m_m \, \delta_{mn} & = \int_{-\pi}^{\pi} \!d\phi\; \bigg \{ m_n \,
    e^{\sigma(\phi)}\, \vec a_m^{\hspace{0.25mm} q\,\dagger}\,
    \bm{S}_m^{\hspace{0.25mm} q}(\phi) \, \bm{S}_n^{\hspace{0.25mm}
      q}(\phi) \,\vec a_n^{\hspace{0.25mm} q} + \vec
    a_m^{Q\,\dagger}\, \bm{S}_m^Q(\phi) \, \bm{S}_n^Q(\phi)\,\vec
    a_n^Q \,
    e^{\sigma(\phi)} \, m_m\\
    & \phantom{x} + \frac{\sqrt2 \pi v}{L}\,
    \delta(|\phi|-\pi)\,e^{\sigma(\phi)} \, \Big[ \vec a_m^{Q \,
      \dagger}\, \bm{C}_m^{Q}(\phi) \, \bm{Y}_{\vec q} \,
    \bm{C}_n^{\hspace{0.25mm} q}(\phi) \, \vec a_n^{\hspace{0.25mm} q}
    - \vec a_m^{\hspace{0.25mm} q\,\dagger}\,
    \bm{S}_m^{\hspace{0.25mm} q}(\phi) \, \bm{Y}_{\vec q}^{\dagger} \,
    \bm{S}_n^{Q}(\phi) \,\vec a_n^{Q} \Big] \bigg\} \,. \hspace{6mm}
  \end{split}
\end{eqnarray}
This result allows to eliminate the term bi-linear in the $Z_2$-even
profiles from (\ref{eq:ghn1n2}) and to express the tree-level Higgs
FCNCs solely in terms of overlap integrals involving $Z_2$-odd fields.

Defining the misalignment $(\Delta g_h^q)_{mn}$ between the SM masses
and the Yukawa couplings via
\begin{equation} \label{eq:mis1}
  (g_h^q)_{mn} \equiv \delta_{mn}\,\frac{m^q_m}{v} - ( \Delta
  g_h^q)_{mn} \,,
\end{equation}
it is then easy to show, by combining the latter definition with
(\ref{eq:ghn1n2}) and (\ref{eq:followingexpression}), that
\begin{equation} \label{eq:Higgscorrection} 
  (\Delta g_h^q)_{mn} = \frac{m^q_m}{v}\,(\Phi_q)_{mn} +
  (\Phi_Q)_{mn}\,\frac{m^q_n}{v} + (\Delta \tilde g_h^q)_{mn}\,,
\end{equation}
where in $t$ notation  
\begin{equation} \label{eq:Phi}
  \begin{split}
    \left( \Phi_q \right)_{mn} = \frac{2\pi}{L\epsilon}
    \int_\epsilon^1\!dt\, \vec a_m^{\hspace{0.25mm} Q
      \dagger}\,\bm{S}_m^{Q}(t)\, \bm{S}_n^{Q}(t)\,\vec
    a_n^{\hspace{0.25mm} Q} \,, \qquad \left( \Phi_Q \right)_{mn} =
    \frac{2\pi}{L\epsilon} \int_\epsilon^1\!dt\, \vec
    a_m^{\hspace{0.25mm} q \dagger}\,\bm{S}_m^{q}(t)\,
    \bm{S}_n^{q}(t)\,\vec a_n^{\hspace{0.25mm} q} \,,
  \end{split}
\end{equation} 
and 
\begin{equation} \label{eq:Deltagtilde}
  (\Delta \tilde{g}_h^q )_{mn} = - \sqrt{2} \,
  \frac{2\pi}{L\epsilon}\int_\epsilon^1\!dt \, \delta(t-1) \, \vec
  a_m^{\hspace{0.25mm} q\,\dagger}\, \bm{S}_m^{\hspace{0.25mm} q} (t) \,
  \bm{Y}_{\vec q}^{\dagger} \, \bm{S}_n^Q(t)\, \vec a_n^Q \,.
\end{equation}
Notice that the latter contribution is absent in
\cite{Casagrande:2008hr}, because this work did not include operators
of the form $(Y_q^{\rm (5D)})_{ij} \left(\bar Q_R^i\right)_{a\alpha}
\hspace{0.25mm} q_L^{c \,j} \, \Phi_{a\alpha}$ in the 5D action.
While omitting these operators is technically possible, since they are
not needed to generate the SM fermion masses, an omission seems
unnatural in the sense that there is no symmetry that would forbid
these $Z_2$-odd Yukawa couplings.

In order to evaluate the contribution $(\Delta \tilde{g}_h^q )_{mn}$
one again has to regularize the $\delta$-function appearing in
(\ref{eq:Deltagtilde}). Following the detailed explanations given in
Section \ref{sec:FermKK} and employing
\begin{equation} \label{eq:magic2} 
    \int_t^1 dt^\prime \,
    \delta^\eta (t^\prime - 1) \left [ \sinh \hspace{0.25mm} \big (
      \bar \theta^\eta (t^\prime - 1) \hspace{0.25mm} {\bm A} \big )
    \right ]^2 = \frac{1}{2} \left [ \sinh \hspace{0.25mm} \big ( \bar
      \theta^\eta (t - 1) \hspace{0.5mm} 2 {\bm A} \big ) \big (2 {\bm
        A} \big )^{-1} - \bar \theta^\eta (t-1) \, {\bm 1} \right ] ,
\end{equation} 
we obtain after some simple algebra the following regularization
independent result
\begin{equation} \label{eq:gtil1} 
  (\Delta \tilde g_h^q)_{mn} =
  \frac{1}{\sqrt{2}} \, \frac{2\pi}{L\epsilon} \, \frac{v^2}{3 M_{\rm
      KK}^2} \, \vec a_m^{Q\,\dagger}\, \bm{C}_m^{Q} (1^-) \,
  \bm{Y}_{\vec q} \, \bm{Y}_{\vec q}^{\dagger} \,\, \bm{g} \left (
    \frac{v}{\sqrt{2} M_{\rm KK}} \, \sqrt{{\bm Y}_{\vec{q}}
      \hspace{0.25mm} {\bm Y}_{\vec{q}}^\dagger} \right ) \bm{Y}_{\vec
    q} \, \bm{C}^{\hspace{0.25mm} q}_n (1^-) \,
  \vec{a}_n^{\hspace{0.25mm} q} \,,
\end{equation} 
with  
\begin{equation} \label{eq:bmg} 
  \bm{g} (\bm{A}) = \frac{3}{2} \left [
    \sinh \big (2 \bm{A} \big ) \big (2 \bm{A} \big )^{-1} 
    - \bm{1} \right ] \left ( \cosh \big (\bm{A} \big ) \bm{A} 
  \right )^{-2} \,.
\end{equation}

It is also straightforward to express (\ref{eq:gtil1}) in
  terms of the rescaled Yukawa matrices introduced in
  (\ref{eq:Yukresc}). Using
\begin{equation} \label{eq:YYYtYt}
  \frac{v}{\sqrt{2} \Mkk}\, \sqrt{\bm{Y}_{\vec q} \bm{Y}_{\vec q}^\dagger}\,=\,
  \tanh^{-1}\left ( \frac{v}{\sqrt{2} \Mkk} \, \sqrt{\bm{\tilde Y}_{\vec{q}} 
      \hspace{0.25mm} \bm{\tilde Y}_{\vec{q}}^\dagger} \right ) \,, 
\end{equation}
we obtain
\begin{equation} \label{eq:gtil2} 
  (\Delta \tilde g_h^q)_{mn} =
  \frac{1}{\sqrt{2}} \, \frac{2\pi}{L\epsilon} \, \frac{v^2}{3 M_{\rm
      KK}^2} \; \vec a_m^{Q\,\dagger}\, \bm{C}_m^{Q} (1^-) \,
  \bm{\tilde Y}_{\vec q} \, \bm{\bar Y}_{\vec q}^{\dagger} \,
  {\bm{\tilde Y}}_{\vec{q}} \, \bm{C}^{\hspace{0.25mm} q}_n (1^-) \,
  \vec{a}_n^{\hspace{0.25mm} q} \,,
\end{equation} 
where  
\begin{equation} \label{eq:bmh}
  \bm{\bar Y}_{\vec q}^\dagger \equiv \bm{\tilde Y}_{\vec q}^\dagger 
  \,\, \bm{h} \left ( \frac{v}{\sqrt{2} \Mkk} 
    \, \sqrt{\bm{\tilde Y}_{\vec{q}} \hspace{0.25mm} \bm{
        \tilde Y}_{\vec{q}}^\dagger} \right ) \,, \quad \bm{h} (\bm{A}) =
  \frac{3}{2} \, \Big [\bm{A}^{-2} + \tanh^{-1} \big (\bm{A} \big) \, 
  \bm{A}^{-1} \left (\bm{1} - \bm{A}^{-2}\right) \Big ] \, . 
\end{equation}
In practice, we compute the relevant matrix-valued functions by
introducing the unitary matrices ${\bm {\cal U}}_{\vec q}$ and ${\bm
  {\cal V}}_{\vec q}$ which diagonalize the hermitian products
$\bm{\tilde Y}_{\vec{q}} \, \bm{\tilde Y}_{\vec{q}}^\dagger$ and
$\bm{\tilde Y}_{\vec{q}}^\dagger \, \bm{\tilde Y}_{\vec{q}}$, 
\begin{equation} \label{eq:diagonalize} 
  \bm{\tilde Y}_{\vec{q}} \, \bm{\tilde Y}_{\vec{q}}^\dagger = {\bm{\cal
      U}}_{\vec q} \; {\bm{\tilde y}}_{\vec q} \, {\bm{\tilde
      y}}_{\vec q}^{\hspace{0.25mm} T} \; {\bm{\cal U}}_{\vec q}^\dagger \,, 
  \qquad
  \bm{\tilde Y}_{\vec{q}}^\dagger \, \bm{\tilde Y}_{\vec{q}} =
  {\bm{\cal V}}_{\vec q} \; {\bm{\tilde y}}_{\vec q}^{\hspace{0.25mm} T} \,
  {\bm{\tilde y}}_{\vec q} \; {\bm{\cal V}}_{\vec q}^\dagger \,, 
\end{equation}
where ${\bm{\tilde y}}_{\vec q}$ is, depending on the value of the
index $\vec q$, a matrix of dimension $3 \times 6$ or $6 \times 9$
containing the non-negative eigenvalues of $\sqrt{\bm{\tilde
    Y}_{\vec{q}} \, \bm{\tilde Y}_{\vec{q}}^\dagger}\,$ on its diagonal.
It then follows that
\begin{equation} 
  \bm{\tilde Y}_{\vec q} \, \bm{\bar Y}_{\vec q}^{\dagger} \,
  {\bm{\tilde Y}}_{\vec{q}} \, = \, {\bm{\cal U}}_{\vec q} \; 
  {\bm{\tilde y}}_{\vec q}\; {\bm{\tilde y}}_{\vec q}^{\hspace{0.25mm} T} 
  \; \bm{h} \left ( \frac{v}{\sqrt{2} \Mkk} \;
    \sqrt{{\bm{\tilde y}}_{\vec q} \, {\bm{\tilde y}}_{\vec q}^{\hspace{0.25mm} T} 
    } \right )  \, {\bm{\tilde y}}_{\vec q} \, {\bm{\cal V}}_{\vec q}^\dagger \,.
\end{equation}

Notice finally that the Yukawa matrices introduced in
(\ref{eq:Yukresc}) and (\ref{eq:bmh}) satisfy ${\bm{\tilde Y}}_{\vec
  q} = {\bm Y}_{\vec q} + {\cal O} (v^2/M_{\rm KK}^2)$ and ${\bm{ \bar
    Y}}_{\vec q}^\dagger = {\bm Y}_{\vec q}^\dagger + {\cal O}
(v^2/M_{\rm KK}^2)$. This implies that as long as one is interested in
the ZMA results for $(\Delta \tilde g_h^q)_{mn}$ only, one can simply
replace $\bm{\tilde Y}_{\vec q} \, \bm{\bar Y}_{\vec q}^{\dagger} \,
{\bm{\tilde Y}}_{\vec{q}}$ by the combination $\bm{Y}_{\vec q} \,
\bm{Y}_{\vec q}^{\dagger} \, {\bm{Y}}_{\vec{q}}$ of original Yukawa
matrices.

It will be useful to derive ZMA results for the elements $(\Delta
g_h^q)_{mn}$. For this purpose, we still need the ${\cal O}
(v^2/\Mkk^2)$ expressions for the rescaled eigenvectors $\hat a_n^A$
with $A = u, u^\prime, u^c, U^\prime, U$. First note that the
relations (\ref{IRBCZMA}), (\ref{eq:effY}), and (\ref{eq:EVeq1}) to
(\ref{eq:singular}) also hold in the up-type quark sector after the
replacements $d\rightarrow u,\, D \rightarrow u^c,\, c_{{{\cal T}}_{2
i}}\rightarrow c_{u^c_i}$ with $\bm{\lambda}_u~=~\sqrt
2/v\,\mbox{diag} \, (m_u,m_c,m_t)$. The remaining $\hat a_n^A$ are
found to satisfy
\begin{equation}\label{eq:upUpU}
  \begin{split}
    \hat a_n^{u^\prime} & = x_n\, {\rm diag} \; \Big (
    F^{-1}(-c_{Q_i})
    \hspace{0.5mm} F^{-1}(c_{Q_i}) \Big ) \, \hat a_n^{u} \,, \\
    \hat a_n^{U^\prime} & = {\rm diag} \; \Big ( F(-c_{{\cal T}_{2
        i}})
    \hspace{0.5mm} F^{-1}(-c_{{\cal T}_{1 i}}) \Big ) \, \hat a_n^{U} \,,\\
    \hat a_n^U & = \frac{x_n}{\sqrt 2}\, {\rm diag} \, \left (
      F^{-1}(-c_{{{\cal T}}_{2i}}) \right ) \, \bm{Y}_d^\dagger \,
    \big [ \bm{Y}_u^{\dagger} \big ]^{-1} \, {\rm diag} \left (
      F^{-1}(c_{u^c_i}) \right ) \, \hat a_n^{u^c} \,.
  \end{split}
\end{equation}
It is also easy to show that the eigenvectors satisfy the sum rules
\begin{equation}
  \begin{split}
    \hat a_n^{u^c \hspace{0.5mm} \dagger} \, \hat a_n^{u^c} + \hat
    a_n^{u^\prime \hspace{0.5mm} \dagger} \, \hat a_n^{u^\prime} = 1
    \,, \qquad \hat a_n^{u \hspace{0.5mm} \dagger} \, \hat a_n^{u} +
    \hat a_n^{U \hspace{0.5mm} \dagger} \, \hat a_n^U + \hat
    a_n^{U^\prime \hspace{0.5mm} \dagger} \, \hat a_n^{U^\prime} = 1
    \, .
  \end{split}
\end{equation}

With these relations at hand, it is straightforward to derive analytic
expressions for the ${\cal O} (v^2/\Mkk^2)$ corrections to ${\bm
\Phi}_{q}$, ${\bm \Phi}_{Q}$, and $\bm{\Delta \tilde{g}}_h^q$.  In the
case of down-type quarks, we find
\begin{equation} \label{eq:PhiD}
  \begin{split} 
    & {\bm \Phi}_d = {\bm x}_d \, {\bm U}_d^\dagger \, \, {\rm diag }
    \left [ \frac{1}{1 - 2 \hspace{0.25mm}c_{Q_i}} \left (
        \frac{1}{F^2(c_{Q_i})} - 1 + \frac{F^2(c_{Q_i})}{3 + 2
          \hspace{0.25mm}
          c_{Q_i}} \right ) \right ] {\bm U}_d \, {\bm x}_d \,, \\
    & {\bm \Phi}_D = {\bm x}_d \, {\bm W}_d^\dagger \, \, {\rm diag }
    \left [ \frac{1}{1 - 2 \hspace{0.25mm}c_{{{\cal T}}_{2 i}}} \left
        ( \frac{1}{F^2(c_{{{\cal T}}_{2 i}})} \left [ 1 + \frac{1 - 2
            \hspace{0.25mm} c_{{{\cal T}}_{2 i}}}{F^2(-c_{{{\cal
                  T}}_{1 i}})} \right ] - 1 + \frac{F^2(c_{{{\cal
                T}}_{2 i}})}{3 + 2 \hspace{0.25mm} c_{{{\cal T}}_{2
              i}}} \right ) \right ]
    {\bm W}_d \, {\bm x}_d \,, \\
    & {\bm{\Delta \tilde{g}}}_h^d  = \frac {\sqrt{2} \, v^2 }{3 M_{\rm
        KK}^2} \; {\bm U}_d^\dagger \; {\rm diag} \left[ F(c_{Q_i})
    \right] \hspace{0.25mm} \bm{Y}_d \hspace{0.25mm} \bm{Y}_d^\dagger
    \hspace{0.25mm} \bm{Y}_d \; {\rm diag } \left[ F(c_{{\cal T}_{2
          i}}) \right] {\bm W}_d\,,
  \end{split}
\end{equation}
while for up-type quarks we obtain 
\begin{eqnarray} \label{eq:PhiU}
  \begin{split} 
    & {\bm \Phi}_u = {\bm x}_u \, {\bm U}_u^\dagger \, \, {\rm diag }
    \left [ \frac{1}{1 - 2 \hspace{0.25mm}c_{Q_i}} \left (
        \frac{1}{F^2(c_{Q_i})} \left [ 1 + \frac{1 - 2 \hspace{0.25mm}
            c_{Q_i}}{F^2(-c_{Q_i})} \right ] - 1 +
        \frac{F^2(c_{Q_i})}{3 + 2 \hspace{0.25mm} c_{Q_i}} \right )
    \right ] \, {\bm U}_u \, {\bm x}_u \,,\\
    & {\bm \Phi}_U = {\bm x}_u \, {\bm W}_u^\dagger \, \, \Bigg \{ \,
    {\rm diag} \left [ \frac{1}{1 - 2 \hspace{0.25mm}c_{u^c_i}} \left
        ( \frac{1}{F^2(c_{u^c_i})} - 1 + \frac{F^2(c_{u^c_i})}{3 + 2
          \hspace{0.25mm} c_{u^c_i}} \right ) \right ] + \frac{1}{2}
    \, {\rm diag} \left ( F^{-1}(c_{u^c_i}) \right ) \!
    \hspace{0.25mm} \bm{Y}_u^{-1} \,
    \bm{Y}_d \hspace{8mm} \\
    & \mbox{} \phantom{xxxxxxxxxi} \times \hspace{0.5mm} {\rm diag}
    \left(\frac{1}{F^2(-c_{{\cal T}_{2i}})} + \frac{1}{F^2(-c_{{\cal
            T}_{1i}})} \right) \bm{Y}_d^\dagger \,\big [
    \bm{Y}_u^{\dagger} \big ]^{-1} \, {\rm diag} \left (
      F^{-1}(c_{u^c_i}) \right ) \Bigg \} \,
    {\bm W}_u \, {\bm x}_u\,, \\
    & {\bm{\Delta \tilde{g}}}_h^u = \frac {\sqrt{2} \, v^2 }{3 M_{\rm
        KK}^2} \; {\bm U}_u^\dagger \; {\rm diag} \left[ F(c_{Q_i})
    \right] \hspace{0.25mm} \bm{Y}_u \hspace{0.25mm} \bm{Y}_u^\dagger
    \hspace{0.25mm} \bm{Y}_u \; {\rm diag } \left[ F(c_{u^c_i})
    \right] {\bm W}_u\,,
  \end{split}
\end{eqnarray}
with ${\bm x}_u \equiv {\rm diag} (m_u, m_c, m_t)/M_{\rm KK}$. Here
$\bm U_u$ ($\bm W_u$) are the left- (right-) handed rotation matrices
diagonalizing the effective up-type Yukawa coupling. 

At this point some comments are in order. First, notice that compared
to the ZMA results in the minimal RS model \cite{Casagrande:2008hr}
the expressions for $\bm \Phi_D$, $\bm \Phi_u$, and $\bm \Phi_U$
contain additional terms, whereas $\bm \Phi_d$ is unchanged. The new
terms in (\ref{eq:PhiD}) and (\ref{eq:PhiU}) are the ones involving
the zero-mode profiles $F(-c_{{{\cal T}}_{1 i}})$, $F(-c_{{{\cal
      T}}_{2 i}})$, and $F(-c_{Q_i})$. They arise from the admixture
of the ${\bm S}^{{\cal T}_1 (-)}_n (t)$, ${\bm S}^{{\cal T}_2 (-)}_n
(t)$, and ${\bm S}^{Q (-)}_n (t)$ profiles in the corresponding
zero-mode wave functions. In each case, the suppression by $v/M_{\rm
  KK}$ due to the admixture is offset by the ${\cal O} (\Mkk/v)$
enhancement of the $Z_2$-odd $(-)$ profile relative to its $(+)$
counterpart. For $c_{Q_i}<1/2$, the leading contribution in ${\bm
  \Phi}_u$ is numerically enhanced by a factor of 2 with respect to
the minimal model. If the $Z d_L^i \bar d_L^j$ vertices are protected
from fermion mixing by (\ref{eq:cbcb}) and $c_{{\cal T}_{1i}} < 1/2$,
then the same is true for ${\bm \Phi}_D$.  Depending on the structure
of the Yukawa matrices and bulk masses, a similar enhancement is
possible in ${\bm \Phi}_U$.  Notice that the extra suppression by
factors of $m^q_n/v$ in ${\bm \Phi}_{d,D,u,U}$ imply that for light
quark flavors the Higgs-boson FCNCs arising from the latter terms are
parametrically suppressed relative to those mediated by the exchange
of a $Z$ boson. This makes the chirally unsuppressed contributions
$\bm{\Delta \tilde{g}}_h^{d,u}$, that arise from the $Z_2$-odd Yukawa
couplings, the dominant sources of flavor violation in the Higgs
sector \cite{Azatov:2009na}. As far as the ${\cal O} (v^2/M_{\rm
  KK}^2)$ corrections to $\bm{\Delta \tilde{g}}_h^{d,u}$ as given in
(\ref{eq:PhiD}) and (\ref{eq:PhiU}) are concerned, we find perfect
agreement with the results presented in the latter article for the
case of a brane-localized Higgs sector. Compared to the minimal RS
model the corrections ${\bm{\Delta \tilde{g}}}_h^{d,u}$ are again
bigger by a factor of 2 in the extended scenario. Notice also that the
factor of $1/3$ arising in the ZMA expressions for $\bm{\Delta
  \tilde{g}}_h^{d,u}$ follows immediately if one applies
(\ref{eq:delthe}) to a composite operator containing two $Z_2$-odd
fermion profiles.

Let us finally mention, that a model-independent analysis of the
flavor misalignment of the SM fermion masses and the Yukawa couplings
has been presented in \cite{Agashe:2009di}. There it has been shown
that at the level of dimension six, chirally unsuppressed
contributions to flavor-changing Higgs-boson vertices generically will
arise from composite operators like $\bar q_L^{\hspace{0.5mm} i}
\hspace{0.25mm} H \hspace{0.25mm} q_R^j \hspace{0.5mm} (H^\dagger H)$
in models where the Higgs is a bound state of a new
strongly-interacting theory. If present, the latter terms will
dominate over the chirally suppressed contributions originating from
operators of the form $\bar q_L^{\hspace{0.5mm} i} \hspace{0.25mm} D
\!\!\!\!/\, \hspace{0.25mm} q_L^j \hspace{0.5mm} (H^\dagger H)$,
because the couplings $y_{q \ast}$ of the composite Higgs to the other
strong interacting states can be large, resulting in $y_{q \ast}^2/(16
\pi^2) \gg m_q/v$. Notice that in our concrete model, considering all
relevant dimension-six operators in lowest-order of the mass insertion
approximation, allows to recover the ZMA results (\ref{eq:PhiD}) and
(\ref{eq:PhiU}) quantifying the misalignment between the Yukawa
couplings and the zero-mode masses (see \cite{Azatov:2009na} for a
illuminating discussion). We emphasize that in our exact solution
(\ref{eq:mis1}), (\ref{eq:Higgscorrection}), (\ref{eq:Phi}), and
(\ref{eq:gtil2}), all new-physics effects induced by the mass
insertions, corresponding to both the non-derivative and derivative
dimension-six operators, the former of which in fact lead to the
chirally unsuppressed terms in (\ref{eq:Higgscorrection}), are
resummed to all orders in $v^2/\Mkk^2$ at tree level. Before
presenting a comprehensive analysis of the phenomenological impact of
Higgs effects in the rare decay $t \to c h$ as well as Higgs-boson
production and decay (see Sections \ref{sec:tch} to
\ref{sec:higgsdecay}), we add that in the case of $\Delta F = 2$
processes, \ie, neutral meson mixing, the importance of Higgs FCNCs
turns out to be limited. The most pronounced effects occur in the case
of the CP-violating parameter $\epsilon_K$, but even here they are
typically smaller than the corrections due to KK gluon exchange
\cite{Duling:2009pj}.

\section{Phenomenological Applications}
\label{sec:pheno}

In the following we consider some applications of our results. We
begin with a discussion of the constraints imposed by the precision
measurements of the bottom-quark pseudo observables. In particular, we
show that the tension in the global $Z\to b\bar b$ fit is not
significantly relaxed for a light Higgs boson with mass $m_h\approx
100 \,{\rm GeV}$. A perfect fit can, however, be obtained for
$m_h\lesssim 1 \,{\rm TeV}$, which is the naturally expected mass
range for the Higgs boson in models with a brane-localized scalar
sector. We furthermore discuss the phenomenology of rare top-quark
decays in the custodial model and compare it to the one of the minimal
RS model. The experimental prospects for observing the rare FCNC
transitions $t\to c Z$ and $t \to ch$ at the LHC turn out to be more
favorable in extended RS scenarios than in the minimal model. The
first complete one-loop calculation of all relevant Higgs-boson
production and decay channels at hadron colliders represents the
highlight of our phenomenological investigations. We find that, due to
the composite nature of the Higgs boson, the top quark, and the KK
gauge bosons and fermions, observable effects in both production and
decay can naturally occur in RS scenarios. This observation could
potentially have a tangible impact on the LHC physics program.

\subsection{\boldmath Bottom-Quark Pseudo Observables}
\label{sec:Zbbnumerics}

In order to derive explicit expressions for $g_L^b \equiv \left( g_L^d
\right)_{33}$ and $g_R^b \equiv \left( g_R^d \right)_{33}$ in the ZMA,
we first need formulas for the leading contributions to the overlap
integrals $\big ( \Delta^{(\prime)}_{D,d} \big)_{33}$. We find to the
order considered
\begin{align}\label{ZMA1}
  \bm{\Delta}_D &= \bm{U}_d^\dagger\,\,\mbox{diag} \left[
    \frac{F^2(c_{Q_i})}{3+2c_{Q_i}} \right] \bm{U}_d \,, \nonumber \\
  \bm{\Delta}_d &= \bm{W}_d^\dagger\,\,\mbox{diag} \left[
    \frac{F^2(c_{{{\cal T}}_{2 i}})}{3+2c_{{{\cal T}}_{2 i}}} \right]
  \bm{W}_d \,, \nonumber \\
  \bm{\Delta}'_D &= \bm{U}_d^\dagger\,\,\mbox{diag} \left[
    \frac{5+2c_{Q_i}}{2(3+2c_{Q_i})^2}\,F^2(c_{Q_i}) \right]
  \bm{U}_d \,, \nonumber \\
  \bm{\Delta}'_d &= \bm{W}_d^\dagger\,\,\mbox{diag} \left[
    \frac{5+2c_{{{\cal T}}_{2 i}}}{2(3+2c_{{{\cal T}}_{2
          i}})^2}\,F^2(c_{{{\cal T}}_{2 i}}) \right] \bm{W}_d \,.
\end{align}
Notice that these expressions have exactly the same form as in the
minimal model \cite{Casagrande:2008hr}.

The matrices $\bm{\varepsilon}_{D,d}^{(\prime)}$ vanish at leading
order in the ZMA, meaning that they are suppressed by an extra factor
of $v^2/\Mkk^2$. We consequently neglect them. The matrices $\bm
\delta_{D,d}$, on the other hand, are of the same order as the $\bm
\Delta_{D,d}^{(\prime)}$ contributions. The ZMA expression for $\bm
\delta_D$ has already been given in (\ref{deltaD2}), and the last
missing ingredient takes the form ${\bm \delta}_d =-1/2\,{\bm \Phi}_d$
and resembles the ZMA result found in the RS model with $SU(2)_L
\times U(1)_Y$ bulk gauge symmetry \cite{Casagrande:2008hr}.

After Taylor expansion of the mixing matrices ${\bm U}_d$ and ${\bm
W}_d$ in powers of the Cabibbo angle $\lambda$
\cite{Casagrande:2008hr}, we finally arrive for $c_{b_R^\prime} ,
\hspace{0.5mm} c_{{{\cal T}}_{1 i}} < 1/2$ at
\begin{align} \label{eq:ZbbRS} 
  g_L^b & = \left( - \frac12 + \frac{s_w^2}{3} \right) \Bigg [ 1 -
  \frac{m_Z^2}{2\Mkk^2}\, \frac{F^2(c_{b_L})}{3 + 2 \mkern+2mu
    c_{b_L}} \left( \omega_Z^{b_L} L - \frac{5+2c_{b_L}}{2 (3 + 2
      \mkern+2mu c_{b_L})} \right)
  \Bigg ] \nonumber \\
  &\mkern+20mu + \frac{m_b^2}{2\Mkk^2}\, \Bigg \{ \frac{1}{1 - 2
    \mkern+2mu c_{b_R}} \left( \frac{1}{F^2(c_{b_R})} \left [ 1 -
      \frac{1 - 2 \mkern+2mu c_{b_R}}{1 - 2 \mkern+2mu
        c_{b^{\prime}_R}} \right ] - 1 + \frac{F^2(c_{b_R})}{3 + 2
      \mkern+2mu c_{b_R}} \right)
  \notag \\
  &\mkern+105mu + \sum_{i=1}^2 \frac{|(Y_d)_{3i}|^2}{|(Y_d)_{33}|^2}
  \frac{1}{1 - 2 \mkern+2mu c_{{\cal T}_{2 i}}} \frac{1}{F^2(c_{b_R})}
  \left [ 1 - \frac{1 - 2 \mkern+2mu c_{{{\cal T}}_{2 i}}}{1 - 2
      \mkern+2mu c_{{{\cal T}}_{1 i}}} \right ]
  \Bigg \} \, , \\[10pt]
  g_R^b & = \frac{s_w^2}{3} \, \Bigg [ 1 - \frac{m_Z^2}{2\Mkk^2}\,
  \frac{F^2(c_{b_R})}{3 + 2 \mkern+2mu c_{b_R}} \left( \omega^{b_R}_Z
    L - \frac{5 + 2 \mkern+2mu c_{b_R}}{2 (3 + 2
      \mkern+2mu c_{b_R})} \right) \Bigg ] \notag \\
  &\mkern+20mu - \frac{m_b^2}{2\Mkk^2}\, \Bigg \{ \frac{1}{1 - 2
    \mkern+2mu c_{b_L}} \left( \frac{1}{F^2(c_{b_L})} - 1 +
    \frac{F^2(c_{b_L})}{3 + 2 \mkern+2mu c_{b_L}} \right) +
  \sum_{i=1}^2 \frac{|(Y_d)_{i3}|^2}{|(Y_d)_{33}|^2} \frac{1}{1 - 2
    c_{Q_i}} \frac{1}{F^2(c_{b_L})} \Bigg \} \, , \hspace{6mm} \notag
\end{align}
where $c_{b_L} \equiv c_{Q_3}$, $c_{b_R} \equiv c_{{{\cal
T}}_{2\hspace{0.05mm}3}}$, and $c_{b^\prime_R} \equiv c_{{{\cal
T}}_{1\hspace{0.05mm}3}}$. Furthermore, $m_b \equiv m_b (\Mkk)$
denotes the bottom-quark $\overline{\rm MS}$ mass evaluated at the KK
scale. Notice that we kept $c_{{{\cal T}}_{1 i}} \neq c_{{{\cal
T}}_{2i}}$, thereby allowing the $P_{LR}$ symmetry to be broken by the
triplet bulk masses. We also retained the parameters $\omega_Z^{b_L}$
and $\omega_Z^{b_R}$. In the custodial RS model one has
$\omega_Z^{b_L} = 0$ and
\begin{equation} \label{eq:omegabR}
  \omega^{b_R}_Z = \frac{3 c_w^2}{s_w^2} \approx 10.0 \,,
\end{equation}
where in order to arrive at the numerical values we have employed
$s_w^2 \approx 0.23$, corresponding to the value of the weak mixing
angle at the $Z$-pole.\footnote{The electromagnetic coupling and the
weak mixing angle are running parameters in the low-energy effective
theory obtained after decoupling the RS contributions at the scale
$\Mkk$. The associated large logarithms can be effectively included by
replacing $s_w^2(\Mkk)$ by $s_w^2(m_Z)$ in the couplings
$g_{L,R}^b$. On the other hand, the value of the bottom-quark mass
entering the matching is frozen at the high scale and does not evolve
in the effective theory.}

By inspection of (\ref{eq:ZbbRS}), we observe that the non-universal
corrections to the $Z b \bar b$ couplings reduce both $g_L^b$ and
$g_R^b$ if the extended $P_{LR}$ symmetry (\ref{eq:extendedPLR}) is at
work. If one allows the $P_{LR}$ symmetry to be broken by $c_{{\cal
T}_{1i}} \neq c_{{\cal T}_{2i}}$, then the shift in $g_L^b$ can also
be positive as a result of fermion mixing. As we will see in a moment,
this always aggravates the quality of the $Z \to b \bar b$ fit. It is
also evident that with respect to the minimal model, where the shift
$\delta g_L^b$ is large and positive while $\delta g_R^b$ is small and
negative \cite{Casagrande:2008hr}, the constraints arising from the
bottom-quark pseudo observables are naively much less stringent. Yet
in order to gauge the improvement and to fully understand the
parameter dependence, in particular the one on the bulk mass
parameters $c_{{\cal T}_{1i}}$, one has to perform a detailed
numerical analysis. Such an exercise is the subject of the remainder
of this subsection.

Consider the ratio of the width of the $Z$-boson decay into bottom
quarks and the total hadronic width, $R_b^0$, the bottom-quark
left-right asymmetry, $A_b$, and the forward-backward asymmetry for
bottom quarks, $A_{\rm FB}^{0,b}$. The dependences of these quantities
on the left- and right-handed bottom-quark couplings are given by
\cite{Field:1997gz}
\begin{equation}\label{eq:bPOtheory}
  \begin{split}
    R_b^0 &= \left [ 1 + \frac{4 \; {\displaystyle \sum}_{q=u,d}
        \left[ (g_L^q)^2 + (g_R^q)^2\right]}%
      {\eta_{\rm QCD}\,\eta_{\rm QED} \left[ (1-6z_b) (g_L^b-g_R^b)^2
          + (g_L^b+g_R^b)^2 \right]}
    \right]^{-1}\! , \\
    A_b &= \frac{2\sqrt{1-4z_b}\,\, {\displaystyle
        \frac{g_L^b+g_R^b}{g_L^b-g_R^b}}}%
    {1-4z_b+(1+2z_b) {\displaystyle \left(
          \frac{g_L^b+g_R^b}{g_L^b-g_R^b} \right)^2}} \,, \qquad
    A_{\rm FB}^{0,b} = \frac34\,A_e\hspace{0.25mm}A_b \,.
  \end{split}
\end{equation}
Radiative QCD and QED corrections are encoded by the factors
$\eta_{\rm QCD}=0.9954$ and $\eta_{\rm QED}=0.9997$, while the
parameter $z_b\equiv m_b^2(m_Z)/m_Z^2=0.997\cdot 10^{-3}$ describes
the effects of the non-zero bottom-quark mass. Since to an excellent
approximation one can neglect the RS contributions to the left- and
right-handed couplings of the light quarks, $g_{L,R}^q$, and to the
asymmetry parameter of the electron, $A_e$, we will fix these
quantities to their SM values $(g_L^u)_{\rm SM}=0.34674$,
$(g_R^u)_{\rm SM} =-0.15470$, $(g_L^d)_{\rm SM}=-0.42434$,
$(g_R^d)_{\rm SM}=0.077345$ \cite{LEPEWWG:2005ema}, and $(A_e)_{\rm
SM} =0.1462$ \cite{Arbuzov:2005ma}. The quoted values correspond to
the SM input parameters given in Appendix~\ref{app:masses}.

Evaluating the relations (\ref{eq:bPOtheory}) using $\big ( g_L^b \big
)_{\rm SM}=-0.42114$ and $\big ( g_R^b \big)_{\rm SM}=0.077420$
\cite{LEPEWWG:2005ema}, we obtain for the central values of the
bottom-quark pseudo observables
\begin{equation}\label{eq:bPOSM}
  \big (R_b^0 \big )_{\rm SM} = 0.21578 \,, \qquad 
  \big ( A_b \big )_{\rm SM} = 0.935 \,, \qquad 
  \big ( A_{\rm FB}^{0,b} \big )_{\rm SM} = 0.1025 \,.
\end{equation}
One should compare these numbers with the experimental results
\cite{LEPEWWG:2005ema}
\begin{equation}\label{eq:bPOsexp}
  \begin{array}{l}
    \big ( R_b^0 \big)_{\rm exp} = 0.21629\pm 0.00066 \,, \\[0.25mm] 
    \big ( A_b \big)_{\rm exp} = 0.923\pm 0.020 \,, \\[1mm]
    \big ( A_{\rm FB}^{0,b} \big )_{\rm exp} = 0.0992\pm 0.0016 \,, 
  \end{array}
  \qquad 
  \rho = \begin{pmatrix}
    1.00 \, & \, -0.08 & \, -0.10 \\ 
    \, -0.08 & \, 1.00 & \, 0.06 \\ 
    -0.10 \, & \, 0.06 & \, 1.00 
  \end{pmatrix} ,
\end{equation}
where $\rho$ is the correlation matrix. While the $R_b^0$ and $A_b$
measurements agree within $+0.8\sigma$ and $-0.6\sigma$ with their SM
predictions for $m_h = 150 \, {\rm GeV}$, the $A_{\rm FB}^{0,b}$
measurement is almost $-2.1\sigma$ away from its SM
expectation.\footnote{For $m_h=115$\,GeV the discrepancy in $A_{\rm
FB}^{0,b}$ would amount to around $-2.5\sigma$.} Shifts of order
$+20\%$ and $-0.5\%$ in the right- and left-handed bottom-quark
couplings relative to the SM could explain the observed
discrepancy. Such a pronounced correction in $g_R^b$ would affect
$A_b$ and $A_{\rm FB}^{0,b}$, which both depend linearly on the ratio
$g_R^b/g_L^b$, in a significant way, while it would not spoil the good
agreement in $R_b^0\propto (g_L^b)^2 + (g_R^b)^2$.

\begin{figure}[!t]
\begin{center}
\vspace{-1cm}
\mbox{\includegraphics[height=2.8in]{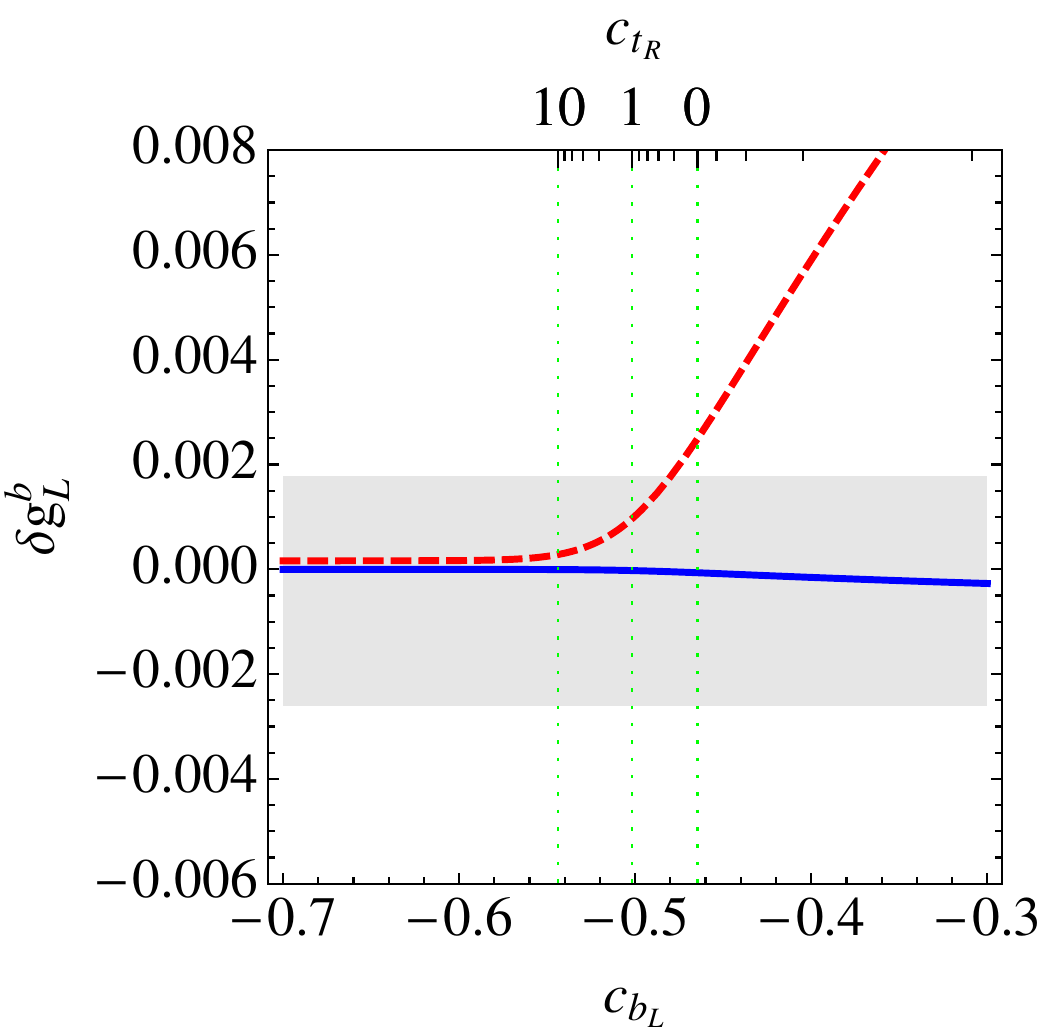}}
\qquad 
\mbox{\includegraphics[height=2.8in]{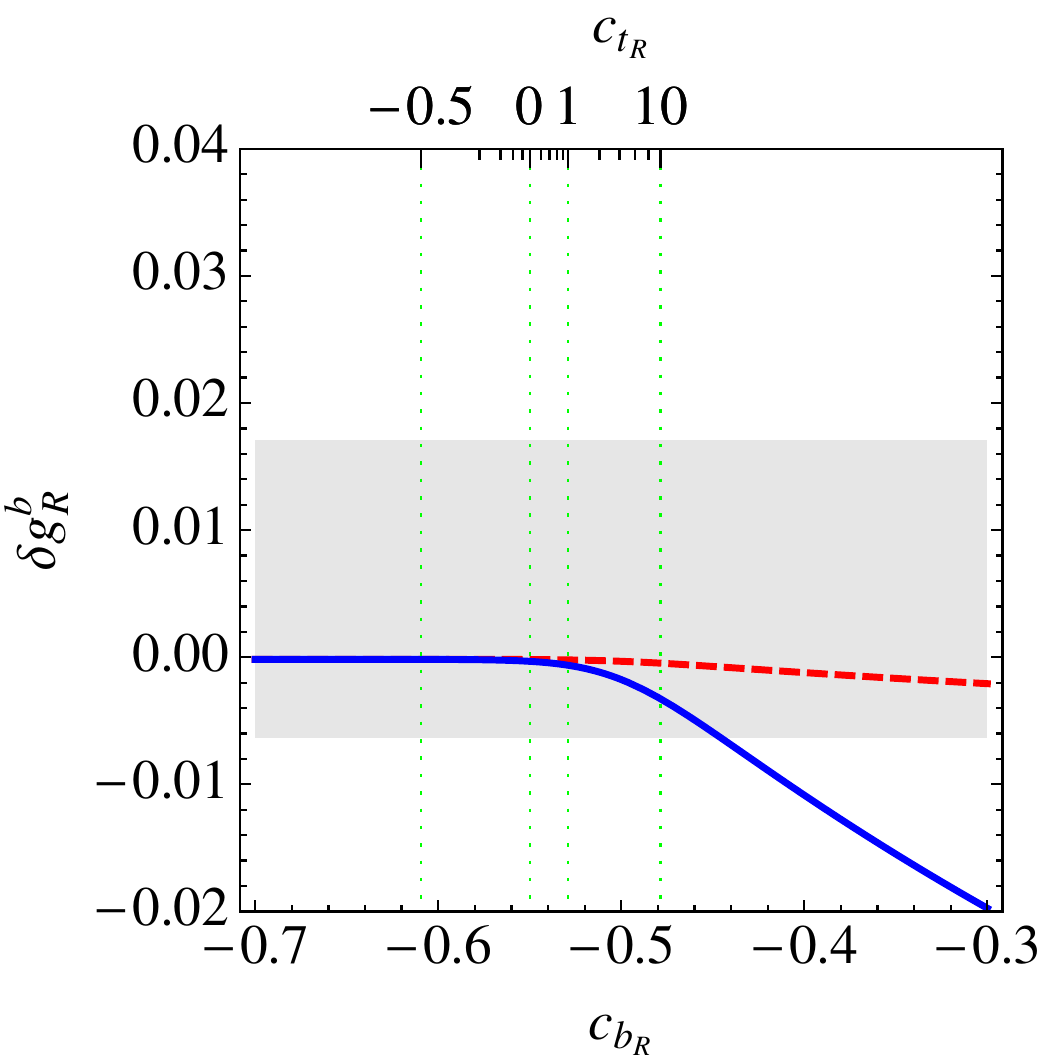}}
\vspace{-2mm}
\parbox{15.5cm}{\caption{\label{fig:dgLRb} Anomalous couplings $\delta
    g_L^b$ (left) and $\delta g_R^b$ (right) as functions of $c_{b_L}$
    and $c_{b_R}$. The blue solid (red dashed) lines correspond to the
    predictions obtained in the RS model with extended $P_{LR}$
    symmetry (minimal RS model). Bulk mass parameters not explicitly
    shown are set to $-1/2$, and all elements of the down-type Yukawa
    matrix are taken to be equal to 1 in magnitude. The light gray
    bands indicate the experimentally allowed 99\%~CL ranges. See text
    for details.}}
\end{center}
\end{figure}

In Figure~\ref{fig:dgLRb} we show our predictions for the anomalous
couplings $\delta g_{L,R}^b \equiv g_{L,R}^b - \big ( g_{L,R}^b \big
)_{\rm SM}$ as functions of the bulk mass parameters
$c_{b_{L,R}}$. Similar plots have been presented in
\cite{Carena:2006bn}. The shown curves correspond to $c_{Q_i} =
c_{{\cal T}_{1i}} = c_{{\cal T}_{2i}} = -1/2$ and $|(Y_d)_{3i}| =
|(Y_d)_{i3}| = |(Y_d)_{33}| = 1$ with $i = 1,2$. We see that compared
to the minimal case (red dashed line) the prediction for $\delta
g_L^b$ in the RS model with extended $P_{LR}$ symmetry (blue solid
line) is, owing to (\ref{eq:custodialQ}), essentially independent of
$c_{b_L}$.\footnote{In order not to induce unacceptably large
corrections to $\delta g_R^b$ induced by effects due to fermion
mixing, one has to require $c_{b_L} \gtrsim -0.55$.} The predictions
for the anomalous coupling $\delta g_{L}^b$ are thus easily within the
experimental 99\% confidence level (CL) bound (light gray band), which
gives a strong motivation to protect the $Z b_L \bar b_L$ vertex
through the mechanism of \cite{Agashe:2006at}. Notice that in the
case of the minimal RS model, $\delta g_L^b$ can be suppressed by
localizing the right-handed top quark very close to the IR brane. This
feature is illustrated by the ticks on the upper border of the frame
in the left panel. The given values of $c_{t_R} \equiv c_{u^c_3}$ have
been obtained by solving $m_t = v/\sqrt{2} \, \left |(Y_u)_{33}
\right| \left |F(c_{b_L}) \hspace{0.25mm} F(c_{t_R}) \right |$ for the
bulk mass parameter $c_{t_R}$, evaluating the top-quark $\overline{\rm
MS}$ mass at $\Mkk = 1 \, {\rm TeV}$ and setting $|(Y_u)_{33}| =
3$. For smaller (larger) values of $|(Y_u)_{33}|$ the ticks are
shifted to the right (left).

In the case of $\delta g_R^b$ we observe instead that, as a result of
(\ref{eq:omegabR}), the corrections to the anomalous coupling are
always larger in the RS model with extended $P_{LR}$ symmetry (blue
solid line) than in the minimal formulation (red dashed
line).\footnote{Notice that in order to reproduce the large top-quark
mass with Yukawa couplings of $\ord (1)$ one has to require $c_{t_R} >
-1/2$, corresponding to $c_{b_R} \gtrsim -0.6$.} It is however
important to remark that even in the former case the $Z b_R \bar b_R$
coupling is predicted to be SM-like, since shifts in $\delta g_R^b$
outside the experimental 99\%~CL range (light gray band) would require
the bulk mass parameter of the right-handed top quark to be
significantly larger than 1. Such a choice appears unnatural, since
$c_{t_R} > 1$ implies that the corresponding bulk mass exceeds the
curvature scale, in which case the right-handed top quark should be
treated as a brane-localized and not a bulk fermion. The latter
feature can be inferred from the ticks on the upper border of the
frame in the right panel. They have been obtained by combining the
equality $m_b = v/\sqrt{2} \, \left |(Y_d)_{33} \right| \left
|F(c_{b_L}) \hspace{0.25mm} F(c_{b_R}) \right |$ with the one for
$m_t$ given earlier, solving again for $c_{t_R}$. The Yukawa
parameters have been fixed to $|(Y_d)_{33}| = 1$ and $|(Y_u)_{33}| =
3$. For smaller (larger) values of $|(Y_d)_{33}|$ the ticks move to
the right (left). Rescaling $|(Y_u)_{33}|$ has the opposite
effect. This observation leads us to the conclusion that,
irrespectively of the bulk gauge group, naturalness in combination
with the requirement to reproduce the observed top- and bottom-quark
masses excludes large corrections to $\delta g_R^b$ in models of
warped extra dimensions in which the left-handed bottom and top quark
reside in the same multiplet. This model-independent conclusion should
be contrasted with the analysis \cite{Djouadi:2006rk}, which finds
sizable corrections in $\delta g_R^b$. The values of the bulk mass
parameters $c_{b_{L,R}}$ and $c_{t_R}$ considered in the latter
article lead however to bottom- and top-quark masses of $m_b \approx
40 \, {\rm GeV}$ and $m_t \approx 75 \, {\rm GeV}$, which are in
conflict with observation. We finally remark that if the left-handed
bottom and top quarks arise as an admixture of the zero-mode fields of
two $SU(2)_L$ doublets, then the bottom- and top-quark masses are
determined by two independent sets of bulk mass parameters, so that it
is possible to account simultaneously for the quark masses and mixings
as well as the $A_{\rm FB}^{0,b}$ anomaly \cite{Bouchart:2008vp}.

\begin{figure}[!t]
\begin{center} 
\hspace{-2mm}
\mbox{\includegraphics[height=2.85in]{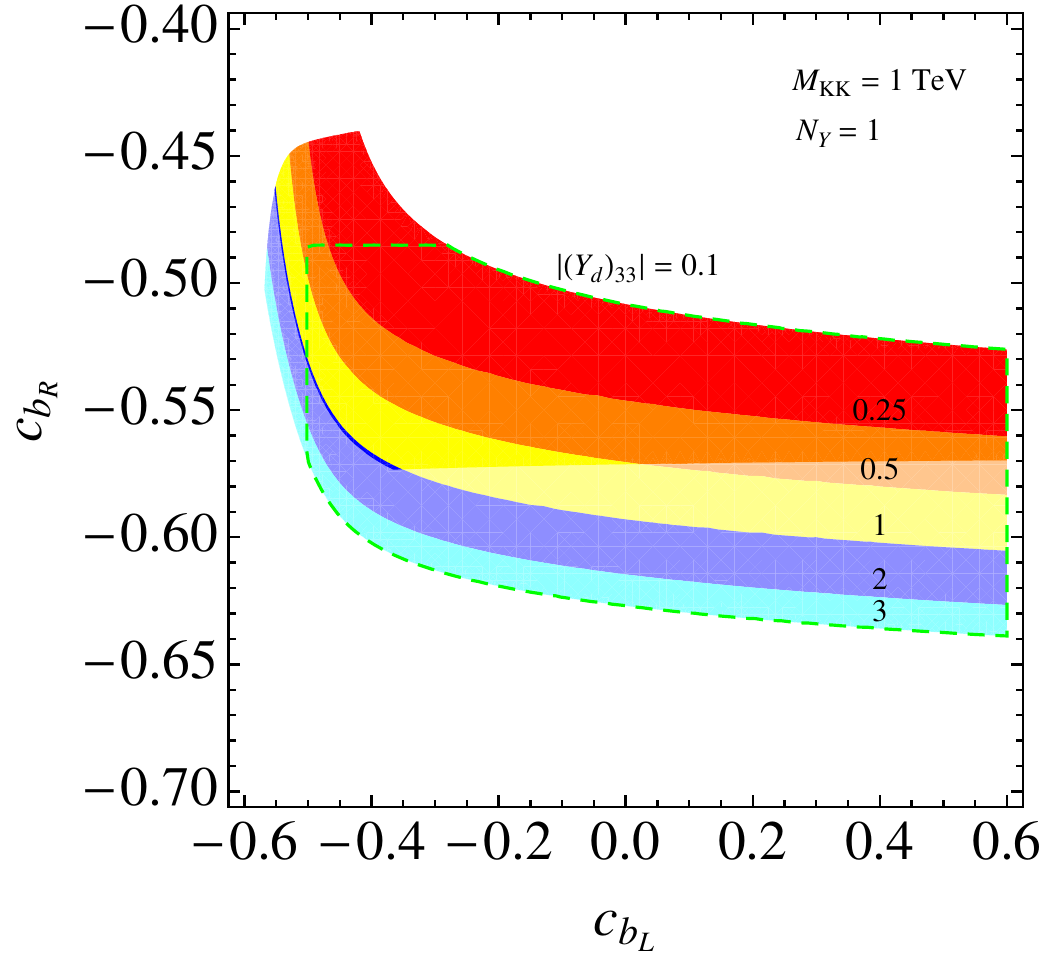}} 
\quad 
\mbox{\includegraphics[height=2.85in]{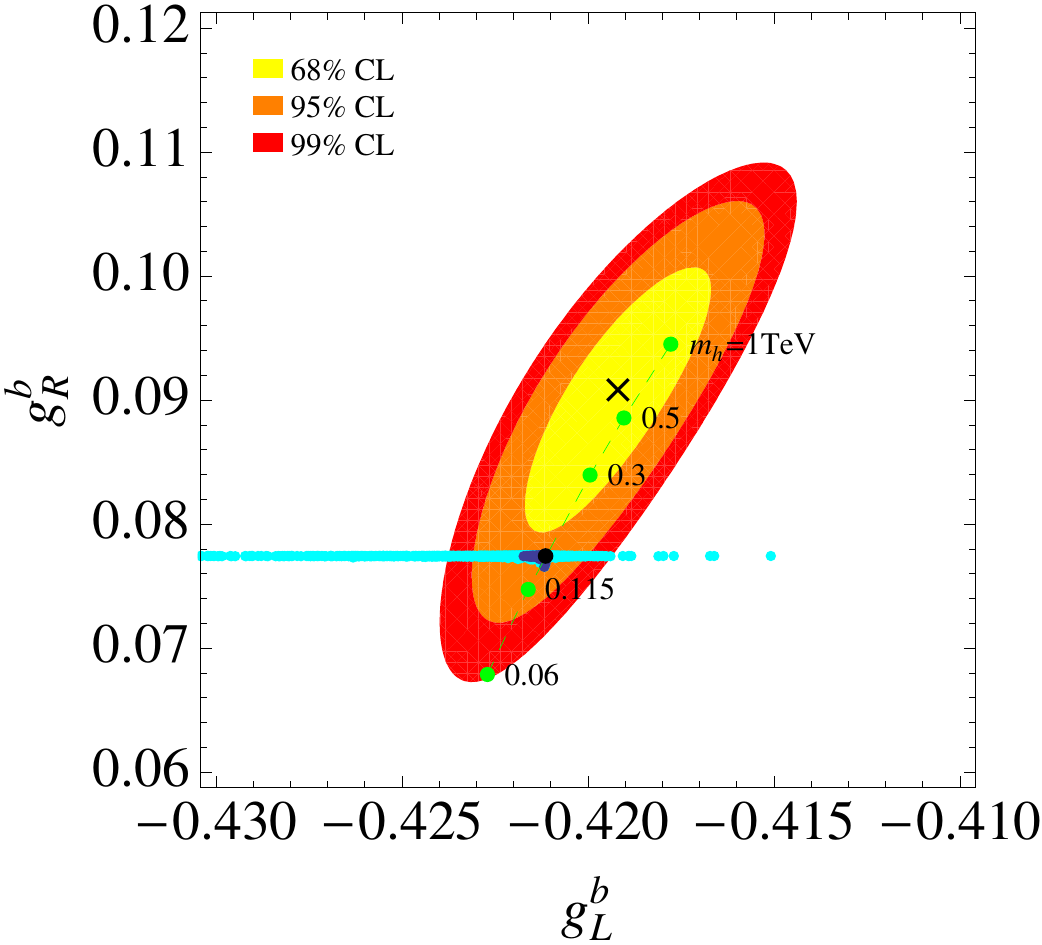}} 
\vspace{-2mm}
\parbox{15.5cm}{\caption{\label{fig:cbgbplots} Left: Region of 99\%
    probability in the $c_{b_L}$--$c_{b_R}$ plane for
    $\Mkk=1$\,TeV. We set $c_{Q_i} = c_{{\cal T}_{1i}} =c_{{\cal
        T}_{2i}} = -1/2$ and $N_Y \equiv |(Y_d)_{ij}|/|(Y_d)_{33}|=1$
    for $i,j = 1,2$. The colored contours indicate the value of
    $|(Y_d)_{33}|$ necessary to reproduce the value of the
    bottom-quark mass. While the whole colored region corresponds to
    the RS results with $c_{b^\prime_R} = c_{b_R}$, only the parameter
    space indicated by bright colors is accessible for $c_{b^\prime_R}
    = 0$. The green dashed, fin-shaped region contains 99\% of the
    parameter points that lead to consistent values of the quark
    masses and mixings. Right: Regions of 68\%, 95\%, and 99\%
    probability in the $g_L^b$--$g_R^b$ plane. The horizontal stripe
    consists of a large number of points in parameter space. The blue
    (cyan) points represent the RS result $c_{b^\prime_R} = c_{b_R}$
    ($c_{b^\prime_R} \neq c_{b_R}$). The black dot is the SM
    expectation for the reference point, and the green dashed line
    indicates the SM predictions for $m_h\in [0.06,1]$\,TeV. See text
    for details.  }}
\end{center}
\end{figure}

The left panel of Figure~\ref{fig:cbgbplots} illustrates the impact of
a possible breaking of the $P_{LR}$ symmetry by the bulk masses
parameters $c_{{\cal T}_{1i}}$. The plot shows the regions of 99\%
probability in the $c_{b_L}$--$c_{b_R}$ plane for $\Mkk=1$\,TeV under
the restriction $0.1<|(Y_d)_{33}|<3$. The colored contours indicate
the magnitude $|(Y_d)_{33}|$ necessary to achieve the correct value of
the bottom-quark mass. Requiring in addition a consistent value of the
quark masses and mixings restricts the parameter space further. This
is indicated by the green dashed, fin-shaped region in the left panel
of Figure~\ref{fig:cbgbplots}, which contains 99\% of the allowed
parameter points. While in the case of an extended $P_{LR}$ symmetry
the whole colored region is accessible, allowing for $c_{b_R^\prime}
\neq c_{b_R}$ can cut away a sizable part of parameter space. This is
demonstrated by the bright colored region, which corresponds to the
choice $c_{b_R^\prime} = 0$. Notice that the $P_{LR}$ breaking
correction to $g_{b_L}$ in (\ref{eq:ZbbRS}) arising from
$c_{b_R^\prime} \neq c_{b_R}$ scales like $-v^2/\Mkk^2 \, | (Y_d)_{33}
|^2 \, F^2 (c_{b_L})$. This explains why values $|(Y_d)_{33}| \gtrsim
1$ are not compatible with the $Z \to b \bar b$ data in the case
$c_{b_R^\prime} = 0$.

The possible size of $P_{LR}$ symmetry-breaking corrections is shown
in the right panel of Figure~\ref{fig:cbgbplots}, which displays the
regions of $68\%$, $95\%$, and $99\%$~CL obtained from a global fit to
the $Z \to b\bar b$ pseudo observables (\ref{eq:bPOsexp}). The
predictions in the RS model with (without) extended $P_{LR}$ symmetry
are superimposed as blue (cyan) scatter points. The shown points
correspond to 10000 random choices of parameters with $\Mkk = [1, 10]
\, {\rm TeV}$, $|(Y_{u,d})_{ij}| \in [0.1, 3]$, ${\rm arg} \left
  ((Y_{u,d})_{ij} \right ) \in [0, 2 \pi[$, and $c_{t_R}\in\,]-1/2,
1]$ that reproduce the quark masses and mixings with a global
$\chi^2/n_{\rm dof}$ better than 11.5/10, corresponding to $68\%$~CL.
The cyan points have been obtained by allowing the bulk mass
parameters $c_{{\cal T}_{1i}}$ with $i = 1, 2, 3$ to take any value in
the range $[-1,0]$. We see that in the former case the small RS
contributions always drive $g_L^b$ to smaller values with respect to
the SM reference point (black dot), while in the latter case moderate
positive and large negative corrections in $g_L^b$ are possible,
leading further away from the best fit values $g_L^b = -0.41918$ and
$g_R^b = 0.090677$ (black cross). In both cases $g_R^b$ remains
essentially unaffected. Thus, like in the minimal model
\cite{Casagrande:2008hr}, the corrections (\ref{eq:ZbbRS}) alone
cannot account for the positive shift in $g_R^b$ needed to explain the
anomaly in $A_{\rm FB}^{0,b}$.\footnote{The corrections to $g_R^b$ are
  always negative but small and hence hardly visible in the figure.}

A perfect fit to the $Z \to b \bar b$ data can however be achieved by
allowing for a heavy Higgs boson, because the shifts
\begin{equation} \label{eq:Zbbmh}
  \Delta g_L^b = 1.77 \cdot 10^{-3} \, \ln \frac{m_h}{m_h^{\rm ref}} \,, \qquad 
  \Delta g_R^b = 0.92 \cdot 10^{-2} \, \ln \frac{m_h}{m_h^{\rm ref}}
\end{equation}
in $g_{L,R}^b$ due to a Higgs-boson mass different from the reference
value $m_h^{\rm ref} = 150 \, {\rm GeV}$ are both positive for $m_h >
m_h^{\rm ref}$. The latter relations parametrize the leading
logarithmic Higgs-mass dependences of $g_{L,R}^b$ and have been
derived with the help of {\tt ZFITTER}
\cite{Arbuzov:2005ma}.\footnote{The default flags of {\tt ZFITTER}
version 6.42 are used, except for setting {\tt ALEM=2} to take into
account the externally supplied value of $\Delta\alpha^{(5)}_{\rm
had}(m_Z)$.} The exact shifts in the $Z b \bar b$ couplings for $m_h
\in [0.06, 1] \, {\rm TeV}$ are indicated by the green dashed line in
the right panel of Figure~\ref{fig:cbgbplots}. One observes that a
Higgs-boson mass in the ballpark of $m_h=0.5$\,TeV would bring the
predictions of $g_{L,R}^b$ so close to the best fit values that
already the small corrections in the RS model with extended $P_{LR}$
symmetry are sufficient to reach the minimum of the $\chi^2$
distribution. Warped models with the Higgs field localized in the IR
might thus indirectly allow for an explanation of the $A_{\rm
FB}^{0,b}$ anomaly, since in these set-ups the Higgs boson is
naturally expected to be heavy, which leads to a good agreement
between the $Z\to b\bar b$ data and theory. Such an option is however
not unproblematic, since the presence of a heavy Higgs boson can
potentially spoil the global electroweak fit in RS models with
custodial protection of the $T$ parameter \cite{Casagrande:2008hr}.

\subsection{Rare Decay \boldmath$t\rightarrow cZ$\unboldmath}
\label{sec:tcZ}

As the top quark, being the heaviest fermion in the SM, is localized
closest to the IR brane, it couples most strongly to the KK
excitations of the gauge bosons. It is thus natural to expect sizable
effects in processes involving flavor-violating top-quark
couplings. Since FCNCs in the up-type quark sector are less
constrained by $K$- and $B$-meson physics than those in the down-type
quark sector, the decay $t\rightarrow cZ$ provides a promising test of
RS models.

From (\ref{eq:Zff}) we derive the branching ratio for this decay,
which is given to excellent approximation by \cite{Casagrande:2008hr}
\begin{equation}
  \begin{split}
    {\cal B}(t\to c Z) &= \frac{2\left( 1-r_Z^2 \right)^2 \left(1+2r_Z^2
      \right)}%
    {\left( 1-r_W^2 \right)^2 \left( 1+2r_W^2 \right)} \\
    &\quad\times \left\{ \left| \left( g_L^u \right)_{23} \right|^2 +
      \left| \left( g_R^u \right)_{23} \right|^2 - \frac{12r_c r_Z^2}
      {\left( 1-r_Z^2 \right) \left( 1+2r_Z^2 \right)}\, \mbox{Re}\big[
      \left( g_L^u \right)_{23}^\ast
      \left( g_R^u \right)_{23} \big] \right\} \\
    &\approx 1.842\,\Big[ \left| \left( g_L^u \right)_{23} \right|^2 +
    \left| \left( g_R^u \right)_{23} \right|^2 \Big] -
    0.048\,\mbox{Re}\big[ \left( g_L^u \right)_{23}^\ast \left( g_R^u
    \right)_{23} \big] \,, 
  \end{split}
\end{equation}
where $r_i\equiv m_i^{\rm pole}/m_t^{\rm pole}$, and for simplicity we
have only kept terms up to first order in $v^2/\Mkk^2$ and the
charm-quark mass ratio $r_c\approx 8.7\cdot 10^{-3}$. The
flavor-changing couplings in the custodial model are given by
\begin{equation}
  \begin{split}
    \left( g_L^u \right)_{23} &= - \frac{m_Z^2}{2\Mkk^2} \left(
      \frac12 - \frac23 s_w^2 \right) \Big[ \omega_Z^{u_L} L \left(
      \Delta_U \right)_{23} - \left( \Delta'_U \right)_{23} \Big]
    - \left( \delta_U \right)_{23} \,, \\
    \left( g_R^u \right)_{23} &= -\frac{m_Z^2}{2\Mkk^2}\,\frac23
    s_w^2\, \left( \Delta'_u \right)_{23} + \left( \delta_u
    \right)_{23} \, .
  \end{split}
\end{equation}
The ZMA expressions for the matrices $\bm{\Delta}_U$,
$\bm{\Delta}'_U$, and $\bm{\Delta}'_u$ are obtained from (\ref{ZMA1})
by the replacements $c_{{\cal T}_{2i}} \to
c_{u^c_i}$, $\bm{U}_d \to \bm{U}_u$, and $\bm{W}_d \to \bm{W}_u$. In
the same approximation one has $\bm{\delta}_U = 1/2 \, \bm{\Phi}_U$
with $\bm{\Phi}_U$ introduced in (\ref{eq:PhiU}) and
\begin{equation} \label{eq:deltau}
  {\bm \delta}_u = \frac{1}{2} \; {\bm x}_u \, {\bm U}_u^\dagger \, \,
  {\rm diag }\left [ \frac{1}{1 - 2 \hspace{0.25mm}c_{Q_i}} \left (
      \frac{1}{F^2(c_{Q_i})} \left [ 1 - \frac{1 - 2 \hspace{0.25mm}
          c_{Q_i}}{F^2(-c_{Q_i})} \right ] - 1 + \frac{F^2(c_{Q_i})}{3 + 2
        \hspace{0.25mm} c_{Q_i}} \right ) \right ] \, {\bm U}_u \, {\bm x}_u
  \,.
\end{equation}
Notice that, compared to the ZMA result in the minimal RS model
\cite{Casagrande:2008hr}, the mixing matrix $\bm{\delta}_u$ contains a
additional term involving the zero-mode profile $F(-c_{Q_i})$.

\begin{figure}[!t]
\begin{center} 
\hspace{-2mm}
\mbox{\includegraphics[height=2.85in]{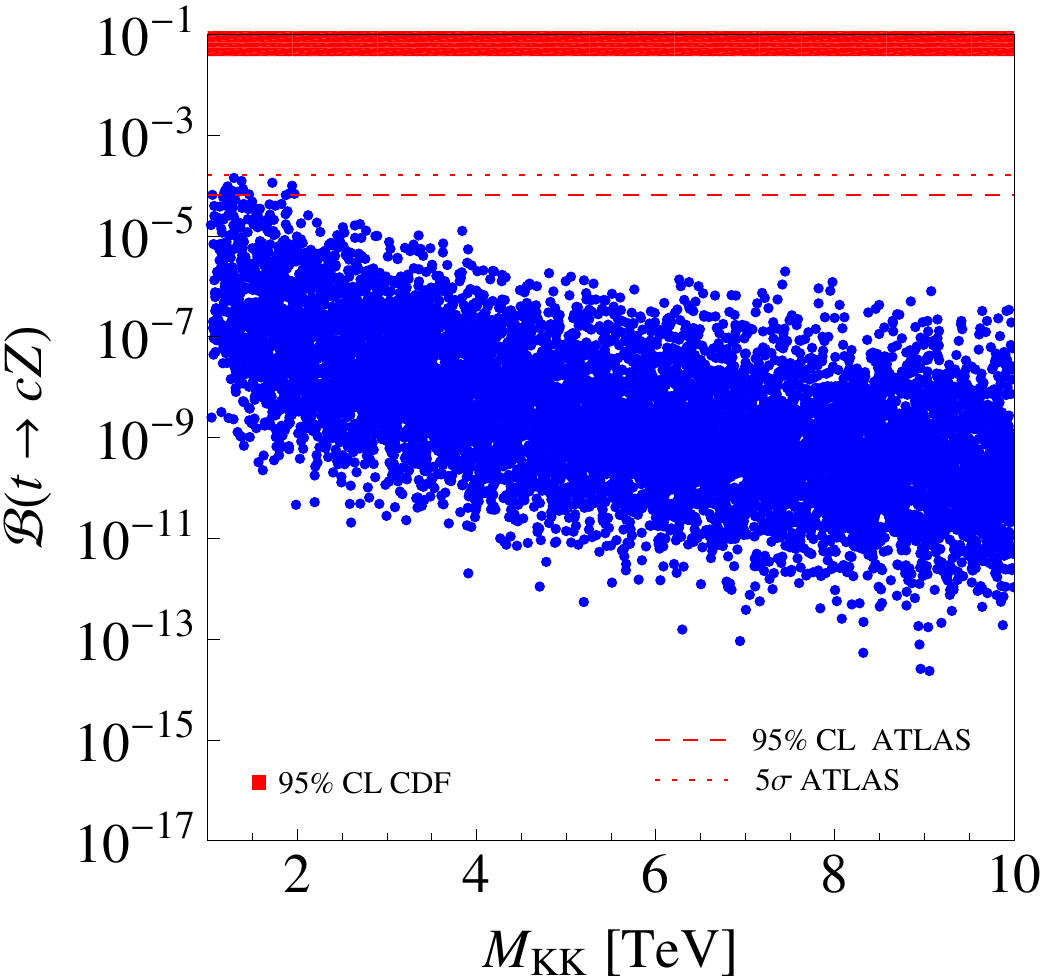}} 
\qquad 
\mbox{\includegraphics[height=2.85in]{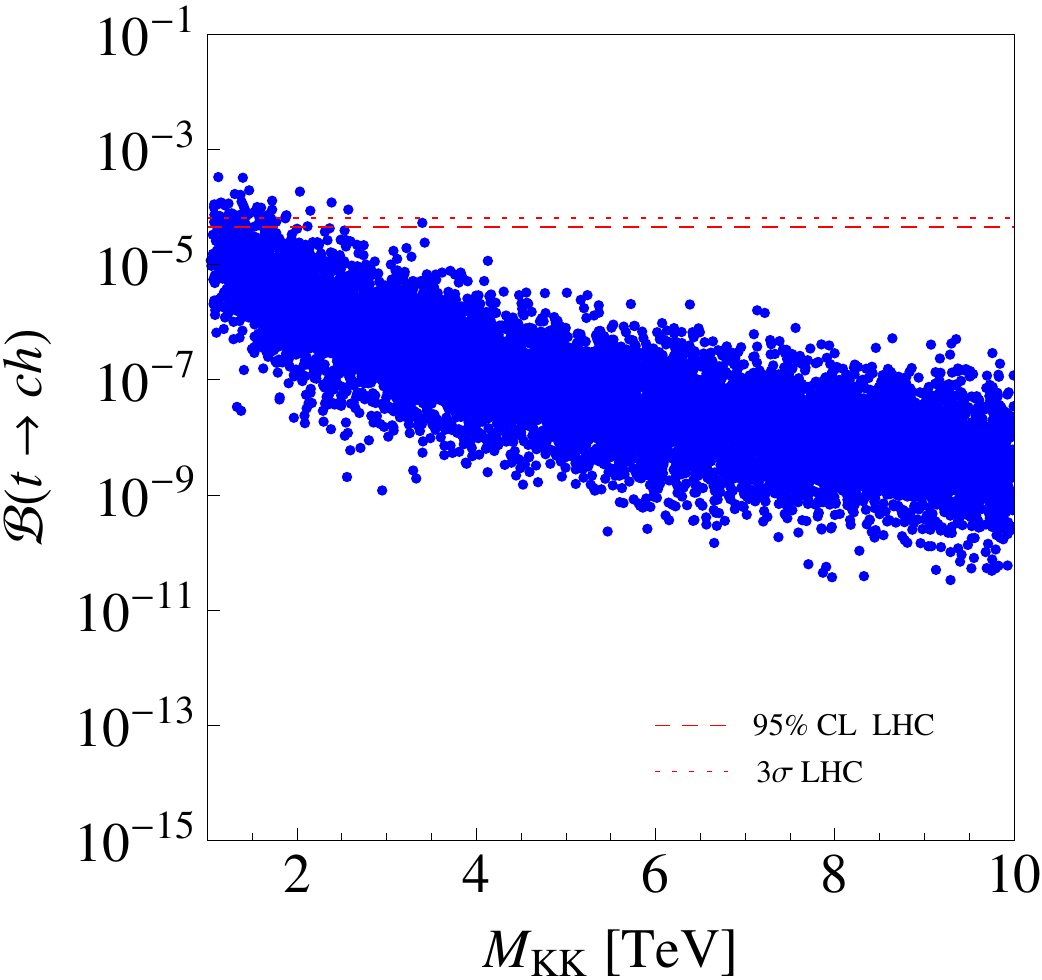}}
\vspace{-2mm}
\parbox{15.5cm}{\caption{\label{fig:tcplots} Branching ratio of the
    rare decays $t\to c Z$ (left) and $t \to c h$ (right) as functions
    of $\Mkk$ in the RS model with extended custodial protection
    $c_{{{\cal T}}_{1 i}} = c_{{{\cal T}}_{2 i}}$. The red band in the
    left panel is excluded at 95\%~CL by the CDF search for $t\to u(c)
    Z$. The red dotted and dashed lines in the left (right) plot
    indicate the expected discovery and exclusion sensitivities of
    ATLAS (LHC) for 100\,fb$^{-1}$ integrated luminosity. All scatter
    points reproduce the correct quark masses, mixing angles, and the
    CKM phase. See text for details.}}
\end{center}
\end{figure}

Inserting the quantum numbers of the representation
(\ref{eq:multiplets}) into (\ref{eq:omegaZ}), we see that the leading
contribution to $\left( g_L^u \right)_{23}$ is enhanced by a factor
\begin{equation} 
\omega_Z^{u_L} = \frac{2 c_w^2}{1 - \frac{4}{3} s_w^2}
\approx 2.2 \,.
\end{equation} 
In contrast to the minimal model \cite{Casagrande:2008hr}, the
right-handed coupling does not receive an $L$-enhanced contribution,
because $\omega_Z^{u_R}=0$. Moreover, the contribution inversely
proportional to $F^2(c_{Q_i})$ in $\bm{\delta}_u$ is highly suppressed
if $c_{Q_i} < 1/2$, since $F^2(-c_{Q_i}) \approx 1 - 2 \hspace{0.25mm}
c_{Q_i}$ in such a case. The leading corrections to the $Z u_R^i \bar
u_R^j$ vertices due to quark mixing are therefore protected by the
custodial symmetry. While these features remove a possible source of
large effects associated with the composite nature of the right-handed
top quark, they imply that the chirality of the $Z tc$ interactions in
the model under consideration is predicted not to be right-handed, as
argued in \cite{Agashe:2006wa}, but left-handed. Of course, other
choices of the quantum numbers of the right-handed up-type quarks than
those in Table~\ref{tab:charges} are possible, so that the RS
framework does not lead to a firm prediction of the chirality of the
$Z tc$ interactions.

The predictions for ${\cal B}(t\to c Z)$ in the custodial RS model
with extended $P_{LR}$ symmetry as a function of $\Mkk$ are shown in
the left panel of Figure~\ref{fig:tcplots}. The experimental upper
bound on FCNC $t\to u(c) Z$ from the CDF experiment amounts to ${\cal
  B}(t\to u(c) Z)<3.7\%$ at 95\%~CL \cite{tcZ:2008aaa} and is shown as
a band. At the LHC, one can search for rare FCNC top-quark transitions
in top-quark production and decays. The ATLAS \cite{Carvalho:2007yi}
and CMS \cite{Ball:2007zza} collaborations have examined this
possibility in simulation studies. The minimal branching ratio ${\cal
  B}(t\to c Z)$ allowing for a $5\sigma$ signal discovery with
100\,fb$^{-1}$ integrated luminosity is expected to be $1.6\cdot
10^{-4}$ at ATLAS. In the absence of a signal, the expected upper
bound at 95\%~CL is $6.5\cdot 10^{-5}$. These values are visualized by
the red dotted and dashed lines in the plot. Our numerical studies
show that for low KK mass scales in the ballpark of 2
TeV,\footnote{Corresponding to masses of the lightest KK gauge bosons
  of around 5 TeV.} which is a realistic possibility in RS models with
custodial protection, the branching ratio ${\cal B}(t\to c Z)$ can
come close to the region which can be probed at the LHC.\footnote{As a
  result of $|F(c_{Q_1})|/|F(c_{Q_2})| \sim \lambda$ the branching
  ratio of $t \to uZ$ is typically suppressed by two orders of
  magnitude compared to $t \to cZ$, rendering the former mode
  unobservable at the LHC. Similar statements apply to the branching
  ratio of $t \to uh$.} In the minimal RS model the possible branching
ratios are smaller, since the strong correlation between the $Ztc$ and
$Zb \bar b$ couplings generically leads to a rejection of points with
large ${\cal B}(t\to c Z)$ \cite{Casagrande:2008hr}. The custodial
protection of the $Z b_L \bar b_L$ vertex thus leads indirectly to
improved prospects of a detection of the decay $t \to c Z$ at the LHC.

\subsection{Rare Decay \boldmath$t\rightarrow ch$\unboldmath}
\label{sec:tch}

The general form of the interactions of fermions with the Higgs boson
has been given in (\ref{eq:hff}). These couplings allow for the
flavor-changing decay $t\to c h$ with a branching ratio
\begin{eqnarray} \label{BRthc} 
  {\cal{B}}(t\to c h) = \frac{2 \left( 1-r_h^2 \right)^2 r_W^2}{\left(
      1-r_W^2 \right)^2 \left( 1+2r_W^2 \right) g^2} \left\{ \left| \left(
        g_h^u \right)_{23} \right|^2 + \left| \left( g_h^u \right)_{32}
    \right|^2 + \frac{4r_c}{1-r_h^2}\, \mbox{Re}\big[ \left( g_h^u
    \right)_{23} \left( g_h^u \right)_{32} \big] \right\} , \hspace{8mm}
\end{eqnarray}
where as before $r_i\equiv m_i^{\rm pole}/m_t^{\rm pole}$, and $g$ is
the $SU(2)_L$ gauge coupling. Again, we have included terms up to
first order in the charm-quark mass. In our numerical analysis we will
use $r_h=0.87$, corresponding to a Higgs-boson mass $m_h=150$\,GeV.

The predictions for ${\cal B}(t\to c h)$ in the custodial RS model
with extended $P_{LR}$ symmetry as a function of $\Mkk$ are shown in
the right panel of Figure~\ref{fig:tcplots}. The LHC is expected to
provide a $3\sigma$ evidence for ${\cal B}(t\to ch)$ larger than
$6.5\cdot 10^{-5}$ or set an upper bound of $4.5\cdot 10^{-5}$ with
95\%~CL if the decay is not observed \cite{AguilarSaavedra:2000aj}.
These limits are indicated by the red dotted and dashed lines in the
plot. We see that for low KK mass scales values of the branching ratio
can even exceed the LHC reach, so that a detection of a possible RS
signal with $t \to ch$ could become reality. Let us add that without
inclusion of the Yukawa couplings involving $Z_2$-odd fermion profiles
the obtained branching fractions would be typically smaller by almost
two orders of magnitude. In the minimal RS model the prospects for an
observation of $t \to c h$ turn out to be less favorable, since the
constraints from $Z \to b \bar b$ typically eliminate those points in
parameter space that would show pronounced effects
\cite{Casagrande:2008hr}.

\subsection{Higgs-Boson Production}
\label{sec:higgsproduction}

\begin{figure}
\begin{center}
\mbox{\includegraphics[width=15.5cm]{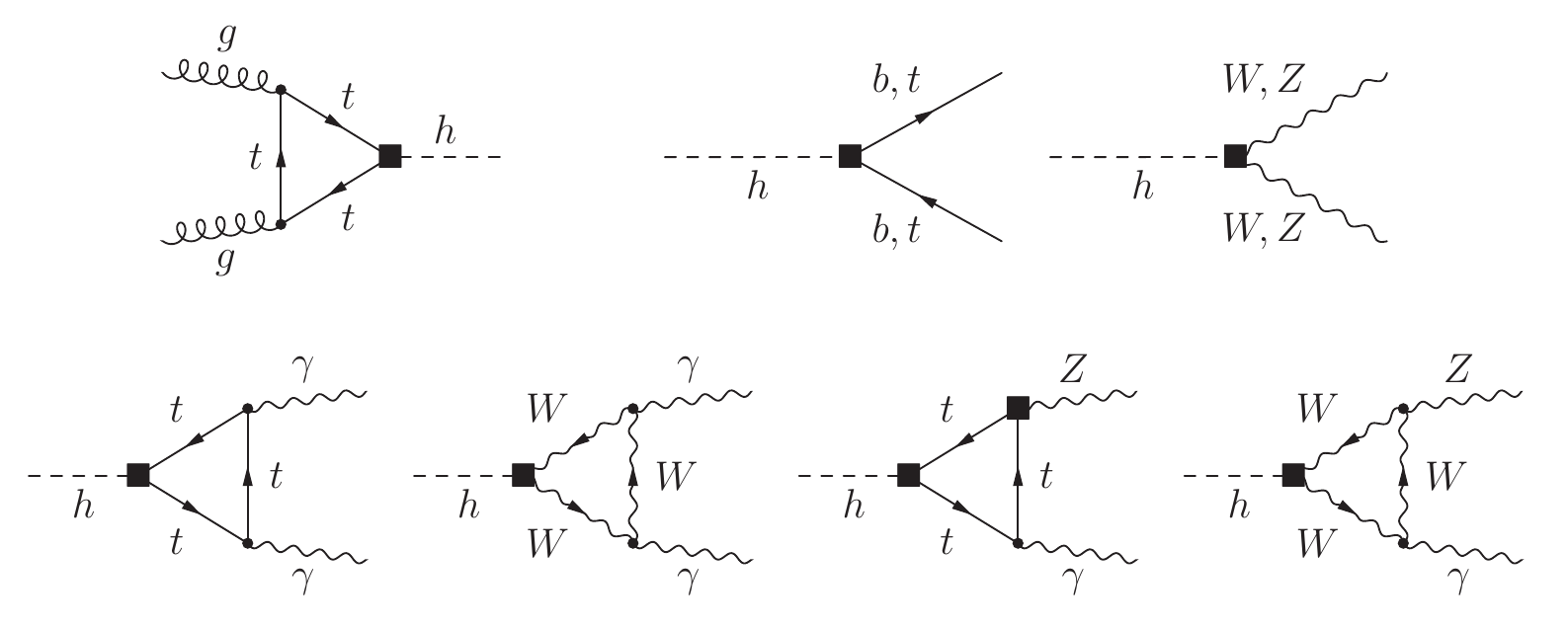}} 
\parbox{15.5cm}{\caption{\label{fig:hprodec} Examples of Feynman
    diagrams involving zero-mode fields only that contribute to the
    production and the decay of the Higgs boson at leading order of
    perturbation theory. Vertices indicated by a black square can
    receive sizable shifts in the RS model relative to the SM
    couplings. See text for details.}}
\end{center}
\end{figure}

\begin{figure}
\begin{center}
\mbox{\includegraphics[width=12.5cm]{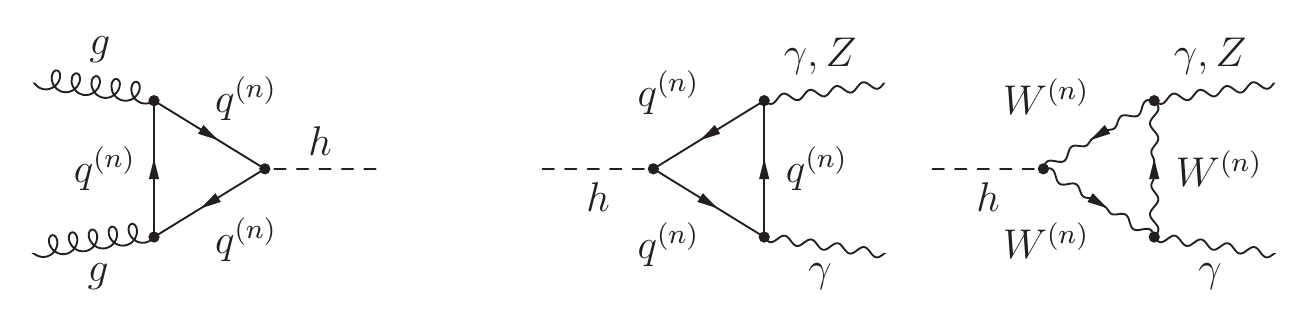}} 
\parbox{15.5cm}{\caption{\label{fig:hkkcontr} Examples of one-loop
    contributions involving KK excitations that contribute to the
    production and the decay of the Higgs boson at leading order of
    perturbation theory. See text for details.}}
\end{center}
\end{figure}

At hadron colliders such as the Tevatron or the LHC the leading
production mechanism of the Higgs boson is gluon-gluon fusion, which
receives its dominant contribution from a top-quark triangle loop.
Within the RS framework, one has to take into account the whole KK
tower of the top as well as the other quark flavors, since all these
modes contribute to the $gg \to h$ amplitude at $\ord
(v^2/\Mkk^2)$. The relevant Feynman diagrams are shown on the very
left in the top row of Figure~\ref{fig:hprodec} and on the left-hand
side of Figure~\ref{fig:hkkcontr}.

In order to calculate the $gg \to h$ production cross section in the
RS model, we rescale the SM prediction, employing
\begin{equation} \label{eq:rescale1}
\sigma (gg \to h)_{\rm RS} = \left | \kappa_{g} \right |^2 \,
\sigma (gg \to h)_{\rm SM} \,, 
\end{equation}
where
\begin{equation} \label{eq:kappagg} 
  \kappa_{g} = \frac{{\displaystyle
      \sum}_{i = t, b} \, \kappa_i \hspace{0.25mm} A_{q}^h
    (\tau_i)\, + \hspace{0.25mm}{\displaystyle \sum}_{j = u,d,\lambda}  
    \, \nu_j }{ {\displaystyle \sum}_{i = t,
      b} \; A_{q}^h (\tau_i)} \,,
\end{equation}
with $\tau_i \equiv 4 \hspace{0,25mm} m_i^2/m_h^2$. The first term in
the numerator encodes the effects due to zero modes running in the
loop and the corresponding sum includes both the virtual top- and
bottom-quark contributions. The form factor $A_{q}^h (\tau_i)$
approaches $1$ for $\tau_i \to \infty$ and vanishes proportional to
$\tau_i$ for $\tau_i \to 0$. Its analytic form is given in
Appendix~\ref{app:formfactors}. As they are power suppressed, the only
phenomenologically relevant correction in $\sigma (gg \to h)_{\rm SM}$
due to lighter quarks is the interference term of the bottom- and
top-quark amplitudes. Its effect can be approximated by multiplying
the cross section $\sigma (gg \to h)_{\rm SM}$ without the
bottom-quark corrections by $\big (1 + 2 \, {\rm Re} \hspace{0.25mm}
A_{q}^h (\tau_b) \big )$. Numerically, this approximate treatment
decreases the SM cross section by about $9\%$, $2\%$, and below $1\%$
for $m_h = 100 \, {\rm GeV}, 300 \, {\rm GeV}$, and $600 \, {\rm
  GeV}$, in good agreement with the next-to-leading order calculation
including the exact mass dependence \cite{Spira:1995mt}. In our
numerical evaluation of the SM Higgs-boson production cross section
via $gg \to h$, the bottom-quark contribution will be included using
the lowest-order approximation.

\begin{figure}[!t]
\begin{center} 
\hspace{-2mm}
\mbox{\includegraphics[height=2.85in]{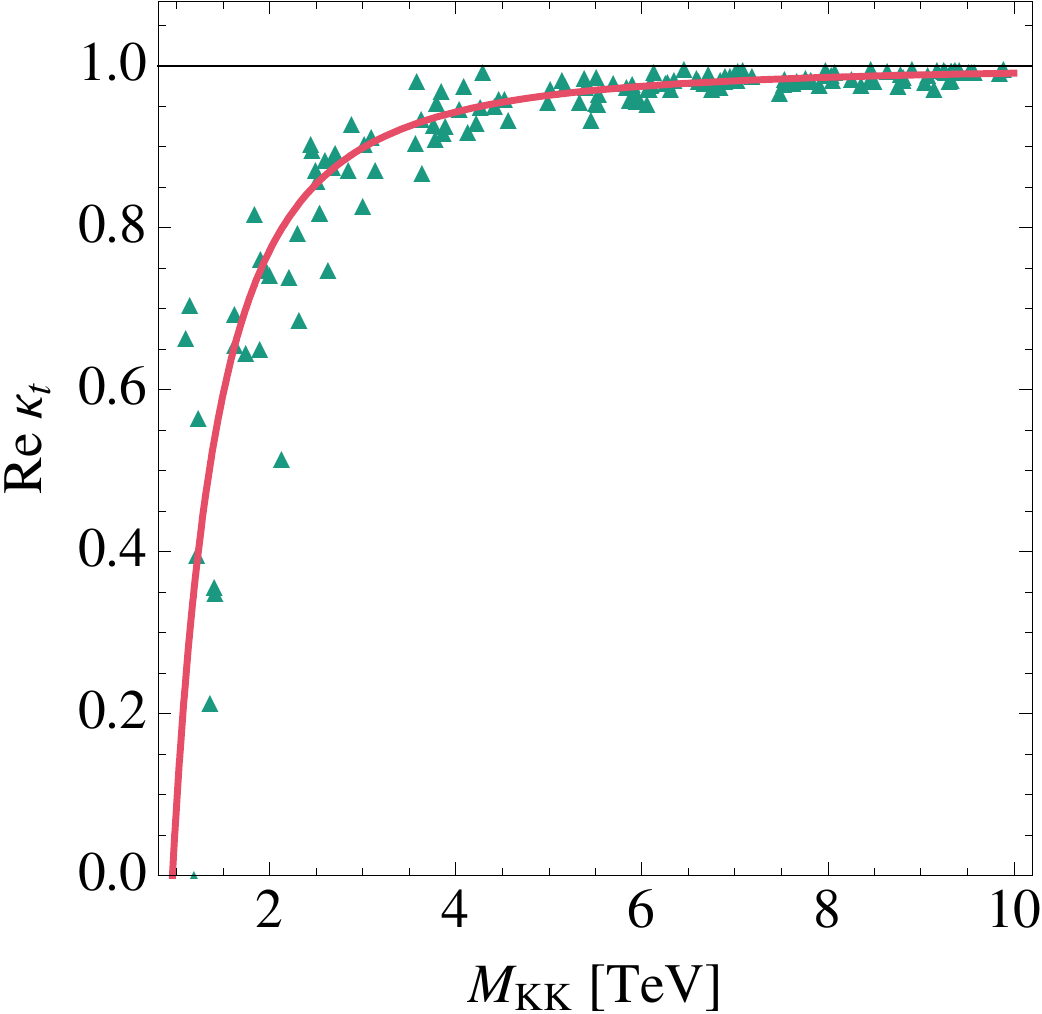}} 
\hspace{4mm}
\mbox{\includegraphics[height=2.85in]{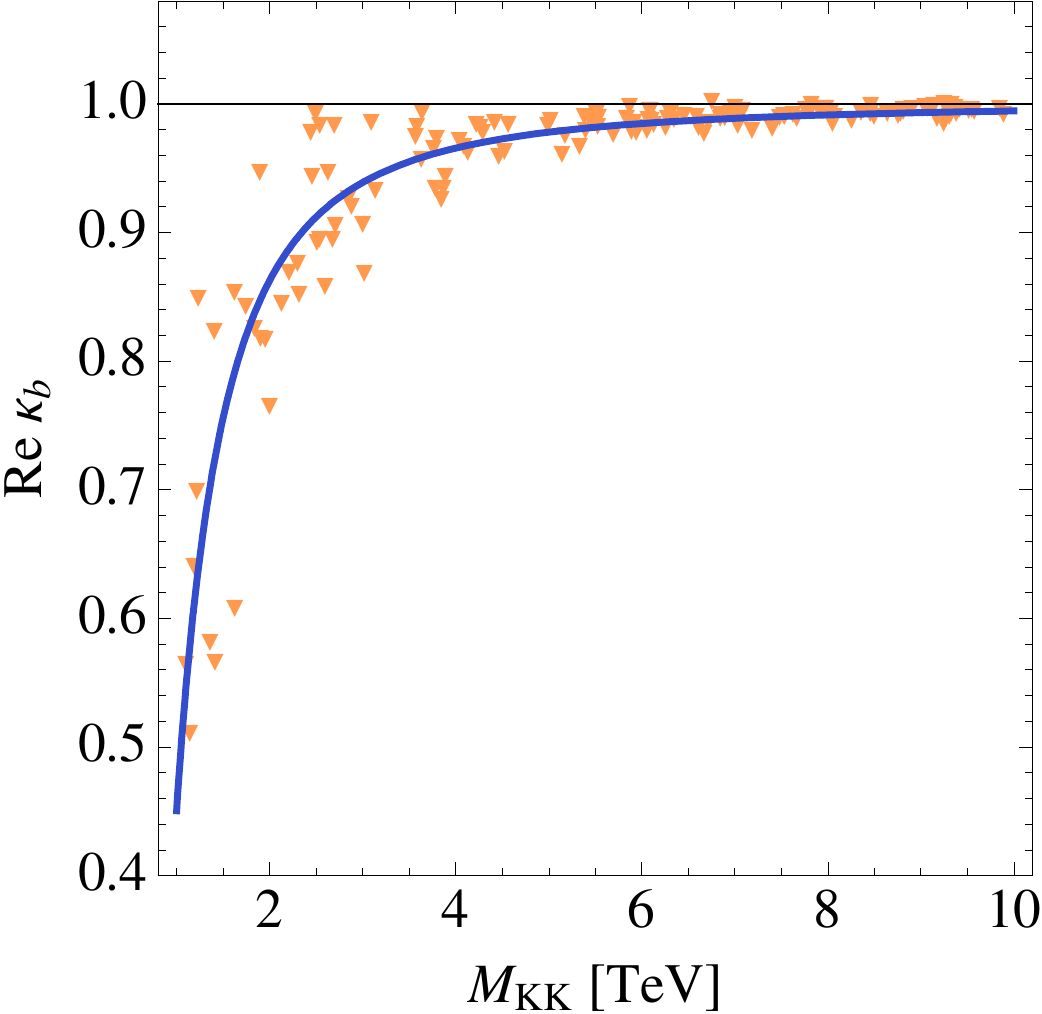}}
\parbox{15.5cm}{\caption{\label{fig:kappa} Predictions for the real
    parts of the ratios of the $ht\bar t$ (left) and $hb\bar b$
    (right) coupling in the custodial RS model relative to the SM
    value. The solid lines show fits to the samples of parameter
    points. See text for details.
  }}
\end{center}
\end{figure}

The quantities
\begin{equation}\label{tbratios}
\kappa_{t} = 1 - \frac{v}{m_{t}} \hspace{0.5mm} 
(\Delta g_h^u)_{33}\,, \qquad 
\kappa_{b} = 1 - \frac{v}{m_{b}} \hspace{0.5mm}
(\Delta g_h^d)_{33}
\end{equation} 
appearing in (\ref{eq:kappagg}) describe the ratios of Higgs-boson
couplings to heavy quarks in the custodial RS model relative to the
corresponding SM values.  In Figure~\ref{fig:kappa} we show the real
parts of these ratios as a function of $\Mkk$ for a set of 150 random
parameter points corresponding to the model with the extended $P_{LR}$
symmetry (\ref{eq:extendedPLR}), which we will always employ in our
numerical analysis. The same sample of model parameter points will be
used in the remainder of this paper.  Since the imaginary parts of
$\kappa_{t,b}$ turn out to be small (the origin of this feature will
be explained below), they are not shown in the figures.  We observe that
both the $ht \bar t$ and the $h b \bar b$ coupling are reduced in the
RS scenario with respect to the SM, resulting in ${\rm Re} \,
\kappa_{t, b} \leq 1$. The same conclusion has been reached in
\cite{Azatov:2009na} for the minimal RS model. Numerically, we find
that for $\Mkk = 2 \, {\rm TeV}$ ($\Mkk = 3 \, {\rm TeV}$) the average
corrections amount to around $-25$\% and $-15$\% ($-10$\% and $-5$\%)
in the top- and bottom-quark sectors, respectively. Since the RS
corrections to the Yukawa couplings scale as $v^2/M_{\rm KK}^2$, the
average value of the ratios $\kappa_{t, b}$ can be parametrized by
$1-a_{t, b} \, v^2/\Mkk^2$ with the coefficients $a_{t,b}$ given in
Table~\ref{tab:kappas}. The quoted values of $a_{t,b}$ have been
obtained from the best fits to the shown sample of scatter points.

The suppression of the Yukawa couplings of the third-generation
quarks, ${\rm Re} \, \kappa_{t,b} \leq 1$, as well as the feature
$|{\rm Im} \, \kappa_{t,b}| \ll 1$ are not difficult to
understand. First, one has $m_3^q/v \, \big ( (\Phi_q)_{33} +
(\Phi_Q)_{33} \big) \geq 0$ since the diagonal elements of the
matrices $\bm{\Phi}_{q,Q}$ introduced in (\ref{eq:Phi}) are absolute
squares. Second, the third term in (\ref{eq:Higgscorrection}) can be
written in the ZMA as
\begin{equation} \label{eq:gtildehZMA}
\begin{split}
  (\Delta \tilde g^u_h)_{33} & = \frac{4 m_t^2}{3 v \Mkk^2} \,
  \sum_{j=1}^3 \, m_j^u \; \Big ( \bm{U}_u^\dagger \; {\rm diag} \left
    [ F^{-2}(c_{Q_i}) \right ] {\bm{U}_u} \Big )_{j3} \, \Big (
  \bm{W}_u^\dagger \; {\rm diag} \left [ F^{-2}(c_{u_i^c}) \right ]
  {\bm{W}_u} \Big )_{3j} \,.
\end{split}
\end{equation}
A similar formula applies to the case of $(\Delta \tilde g^d_h)_{33}$.
Because the diagonal elements of the matrices $\bm{U}_u^\dagger \;
{\rm diag} \left [ F^{-2}(c_{Q_i}) \right ] {\bm{U}_u}$ and
$\bm{W}_u^\dagger \; {\rm diag} \left [ F^{-2}(c_{u_i^c}) \right ]
{\bm{W}_u}$ are absolute squares, the term with $j = 3$ is obviously
positive semi-definite. The terms with $j=1,2$, on the other hand, can
have an arbitrary complex phase. Yet, due to the strong chiral
suppression, $m_c/m_t \approx 1/275$ and $m_u/m_t \approx 10^{-5}$,
the imaginary part of (\ref{eq:gtildehZMA}) turns out to be negligibly
small, leaving us with $(\Delta \tilde g^u_h)_{33} \geq 0$. The same
holds true for $(\Delta \tilde g^d_h)_{33}$, although the chiral
suppression is weaker in this case, $m_s/m_b \approx 1/50$ and
$m_d/m_b \approx 1/800$. Recalling that $(\Delta g_h^q)_{33} = m_3^q/v
\, \big ( (\Phi_q)_{33} + (\Phi_Q)_{33} \big) + (\Delta \tilde
g^q_h)_{33} \geq 0$ enters (\ref{eq:mis1}) with a minus sign, we
conclude that the $ht\bar t$ and $h b \bar b$ couplings are predicted
to be suppressed relative to their SM values in both the minimal and
the extended RS models. We believe that this finding is
model-independent and holds in a wide class of RS set-ups.  The same
conclusion has been drawn in the context of models where the Higgs
arises as a pseudo Nambu-Goldstone boson \cite{Falkowski:2007hz,
  Low:2009di}.

The second term in the numerator of (\ref{eq:kappagg}) represents the
contribution to the $gg \to h$ amplitude arising from the virtual
exchange of KK quarks. The corresponding Feynman graph is shown on the
very left in Figure \ref{fig:hkkcontr}. In the up-type quark sector
the associated coefficient takes the form
\begin{equation}\label{eq:nug}
  \begin{split}
    \nu_u & = \,v \;\sum_{n = 4}^\infty \; \frac{(g_h^u)_{nn}}{m_n^u}
    \hspace{0.5mm} A_{q}^h (\tau_n^u) \\ & = \frac{2 \pi}{\epsilon L}
    \, \sum_{n = 4}^\infty \; \frac{\vec a_{n}^{\hspace{0.25mm}
        U\dagger} \,\bm{C}_{n}^{U} (\pi^-) \left(\displaystyle
        \bm{1}-\frac{v^2}{3\hspace{0.25mm}\Mkk^2} \, \bm{\tilde
          Y}_{\vec u}\bm{\bar Y}_{\vec u}^\dagger\right)
      \bm{S}_{n}^U(\pi^-) \, \vec a_{n}^{\hspace{0.25mm} U} }{x_n^u}
    \hspace{0.5mm} A_{q}^h (\tau_n^u) \,.
  \end{split}
\end{equation}
Similar relations hold in the sector of down-type and $\lambda$
quarks.\footnote{With $\lambda$ quarks we denote all fermionic KK
  excitations with electric charge $5/3$.} Since the mass of the first
KK up-type quark is already much larger than the Higgs-boson mass,
$m_4^u/\Mkk = \ord({\rm a\ few}) \hspace{0.25mm} \gg m_h/\Mkk$, it is
an excellent approximation to replace the function $A_{q}^h
(\tau_n^u)$ by its asymptotic value of 1 obtained for $\tau_n^u \equiv
4 \left (m_n^u \right )^2/m_h^2  \, \to \, \infty$.

Before presenting our numerical results for these contributions, we
would like to add some comments about the convergence of the sum in
(\ref{eq:nug}). In the SM, the top-quark contribution to the $gg \to
h$ amplitude is proportional to $y_t/m_t$ in the decoupling limit. In
this limit the amplitude can be described by the effective operator
$h/v \, G^a_{\mu \nu} G^{a \, \mu \nu}$, whose Wilson coefficient is
related to the QCD $\beta$-function. This relationship arises through
low-energy theorems appropriate to external Higgs bosons with
vanishing momentum \cite{Low:2009di, Ellis:1975ap, Shifman:1979eb,
  Kniehl:1995tn}, which apply to any quantum field theory. In the
context of the RS framework they imply that the sum in (\ref{eq:nug})
must be convergent, because the running of $\alpha_s$ can be shown to
be logarithmic in warped extra-dimension models \cite{Randall:2001gb,
  Pomarol:2000hp, Goldberger:2002cz, Agashe:2002bx, Goldberger:2002hb,
  Contino:2002kc, Choi:2002ps, Goldberger:2003mi}. While the
finiteness of the effective $hgg$ coupling is thus guaranteed on
general grounds, an explicit calculation of (\ref{eq:nug}) in the KK
decomposed 4D theory turns out to be non-trivial. This is due to the
fact that the Higgs VEV induces $\ord(1)$ mixings between the various
modes of a single KK level \cite{Casagrande:2008hr}. For example, in
the up-type quark sector there are five types of fields, namely $u$,
$u^\prime$, $u^c$, $U^\prime$, and $U$. Each of them exists in three
different flavors, so that there are altogether 15 KK modes of similar
mass in each level. In the down-type quark sector, one instead ends up
with nine modes, while in the minimal RS model one has six states per
KK level in both the up- and the down-type quark sectors
(corresponding to $SU(2)_L$ doublets and singlets). Finally, in the
$\lambda$-type quark sector one again faces nine KK excitations per
level. In contrast, exotic matter is not present in the minimal RS
model. Since the mixing effects among the states of the same KK level,
encoded in the eigenvectors $\vec a_n^A$, are large, they cannot be
treated perturbatively, and one has to resort to numerical methods as
long as one is interested in the case of three families. However,
restricting oneself to the simpler case of a single generation, it
turns out to be possible to derive an analytic expression for
(\ref{eq:nug}). This formula will be given in a companion paper.

\begin{figure}[!t]
\begin{center}
\mbox{\includegraphics[height=2.8in]{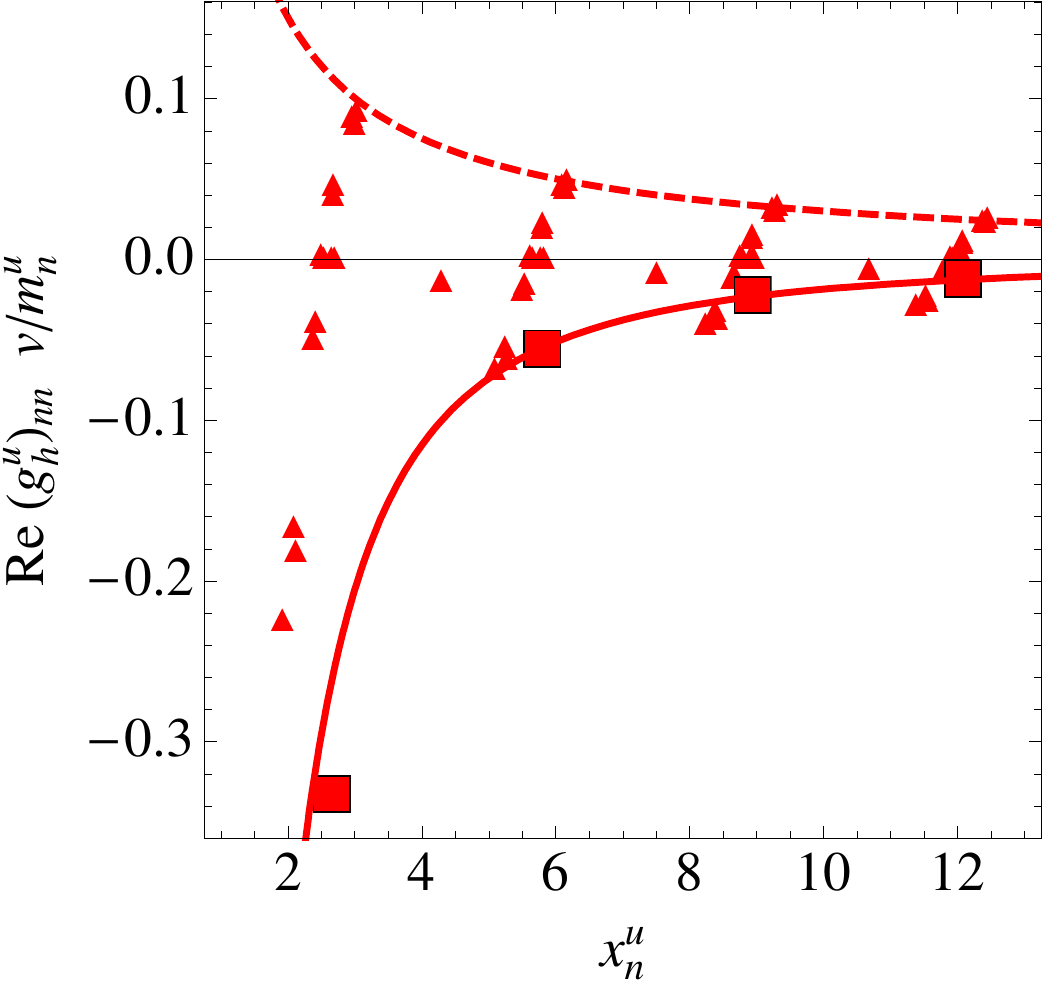}}
\qquad
\mbox{\includegraphics[height=2.8in]{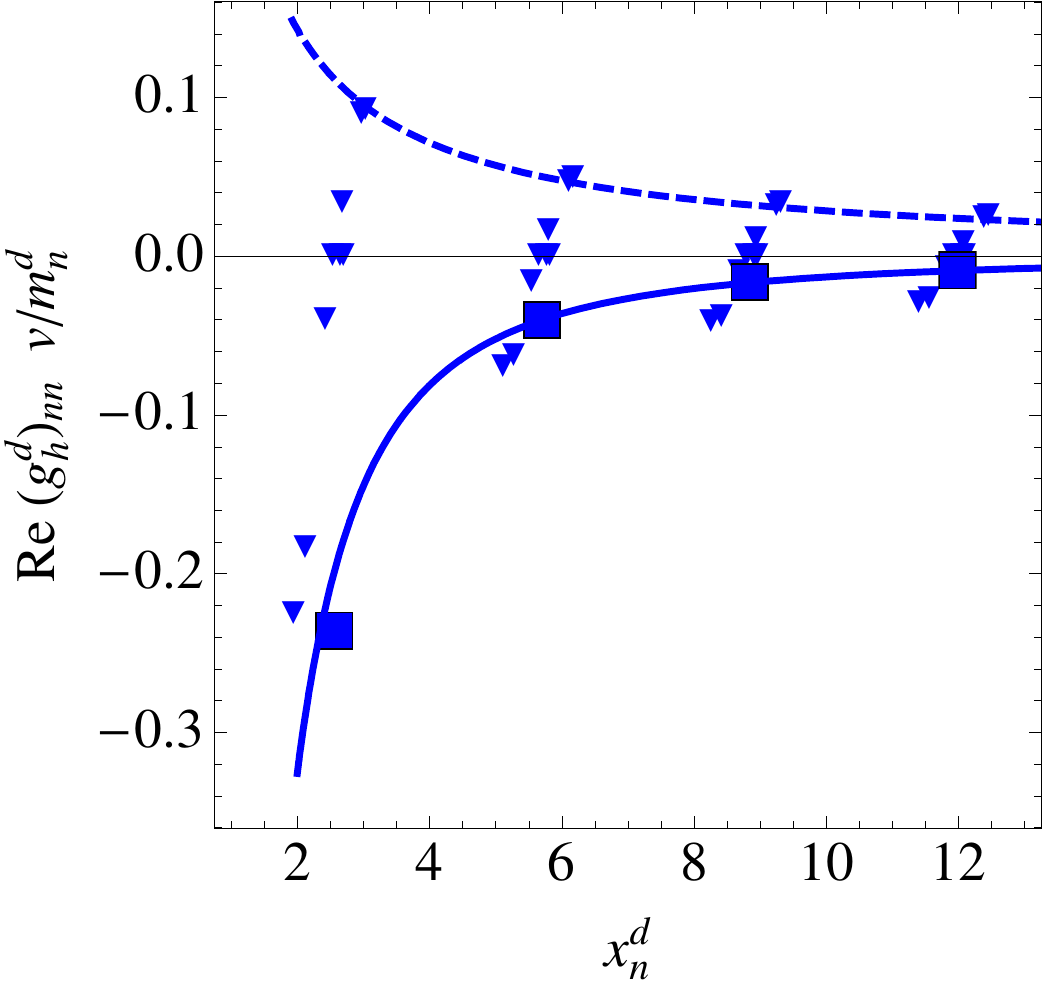}}
\parbox{15.5cm}{\caption{\label{fig:nug} Numerical results for the
    real parts of the coefficients $\nu_{u,d}$ corresponding to a
    specific parameter point with $M_{\rm KK} = \, 2 \, {\rm
      TeV}$. The red (blue) dots in the left (right) panel display the
    first 60 (36) terms in the KK sum for up- (down-type) quarks,
    while the red (blue) filled boxes indicate the sums over a
    complete KK level. See text for details.  }}
\end{center}
\end{figure}

In order to calculate the KK sum numerically, one first has to find
the solutions to the eigenvalue equation
(\ref{fermeigenvals}).\footnote{In the absence of soft $P_{LR}$
  breaking, $c_{{{\cal T}}_{1 i}} = c_{{{\cal T}}_{2 i}}$, three out
  of the 15 (9) states in each up-type ($\lambda$-type) quark KK
  level will have masses that resemble the ones found in the spectrum
  for $(Y_{u,d})_{ij} = 0$. This feature is easy to understand,
  because a unitary transformation ${\cal U}$ acting on the quarks
  $(U,U')^T \rightarrow \, \mathcal{U}\, (U,U')^T$ reshuffles only the
  Yukawa interactions but leaves all other bi-linear terms in the
  action (\ref{Sferm2}) as well as the BCs invariant. The combinations
  $(U-U')/\sqrt{2}$ and $(\Lambda-\Lambda')/\sqrt{2}$ are thus
  unaffected by the Higgs mechanism.} In the case of the up-type quark
sector, this requires determining the roots of a $6\times 6$
determinant, which in practice turns out to be intricate, because one
needs to find suitable starting points to search for the roots.  We
obtain these starting values by diagonalizing a truncated mass matrix
obtained in the perturbative approach \cite{Huber:2003tu,
  Goertz:2008vr, delAguila:2000kb}. In Figure~\ref{fig:nug} we display
the results of our numerical calculations for one parameter point with
$M_{\rm KK} = 2 \, {\rm TeV}$. The dots correspond to the real parts
of individual terms in the sum (\ref{eq:nug}) for up- and down-type
quarks, while the filled boxes indicate the values obtained by summing
up the contributions of one KK level. Results for the exotic
$\lambda$-type quarks are not shown, since they resemble those found
in the down-type quark sector. By inspection of the two panels one
immediately notices two important features of the KK
contributions. First, even though the contribution of an individual
mode can be positive and negative, the sum over an entire KK level is
strictly negative. Second, the importance of higher-level KK sums
decreases quadratically, ensuring that (\ref{eq:nug}) converges to a
finite value. This feature is indicated by the solid lines, which
represent the best fits to $1/x_{n}^2$ including the results of the
second, third, and fourth KK-level sums.\footnote{The dashed lines depict
the $1/x_{n}$ behavior of the sum over a single fermion tower. The convergence 
of the total sum is guaranteed by cancellations between different modes
of the same KK level.} In order to calculate
(\ref{eq:nug}), we then evaluate the corresponding fit at $\bar
x_{1}^{u} + (k - 1) \, \pi$ and resum the resulting series ($k = 1,
\ldots, \infty$) into a trigamma function, $\psi^{(1)} (\bar x_{1}^u/\pi)$.
Here $\bar x_{1}^u$ denotes the mean mass value of the
first up-type quark KK level in units of the KK scale. In this way, we
effectively include the entire tower of KK quarks in our result for
$\nu_{u}$. The same procedure is applied in the case of the
coefficients $\nu_d$ and $\nu_\lambda$.

\begin{figure}[!t]
\begin{center} 
\hspace{-2mm}
\mbox{\includegraphics[height=2.85in]{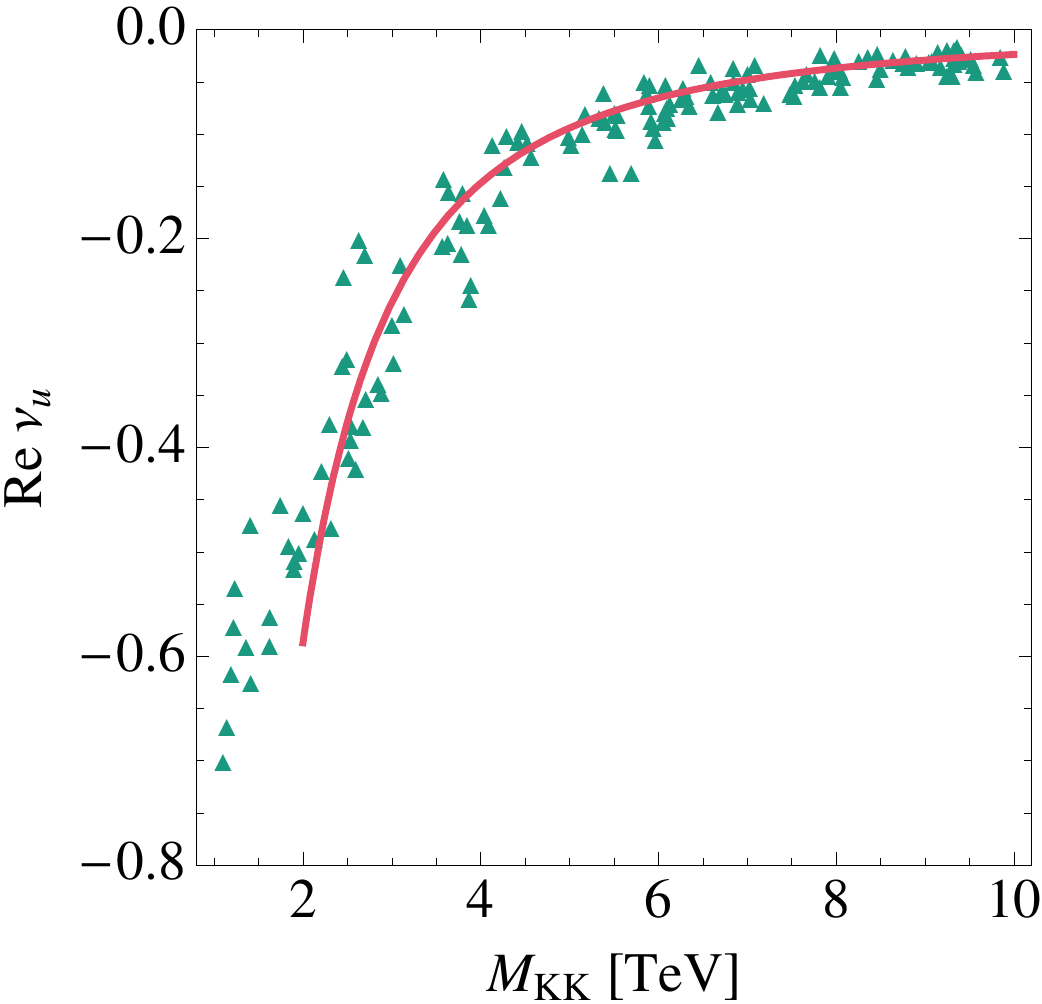}} 
\hspace{4mm}
\mbox{\includegraphics[height=2.85in]{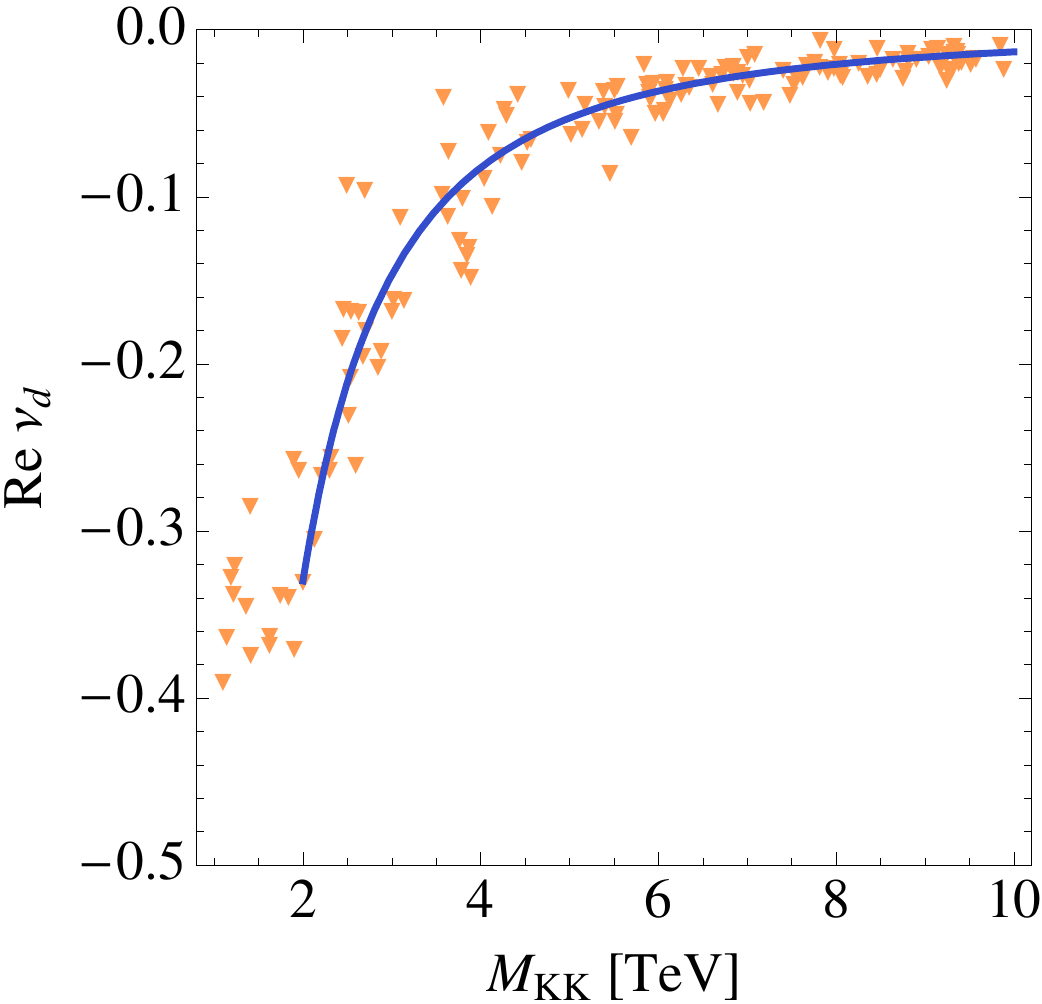}}
\parbox{15.5cm}{\caption{\label{fig:nu} Predictions for the real parts
    of the coefficients $\nu_{u}$ and $\nu_{d}$ in the custodial RS
    model. The solid lines indicate the best fits to the shown sample
    of parameter points lying in the range $\Mkk = [2, 10] \, {\rm
      TeV}$. See text for details.  }}
\end{center}
\end{figure}

In Figure~\ref{fig:nu} we display the real parts of the coefficients
$\nu_u$ and $\nu_d$ as a function of the KK scale for a set of 150
randomly chosen parameter points. The results for the coefficient
$\nu_\lambda$ are almost identically to those of $\nu_d$, and we thus
do not show them explicitly. We see that the corrections to the
effective $hgg$ coupling arising from triangle diagrams involving KK
quarks are all strictly negative. In the up-type quark sector the
corrections are almost a factor of 2 larger than those appearing in
the down- and $\lambda$-type quark sectors. This feature can be traced
back to the higher multiplicity of states in the former relative to
the later sectors, which suggests that $\nu_u/\nu_{d,\lambda} = 15/9
\approx 1.7$. Numerically, we find that for $\Mkk = 2 \, {\rm TeV}$
($\Mkk = 3 \, {\rm TeV}$) the average value of the real parts of
$\nu_u$ and $\nu_{d,\lambda}$ amounts to about $-0.59$ and $-0.34$
($-0.26$ and $-0.15$) with the ratio of the values being quite close
to the naive estimate. We also observe that the imaginary parts of the
coefficients $\nu_{u,d,\lambda}$ are orders of magnitude smaller than
the real parts. This feature can be understood from (\ref{eq:mis1})
and (\ref{eq:Higgscorrection}). Obviously, the only term in these
equations that has a phase is $(\Delta \tilde g_h^q)_{nn}$. This
contribution is however suppressed by $v^2/\Mkk^2$ relative to the
leading term of ${\cal O} (1)$. Since the leading KK quark corrections
to the effective $hgg$ vertex decouple as $v^2/\Mkk^2$, we parametrize
the average values of $\nu_{u, d, \lambda}$ as $a_{u, d, \lambda} \,
v^2/\Mkk^2$ and determine the values of $a_{u, d, \lambda}$ from the
best fit to the shown sample of points restricted to the range $\Mkk =
[2, 10] \, {\rm TeV}$.  The obtained values for the coefficients
$a_{u, d, \lambda}$ are collected in Table~\ref{tab:kappas}. Points
with KK scale below $2 \, {\rm TeV}$ have been excluded in the fit,
since they depend sensitively on higher-order terms in $v/\Mkk$. This
feature is noticeable in the plots, which show that for very low KK
scale the exact results for $\nu_{u,d}$ are typically above the solid
lines indicating our fits. This should be kept in mind when using the
parameterizations $a_{u,d,\lambda} \, v^2/\Mkk^2$ to calculate
$\nu_{u,d,\lambda}$ for KK scales below 2 TeV.

\begin{figure}[!t]
\begin{center} 
\hspace{-2mm}
\mbox{\includegraphics[height=2.85in]{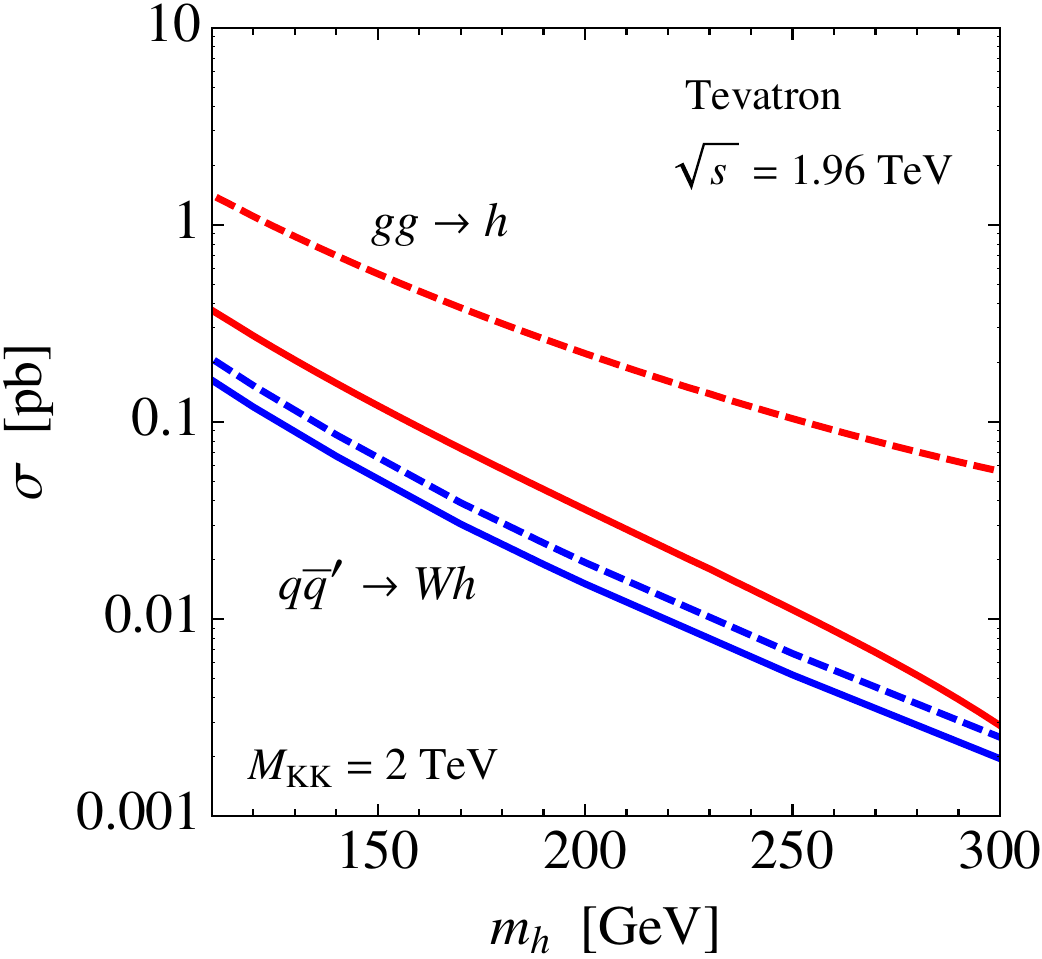}} 
\hspace{4mm}
\mbox{\includegraphics[height=2.85in]{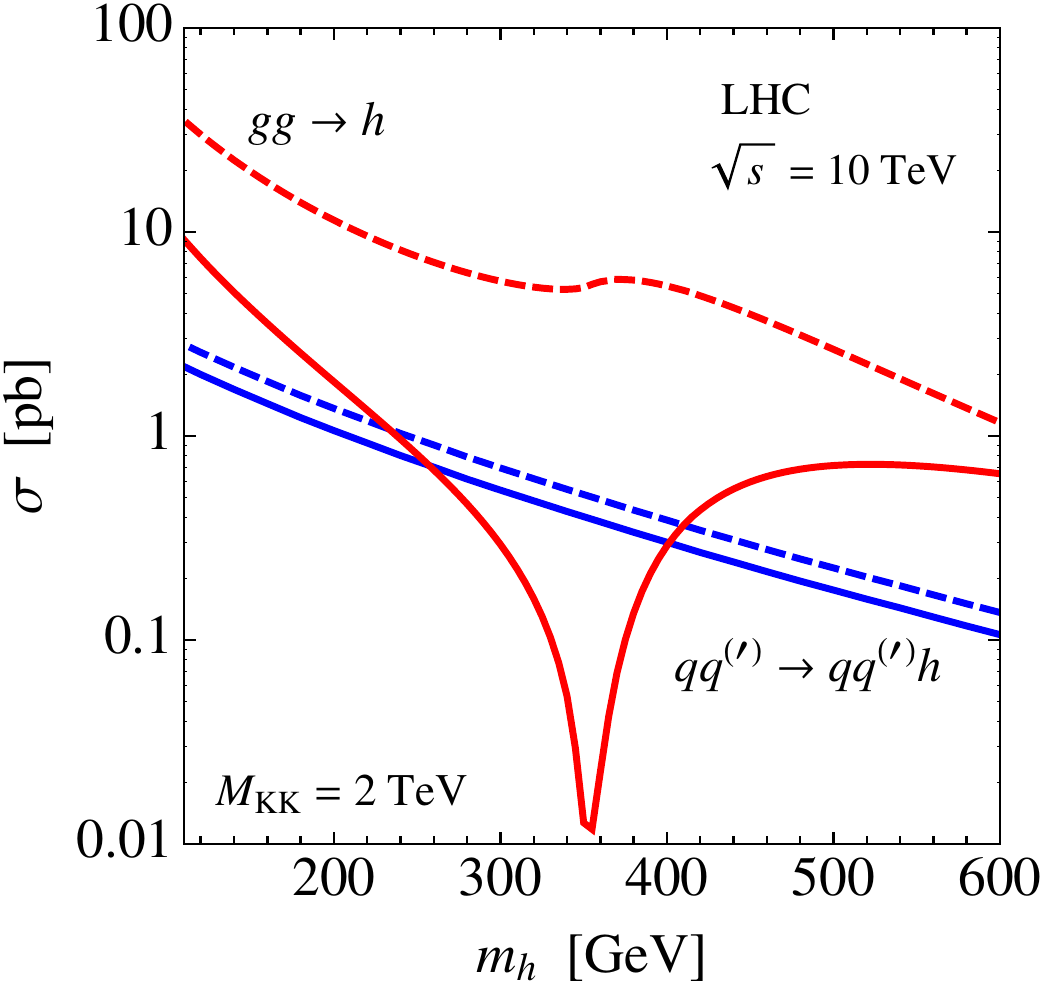}}
\vspace{2mm}
\hspace{-2mm}
\mbox{\includegraphics[height=2.85in]{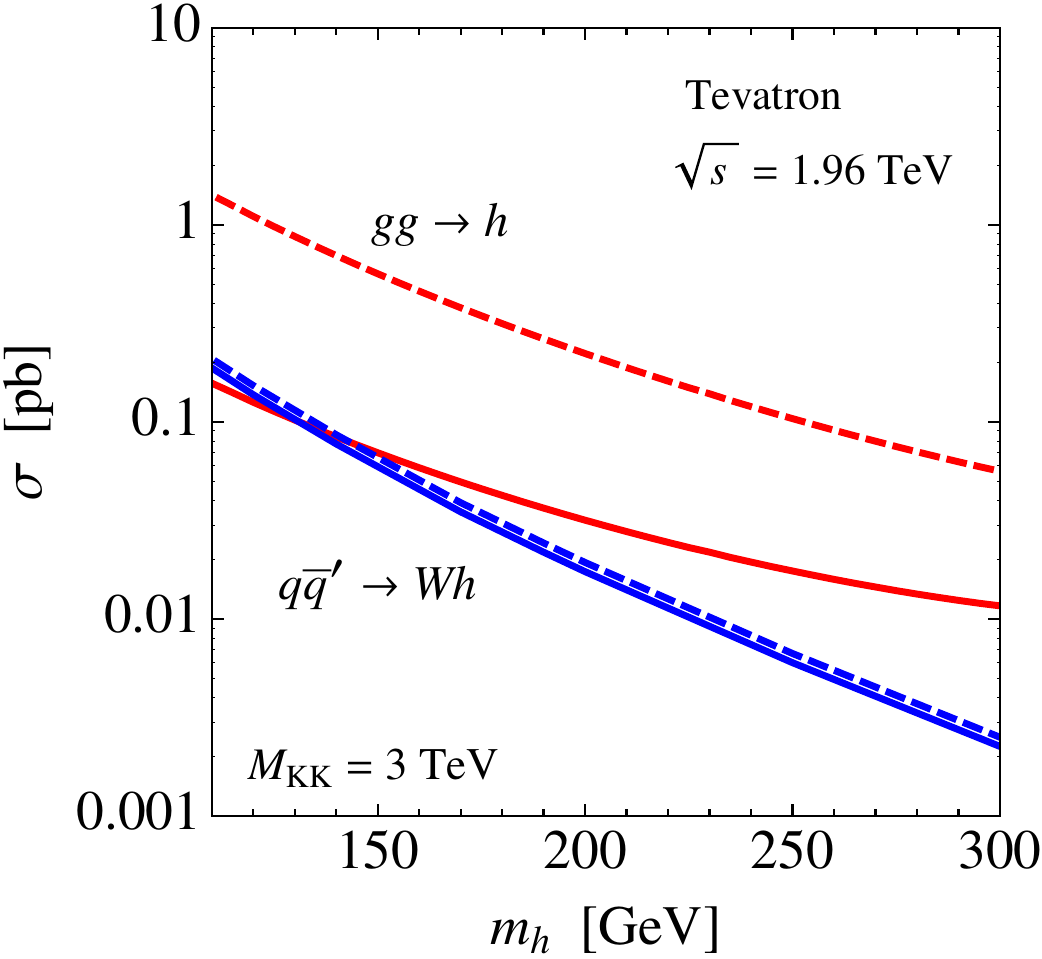}} 
\hspace{4mm}
\mbox{\includegraphics[height=2.85in]{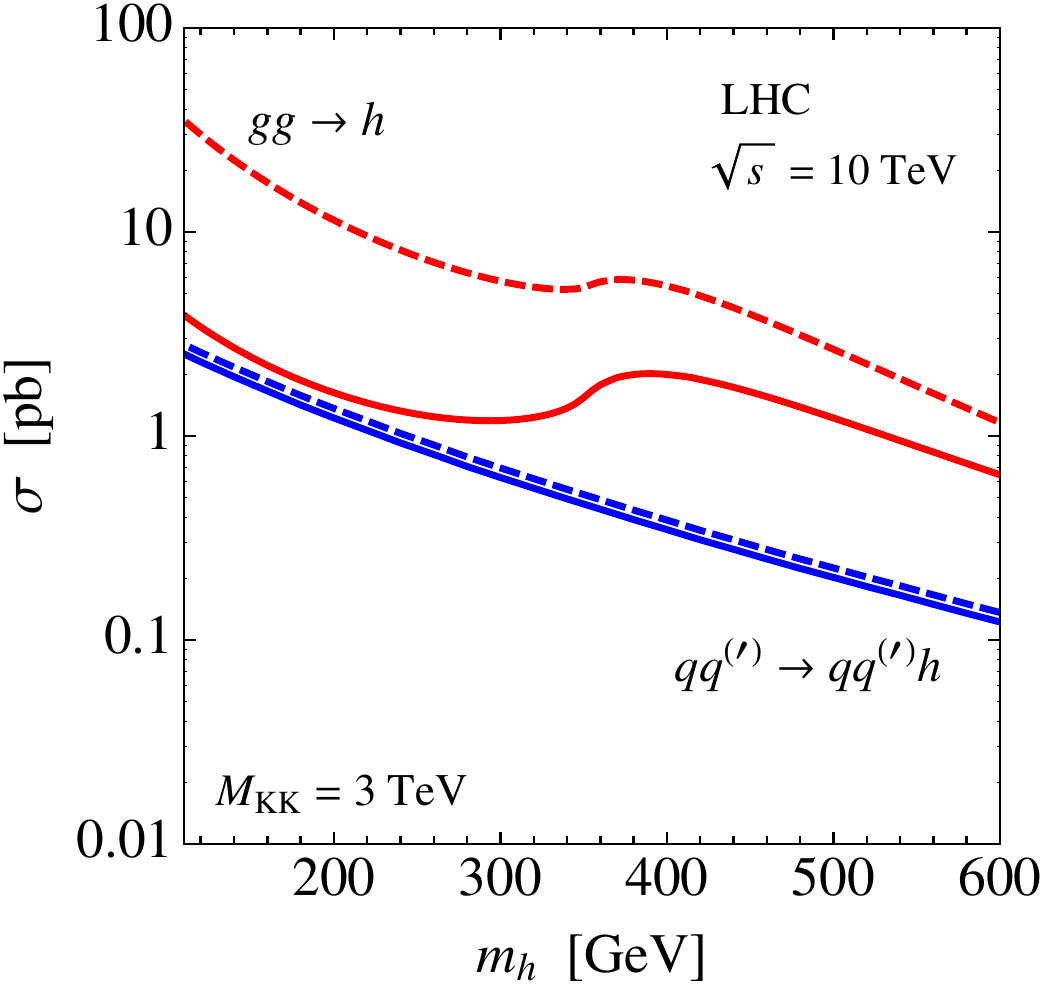}}
\parbox{15.5cm}{\caption{\label{fig:prodplots} Main Higgs-boson
    production cross sections at the Tevatron (left) and the LHC
    (right) for center-of-mass energies of $\sqrt{s} = 1.96 \, {\rm
      TeV}$ and $\sqrt{s} = 10 \, {\rm TeV}$, employing $\Mkk=2$ TeV
    (upper row) and $\Mkk=3$ TeV (lower row).  In the case of the
    Tevatron the panels show gluon-gluon fusion (red) and associated
    $W$-boson production (blue), while for the LHC the dominant
    channels are gluon-gluon (red) and weak gauge-boson fusion
    (blue). The dashed lines illustrate the SM predictions, while the
    solid lines indicate the results obtained in the custodial RS
    model. See text for details.}}
\end{center}
\end{figure}

Our numerical results for the Higgs-boson production cross sections at
the Tevatron and LHC for the center-of-mass energy $\sqrt{s} = 1.96 \,
{\rm TeV}$ and $\sqrt{s} = 10 \, {\rm TeV}$ are shown in
Figure~\ref{fig:prodplots}. The calculation of $\sigma (gg \to h)_{\rm
  SM}$ is based on \cite{Ahrens:2008nc}, which combines the
next-to-next-to-leading fixed-order corrections
\cite{Harlander:2002wh, Anastasiou:2002yz, Ravindran:2003um} with
resummation of both threshold logarithms from soft-gluon emission
\cite{Moch:2005ky, Laenen:2005uz, Idilbi:2005ni, Ravindran:2006cg,
  Idilbi:2006dg} and terms of the form $(N_c \pi \alpha_s )^n$
\cite{Ahrens:2008qu}. In the evaluation of the SM Higgs-boson
production cross section, the {\tt MRST2006NNLO} parton distribution
functions \cite{Martin:2007bv} and the associated normalization
$\alpha_s(m_Z) = 0.1191$ for the strong coupling constant are used.
The SM predictions are depicted by red dashed lines, whereas the solid
red lines correspond to the RS results. The latter predictions have
been obtained by employing (\ref{eq:rescale1}) and (\ref{eq:kappagg})
using the fit formulas for $\kappa_{t,b}$ and $\nu_{u, d, \lambda}$
discussed before. The relevant values for $a_{t,b,u,d,\lambda}$ can be
found in Table~\ref{tab:kappas}. All four panels show clearly that the
Higgs production cross sections in gluon-gluon fusion experience a
significant reduction in the custodial RS model. We emphasize that the
destructive interference between the SM and the KK-quark contributions
to the $gg \to h$ amplitude has been conjectured in
\cite{Falkowski:2007hz, Low:2009di} to be a general feature of
new-physics models where new colored fermions add to the quadratic
divergence of the Higgs-boson mass (which is the case in RS set-ups
\cite{Casagrande:2008hr}). For the considered Higgs-boson masses, we
find in the case of $\Mkk = 2 \, {\rm TeV}$ ($\Mkk = 3 \, {\rm TeV}$)
suppressions that range between $-65\%$ and $-95\%$ ($-80\%$ and
$-90\%$) and from $-45\%$ to almost $-100\%$ ($-45\%$ to $-90\%$) at
the Tevatron and LHC, respectively (see also
Figure~\ref{fig:kappas}). Interestingly, the found depletions survive
even at $\Mkk = 5 \, {\rm TeV}$, still reaching up to $-40 \%$ at both
colliders. Since both the theoretical accuracy \cite{Ahrens:2008nc,
  Harlander:2002wh, Anastasiou:2002yz, Ravindran:2003um} and the
expected experimental precision \cite{Ball:2007zza, Aad:2009wy} are
at the level of 10\%, such pronounced reductions in Higgs events from
gluon-gluon fusion should be observable at the LHC.  The non-trivial
Higgs-mass dependence of the displayed RS curves results from the
interplay between the RS zero- and KK-mode contributions.  The real
part of the zero-mode amplitude increases until the $t \bar t$
threshold is reached and decreases above threshold quadratically with $m_h$
(modulo logarithmic effects). It is positive for all values of the
Higgs-boson mass. On the contrary, the real part of the amplitude
associated to the virtual exchange of KK quarks is negative and a
constant in the heavy-mass limit.  Since for $\Mkk \lesssim 2 \,{\rm
  TeV}$ the latter contribution is always dominant, the correction
arising from KK-quark triangle diagrams effectively flips the sign of
the real part of the total $gg \to h$ amplitude with respect to the SM
expectation for small and high Higgs masses.  In the threshold region,
$m_h \approx 2 \hspace{0.25mm} m_t$, the destructive interference
between the individual contributions can, on the other hand, become
almost perfect, leading to a strong suppression of Higgs production
via gluon-gluon fusion. This feature is nicely illustrated by the
upper right panel of Figure~\ref{fig:prodplots}. Because the RS
contributions decouple rapidly for increasing KK scale, a complete
extinction of the sum of individual amplitudes is not possible for
$\Mkk \gtrsim 2 \,{\rm TeV}$. In this case, the zero-mode contribution
to $gg \to h$ dominates, and the Higgs-mass
dependence of the RS prediction is similar to the one of the SM
result. We emphasize that in spite of the many parameters in the
fermion sector of the custodial RS model, the shown results for the
Higgs-boson production cross section depend to first order only on the
overall KK-mass scale. This claim is supported by the narrow spread of
scatter points depicted in the two panels of Figure~\ref{fig:nu}.
 
\begin{table}[!t]
\begin{center}
\begin{tabular}{|c|c|c|} 
  \hline
  $a_t$ & 
  $a_b$ & 
  $a_t^V$ \\ \hline
  $15.08 - 0.79 \hspace{0.25mm} i$ & 
  $9.08 + 0.43 \hspace{0.25mm} i$ & 
  $3.63$ \\
  \hline 
\end{tabular}

\vspace{4mm}

\begin{tabular}{|c|c|c|c|c|c|c|} 
  \hline
  $a_u$ & 
  $a_d$ &
  $a_\lambda$ &
  $a_{\gamma Z}^u$ & 
  $a_{\gamma Z}^d$ &
  $a_{\gamma Z}^\lambda$ &
  $a_{\gamma Z}^W$\\ \hline
  $-38.80$ & 
  $-21.80$ & 
  $-22.58$ & 
  $-46.46-0.02\hspace{0.25mm}i$ & 
  $17.98$ & 
  $-6.38$ & 
  $10.76$ \\
  \hline 
\end{tabular}
\end{center}
\parbox{15.5cm}{\caption{\label{tab:kappas} Fit coefficients in units of 
    $v^2/\Mkk^2$ entering the various contributions to Higgs-boson 
    production and decay. Corrections due to zero and KK modes are 
    displayed in the upper and lower table, respectively.
    See text for details.}
}
\end{table}

Compared to gluon-gluon fusion, Higgs-boson production through weak
gauge-boson fusion, $q q^{(\prime)} \to qq^{(\prime)} V^{\ast}
V^{\ast} \to qq^{(\prime)} h$ with $V = W,Z$, which is known to be
extremely useful for discovery at the LHC, receives only moderate
corrections of around $-20\%$ ($-10 \%$) for $\Mkk = 2 \, {\rm TeV}$
($\Mkk = 3 \, {\rm TeV}$). The same reduction will affect associated
$W$-boson production, $q \bar q^{\hspace{0.25mm} \prime} \to W^\ast
\to Wh$, which is the only channel that in principle would allow for a
Higgs discovery at the Tevatron. The RS predictions for the production
cross section for $q \bar q^\prime \to Wh$ at the Tevatron and for
$qq^{(\prime)} \to qq^{(\prime)} h$ at the LHC are illustrated by the
solid blue lines in the left and right panels of
Figure~\ref{fig:prodplots}, respectively. The corresponding SM
predictions are taken from \cite{Aglietti:2006ne} and represented by
the blue dashed lines. Finally, the cross section of associated
top-quark pair production, $q \bar q \to t \bar t^\ast \to t \bar t
h$, will also experience a reduction. For values of the KK scale in
the ballpark of $2 \, {\rm TeV}$, this suppression can amount up to
$-40\%$.  Since $q q^{(\prime)} \to qq^{(\prime)} h$, $q \bar
q^{\hspace{0.25mm} \prime} \to Wh$, and $q \bar q \to t \bar t h$ are
tree-level processes, their RS predictions have all been obtained by a
simple rescaling of the corresponding SM results.

In summary, we find that the main Higgs-boson production modes at
hadron colliders are suppressed in the custodial RS model relative to
the SM. Suppression effects in $gg \to h$ were also reported in
\cite{Azatov:2009na, Falkowski:2007hz,
  Cacciapaglia:2009ky}.\footnote{See also \cite{Espinosa:2010vn} for
  a recent detailed analysis of Higgs-boson production cross sections
  and decay rates in a related context.}  A direct numerical
comparison with our findings is however not possible, since
\cite{Azatov:2009na} only included zero-mode corrections, while
\cite{Falkowski:2007hz, Cacciapaglia:2009ky} studied RS variants that
differ from the specific set-up considered here. In
\cite{Djouadi:2007fm} the authors studied corrections to gluon-gluon
fusion arising from virtual exchange of light fermionic KK
modes. There it has been claimed that for a heavy bottom-quark partner
with a mass $m_{b^\prime}$ of a few hundred GeV the Higgs-boson
production cross section via $gg \to h$ can be significantly enhanced.
We would like to point out in this context that in order to achieve
$m_{b^\prime} \ll \Mkk$ with the embedding of quarks as chosen in
(\ref{eq:multiplets}), the $P_{LR}$ symmetry has to be broken strongly
via the bulk mass parameters of the ${\cal T}_{1}$ multiplets by
choosing $c_{{\cal T}_{1i}}$ rather far away from $c_{{\cal T}_{2i}}$.
While for $c_{{\cal T}_{1i}} > 1/2$ it is possible to achieve ${\rm
  Re} \hspace{0.5mm } \nu_d > 0$ and thus an enhancement of the $gg
\to h$ cross section, such choices of parameters need to be fine-tuned
to reproduce the measured mass spectrum of the SM quarks
for anarchic Yukawa couplings. If, on the other hand, 
$c_{{\cal T}_{1i}} < 1/2$, we find that ${\rm Re}
\hspace{0.5mm } \nu_d$ remains strictly negative, and as a result the
$gg \to h$ channel experiences a reduction. We furthermore add that
choices of $c_{{\cal T}_{1i}}$ corresponding to a strong breaking of
the $P_{LR}$ symmetry lead, barring an accidental cancellation, to a
sizable negative shift in the $Z b_L \bar b_L$ coupling through
(\ref{deltaD2}), which is problematic in view of the stringent
constraints arising from the $Z \to b \bar b$ pseudo observables. To
which extent electroweak precision measurements constrain the masses
of light fermionic KK partners deserves further study.

\subsection{Higgs-Boson Decay}
\label{sec:higgsdecay}

We now move on to study the decay modes of the Higgs boson. In this
context, we will consider all processes with quarks and gauge bosons
in the final state that can receive important RS corrections and have
a branching fraction larger than $10^{-4}$. As we have not explicitly
specified the embedding of the fermions in the lepton sector, we
ignore decays into taus and muons. Due to the UV localization of the
leptonic fields, we however expect that the decay widths of the Higgs
into charged leptons are all SM-like. Furthermore, we will not include
loop contributions of KK leptons in our analysis of the $h \to
\gamma\gamma$ and $h \to \gamma Z$ decay channels. We will comment on
the potential impact of this omission below.

In order to be able to calculate the decay rates of the Higgs boson
into massive gauge bosons, we still need to evaluate the RS
corrections to the $WWh$, $ZZh$, and $WWZ$ tree-level vertices. Due to
the unbroken $U(1)_{\rm EM}$ gauge group, the $WW\gamma$ coupling is
unchanged with respect to the SM to all orders in $v^2/\Mkk^2$.  The
weak couplings involving the Higgs boson are derived from the cubic
and quartic interactions due to (\ref{eq:D}). In unitary gauge, the
relevant terms in the Lagrangian read
\begin{equation} 
  {\cal L}_{\rm 4D} \ni \left(h^2+2\,v\,h\right)
  \left [ \, \frac{g_L^2}4 \left (1 - \Delta g_h^W \right ) W^+_\mu 
    \hspace{0.25mm} W^{-\mu}+\frac{g_L^2+g_Y^2}{8} \left (1 - 
      \Delta g_h^Z \right ) Z_\mu \hspace{0.25mm} Z^{\mu} \, \right ] ,
\end{equation}
where 
\begin{align}
  \begin{split}\label{eq:Hbosoncoupl}
    \Delta g_h^V =x_V^2\left[L\left(1+\frac{s_V^2}{c_V^2}\right)
      -1+\frac 1{2L}\right]+\ord\left(x_V^4\right) ,
  \end{split}
\end{align}
and $x_V \equiv m_V/\Mkk$ for $V = W,Z$. In the case of the $P_{LR}$
symmetry (\ref{PLR}), one has $s_W^2/c_W^2 = 1$ and $s_Z^2/c_Z^2 = 1 -
2 \hspace{0.25mm} s_w^2\,$, which implies that the leading correction
due to $\Delta g_h^{W,Z}$ takes the form $-2 \hspace{0.5mm}
m_W^2/\Mkk^2 \hspace{0.5mm} L$. For $\Mkk = 2 \, {\rm TeV}$ ($\Mkk = 3
\, {\rm TeV}$) these terms lead to a suppression of the $WWh$ and
$ZZh$ couplings by about $-10\%$ ($-5\%$) compared to the SM. Notice
that in the minimal RS model the expressions (\ref{eq:Hbosoncoupl})
hold in the limit $s_{W,Z} \rightarrow 0$, and consequently the
corrections to the couplings of the Higgs to massive gauge bosons are
smaller by about a factor of 2.  Our finding that the couplings $WWh$
and $ZZh$ experience a reduction from their SM expectations confirms
the model-independent statements made in \cite{Low:2009di}.

The partial decay widths $\Gamma (h \to f)$ of the Higgs boson
decaying to a final state $f$ are again obtained by rescaling the SM
decay widths. We use
\begin{equation} \label{eq:Gammahtof}
  \Gamma (h \to f)_{\rm RS} = \left | \kappa_f \right |^2 \,
  \Gamma (h \to f)_{\rm SM} \,, 
\end{equation}
with 
\begin{equation} 
  \kappa_{W} = 1 - \Delta g_h^W \,, \qquad 
  \kappa_{Z} = 1 - \Delta g_h^Z \,,
\end{equation}
in the case of the decay of the Higgs boson into a pair of $W$ and $Z$
bosons, respectively. The relevant $\kappa_{g, t, b}$ parameters for
decays into two gluons, top or bottom quarks have already been given
in (\ref{eq:kappagg}) and (\ref{tbratios}).  In
Figure~\ref{fig:hprodec} the diagrams inducing the decay into a pair
of heavy quarks and massive gauge bosons are shown on the right in the
top row. Apart from the change in the $ht \bar t$ coupling, we neglect
RS corrections to the three-body decay $h \to tt^\ast \hspace{0.5mm}
(WW^\ast) \to tbW$, which relative to the two-body mode $h \to t \bar
t$ amounts to a correction of (far below) 1\% in the SM. Given the
smallness of this effect, the omission of possible new-physics effects
in the $Wtb$ coupling that would affect the $h \to tbW$ channel is for
all practical purposes irrelevant.

In the case of the final state with two photons, we employ
\begin{equation} \label{kappagamma} 
\kappa_{\gamma} =
  \frac{{\displaystyle \sum}_{i = t, b} \; N_c \hspace{0.5mm} Q_i^2
    \hspace{0.5mm} \kappa_i \hspace{0.25mm} A_{q}^h (\tau_i) +
    \kappa_{W} \hspace{0.25mm} A_W^h (\tau_W) + {\displaystyle
      \sum}_{j = u, d, \lambda} \; N_c \hspace{0.5mm} Q_j^2
    \hspace{0.5mm} \nu_j + \nu_\gamma^W}{{\displaystyle \sum}_{i = t,
      b} \; N_c \hspace{0.5mm} Q_i^2 \hspace{0.5mm} A_{q}^h (\tau_i) +
    A_W^h (\tau_W)} \,,
\end{equation}
in (\ref{eq:Gammahtof}), where $N_c =3$, $Q_{t,u} = 2/3$, $Q_{b,d} =
-1/3$, $Q_\lambda = 5/3$, $\tau_W \equiv 4 \hspace{0.25mm} m_W^2/
m_h^2$, and the explicit expression for the form factor $A_{W}^h
(\tau_W)$, encoding the $W$-boson contribution, can be found in
Appendix~\ref{app:formfactors}. The first, second, and third terms in
the numerator describe the effects of virtual heavy-quark, $W$-boson,
and KK-quark exchange, respectively. The corresponding one-loop graphs
are shown on the left in the bottom row of Figure~\ref{fig:hprodec}
and in the center plot of Figure~\ref{fig:hkkcontr}. The amplitude
$A_{W}^h (\tau_W)$ interferes destructively with the quark
contribution $A_{q}^h (\tau_i)$, falling from $-21/4$ for $\tau_W \to
\infty$ to $-15/4 - 9 \pi ^2/16$ at the $WW$ threshold $\tau_W = 1$
and finally approaching $-3/2$ in the limit $\tau_W \to 0$. Comparing
these numbers with the ones for $A_{q}^h (\tau_i)$ quoted earlier, one
observes that within the SM the $W$-boson contribution to the $h\to
\gamma \gamma$ decay amplitude is always dominant below threshold.

We emphasize that in (\ref{kappagamma}) contributions from leptonic KK
modes are not included. While the precise impact of these effects
depends on the exact realization of the lepton sector (which we have
not specified), it is possible to predict their relative sign as well
as estimate their size. Generalizing the result (\ref{kappagamma}) to
include contributions from triangle diagrams with KK leptons only
requires to perform the replacement
\begin{equation} \label{kkleptonsinkappaA}
  {\displaystyle \sum}_{j = u, d, \lambda} \; N_c
  \hspace{0.5mm} Q_j^2 \hspace{0.5mm} \nu_j \; \to \; {\displaystyle
    \sum}_{j = u, d, \lambda} \; N_c \hspace{0.5mm} Q_j^2
  \hspace{0.5mm} \nu_j + Q_l^2 \nu_l \, = \, \frac{4 \hspace{0.25mm} \nu_u}{3}  +  
  \frac{\nu_d}{3} +\frac{25 \hspace{0.25mm} \nu_\lambda}{3} + \nu_l \,,
\end{equation}
where $\nu_u \approx 2 \hspace{0.25mm} \nu_d \approx 2 \hspace{0.25mm}
\nu_\lambda$ and the parameter $\nu_l$ encodes the effects due to 
KK-lepton loops. Under the reasonable assumption that $\nu_l \approx
\nu_u/2$, we conclude from (\ref{kkleptonsinkappaA}) that the KK
lepton contribution to the $h \to \gamma \gamma$ amplitude amounts to
approximately $10\%$ of the KK quark corrections and interferes
constructively with the latter. Based on this estimate we expect that
an omission of KK lepton effects in the calculation of $\kappa_\gamma$
has only a minor numerical impact on the obtained Higgs-boson
branching fractions.

The quantity $\nu_\gamma^W$ representing the one-loop contribution of
the $W$-boson KK modes can be calculated analytically in the
decoupling limit. The corresponding Feynman diagram is displayed on
the very right in Figure \ref{fig:hkkcontr}. Employing the results
for the KK sums derived in Section \ref{sec:kksum}, we obtain
\begin{equation}\label{eq:nugammaW}
\begin{split}
  \nu_\gamma^W & = \frac{2 \pi x_W^2 \left ( g_L^2 + g_R^2 \right
    )}{g_L^2} \, \sum_{n = 1}^\infty \, \frac{\vec{d}_W^{\, T}
    \hspace{0.5mm} \vec{\chi}_n^{\, W} (1) \, \vec{\chi}_n^{\, W \, T}
    (1) \hspace{0.5mm} \vec{d}_W}{\big (x_n^W \big)^2} \,
  A_W^h (\tau_n^W) \\
  & = \frac{2 \pi x_W^2 \left ( g_L^2 + g_R^2 \right )}{g_L^2} \;
  \vec{d}_W^{\, T} \hspace{0.5mm} \big [ {\bm \Sigma}_W (1,1)- {\bm
    \Pi}_W (1,1)\big ] \hspace{0.5mm} \vec{d}_W \, \left (
    -\frac{21}{4} + {\cal O} \left (1/\tau_n^W \right ) \right ) \\
  & =-\,\frac{21}8\, \Delta g_h^W \left(1+{\cal O} \left (1/\tau_n^W
    \right )\right) ,
\end{split}
\end{equation}
where $\vec{d}_W = (c_W, -s_W )^T$ and $\tau_n^W \equiv 4
\hspace{0.25mm} \big ( m_n^{W} \big)^2/m_h^2$. Since already $m_1^W
\approx 2.5 \hspace{0.25mm} {\Mkk} \gg m_h$, the terms suppressed by
powers of $\tau_n^W$ in (\ref{eq:nugammaW}) can be ignored in
practice. The result for $\Delta g_h^W$ can be found in
(\ref{eq:Hbosoncoupl}).

\begin{figure}[!t]
\begin{center} 
\hspace{-2mm}
\mbox{\includegraphics[height=2.85in]{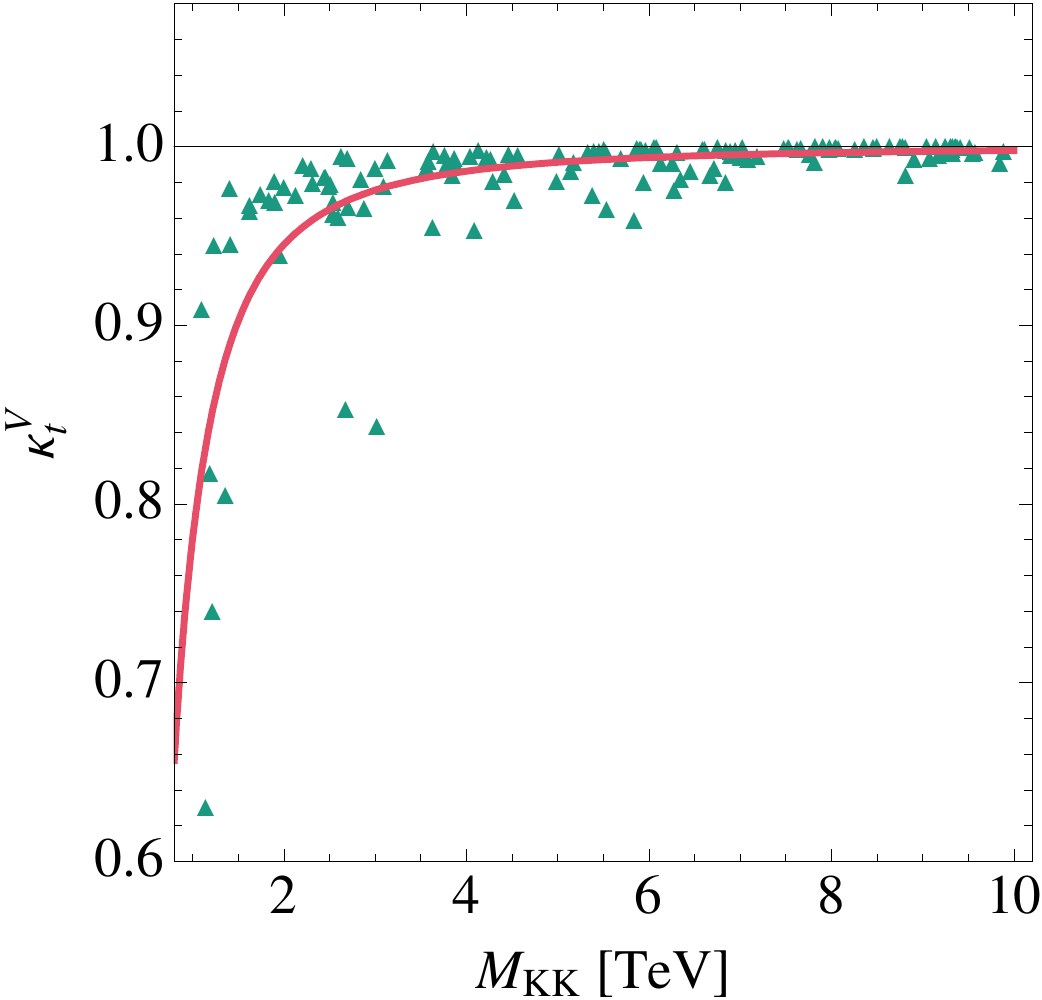}} 
\hspace{4mm}
\mbox{\includegraphics[height=2.85in]{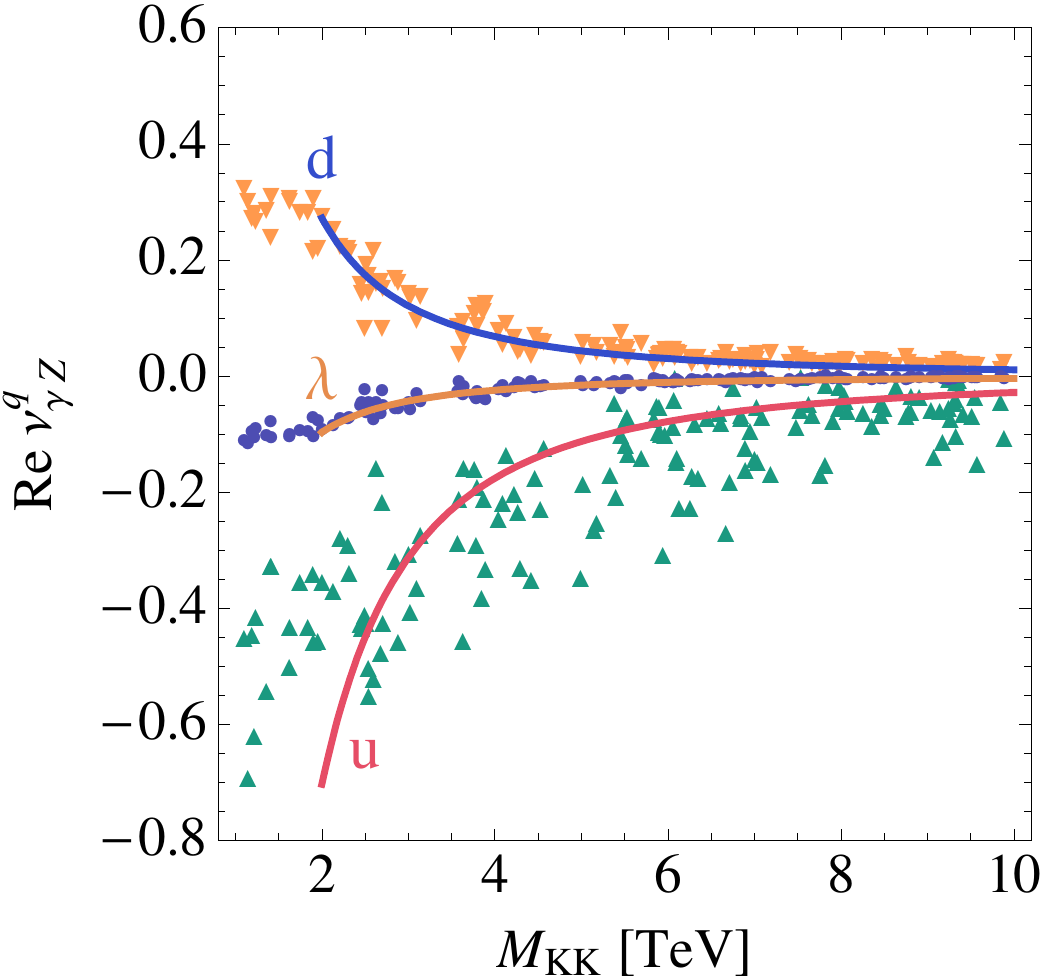}}
\parbox{15.5cm}{\caption{\label{fig:vectorZplot} Predictions for the
    vector couplings of the $Z$ boson to top and bottom quarks (left)
    and for the various types of KK-quark contributions to the
    effective $h\gamma Z$ coupling (right) in the custodial RS
    model. The solid lines show fits to the scatter points. See text
    for details.  }}
\end{center}
\end{figure}

In order to compute the last missing decay channel, namely $h \to
\gamma Z$, we use
\begin{equation} \label{kappagammaZ} 
  \kappa_{\gamma Z} = \frac{{\displaystyle \sum}_{i = t, b} \; N_c \,
    \displaystyle \frac{2 \hspace{0.5mm} Q_i \hspace{0.25mm} v_i}{c_w}
    \hspace{0.5mm} \kappa_i \hspace{0.5mm} \kappa_i^V \hspace{0.25mm}
    A_{q}^h (\tau_i, \lambda_i) + \kappa_{W} \hspace{0.25mm} A_W^h
    (\tau_W, \lambda_W) + {\displaystyle \sum}_{j = u, d,\lambda} \; N_c
    \, \displaystyle \frac{2 \hspace{0.5mm} Q_j \hspace{0.25mm} v_j}{c_w}
    \hspace{0.5mm} \nu^j_{\gamma Z} + \nu_{\gamma Z}^W}{{\displaystyle
      \sum}_{i = t, b} \; N_c \, \displaystyle \frac{2 \hspace{0.5mm} Q_i
      \hspace{0.25mm} v_i}{c_w} \, A_{q}^h (\tau_i, \lambda_i) + A_W^h
    (\tau_W, \lambda_W)} \,,
\end{equation}
in (\ref{eq:Gammahtof}). Here $v_i \equiv T_L^{3\, i} - 2
\hspace{0.25mm} s_w^2 \hspace{0.25mm} Q_i$, and $\lambda_i \equiv 4
\hspace{0.25mm} m_i^2/m_Z^2$ for $i = t, b, W$. The amplitudes
$A_{q,W}^h (\tau_i, \lambda_i)$ encoding the effects of virtual quarks
and $W$ bosons in $h \to \gamma Z$ are collected in
Appendix~\ref{app:formfactors}. The corresponding Feynman diagrams are
shown on the right in the bottom row of Figure~\ref{fig:hprodec}. Like
in the case of $h \to \gamma \gamma$, the SM decay rate for $h \to
\gamma Z$ is in large parts of the parameter space dominated by the
$W$-boson loop contribution.  One has $A_q^h (\tau_i, \lambda_i) =
-1/3$ for $\tau_i , \lambda_i \to \infty$ and $A_q^h (\tau_i,
\lambda_i) = 0$ for $\tau_i , \lambda_i \to 0$.  On the other hand,
the function $A_W^h (\tau_W, \lambda_W)$ rises from around $4.6$ to
$9.8$ between $\tau_W \to \infty$ and $\tau_W = 1$, and then falls to
approximately $0.6$ in the limit $\tau_W \to 0$.
 
The first term in the numerator of (\ref{kappagammaZ}) depends on the
ratios
\begin{equation} \label{eq:kappaiV}
  \kappa_{t}^V = \frac{\big ( g_L^u \big )_{33} + \big ( g_R^u \big )
    _{33}}{v_t} \,, \qquad \kappa_{b}^V = \frac{\big ( g_L^d \big )_{33} +
    \big ( g_R^d \big ) _{33}}{v_b} \,,
\end{equation}
which quantify the relative shift in the vector coupling of the $Z$
boson to top and bottom quarks. In the left panel of
Figure~\ref{fig:vectorZplot} we show the predictions for
$\kappa^V_{t}$ versus $\Mkk$ for 150 randomly chosen model parameter
points. It is evident from the plot that the vector coupling of the
$Z$ boson to top quarks is always reduced in the custodial RS model
relative to the SM. Numerically, the suppression amounts to a moderate
effect of $-5\%$ ($-2.5\%$) for $\Mkk = 2 \, {\rm TeV}$ ($\Mkk = 3 \,
{\rm TeV}$). In contrast, the $Z$-boson coupling to bottom-quark pairs
is larger than its SM value, but numerically the resulting
effects turn out to be negligibly small due to the custodial
protection mechanism. Consequently, we will set $\kappa_b^V$ to 1 in
our numerical analysis. Parameterizing the average value of the
relative shift $\kappa_{t}^V$ by $(1-a_{t}^V \, v^2/\Mkk^2)$ the
coefficient $a_{t}^V$ can again be determined through a fit. Employing
the shown set of parameter points, we obtain the value for $a_{t}^V$
given in Table~\ref{tab:kappas}.

The second term in the numerator of (\ref{kappagammaZ}) encodes the
contribution to the $h \to \gamma Z$ transition arising from the
$W$-boson triangle graph. The calculation of this zero-mode
contribution is greatly simplified by the following two
observations. First, one has
\begin{equation} \label{eq:one}
  \frac{2 \pi}{L} \int_{\epsilon}^1 \!  \frac{dt}{t} \, \chi^{(+)}_0
  (t) \, = \sqrt{2 \pi} + {\cal O} \left ( \frac{v^4}{\Mkk^4} \right
  )\,,
\end{equation} 
and second $\big [(\vec{A}_0^a)_2 \hspace{0.5mm} \chi_0^{(-)} (t) \big
]^2 = {\cal O} (v^4/\Mkk^4)$. The expressions for $\chi_0^{(\pm)} (t)$
and $\vec{A}_0^a$ necessary to derive these results can be found in
(\ref{eq:expprof}) and (\ref{eq:vecA0a}). In combination these two
relations imply that the triple gauge-boson vertex involving two $W$-
and one $Z$-boson fields does not receive corrections at ${\cal O}
(v^2/\Mkk^2)$ in the RS model, regardless of the specific gauge
group. By the same line of reasoning, it is also readily seen that all
quartic gauge-boson vertices first differ at order $v^4/\Mkk^4$ from
the corresponding SM expressions. In view of this extra suppression,
we will set the triple gauge-boson couplings of the zero modes to
their SM values when evaluating the Higgs-boson branching
fractions. In this approximation the effect of virtual $W$-boson
exchange to (\ref{kappagammaZ}) is simply given by the combination
$\kappa_W A_W^h (\tau_W, \lambda_W)$, which up to the different form
factor resembles the form of the corresponding term in
(\ref{kappagamma}).

The third term in the numerator of (\ref{kappagammaZ}) describes the
contribution to the $h \to \gamma Z$ amplitude stemming from the
virtual exchange of KK quarks. The corresponding one-loop diagram is
displayed in the middle of Figure \ref{fig:hkkcontr}. In the up-type
quark sector we find
\begin{equation} \label{eq:nugammaZu}
\begin{split}
  \nu_{\gamma Z}^u =& \,v \;\sum_{n = 4}^\infty \;
  \frac{(g_h^u)_{nn}}{m_n^u} \, \kappa_n^{u, V} A_{q}^h (\tau_n^u,
  \lambda_n^u) \\ 
  =& \, \frac{2 \pi}{\epsilon L} \sum_{n = 4}^\infty \; \frac{\vec
    a_{n}^{U\dagger}\, \bm{C}_{n}^{U} (\pi^-) \left(\displaystyle
      \bm{1}-\frac{v^2}{3\,\Mkk^2} \bm{\tilde Y}_{\vec u}\bm{\bar
        Y}_{\vec u}^\dagger\right) \bm{S}_{n}^U(\pi^-)\, \vec
    a_{n}^{\hspace{0.25mm} U} }{x_n^u}\hspace{0.5mm} \,\kappa_n^{u, V}
  A_{q}^h (\tau_n^u, \lambda_n^u) \,, \hspace{6mm}
\end{split}
\end{equation}
where $\kappa_n^{u, V}$ denotes the relative strength of vector
coupling of the $Z$ boson to the $n^{\rm th}$ up-type quark KK mode
defined in analogy to (\ref{eq:kappaiV}), and $\lambda_n^u \equiv 4
\hspace{0.25mm} \big(m_n^u\big)^2/m_Z^2$. Analog expressions apply in
the case of down- and $\lambda$-type quark KK modes. Since an analytic
calculation of (\ref{eq:nugammaZu}) turns out to be impractical, we
resort to a numerical evaluation of the KK sum employing the method
described in Section~\ref{sec:higgsproduction}. The predictions for
the real parts of $\nu_{\gamma Z}^{u}$, $\nu_{\gamma Z}^{d}$, and
$\nu_{\gamma Z}^{\lambda}$ corresponding to a set of 150 random model
parameter points are depicted in the right panel of
Figure~\ref{fig:vectorZplot}. The solid lines displayed there indicate
the best fit of the form $a_{\gamma Z}^{u, d, \lambda} \hspace{0.5mm}
v^2/\Mkk^2$ to the sample of points with KK scales in the range $[2,
10] \, {\rm TeV}$.  As before, points with $\Mkk < 2 \, {\rm TeV}$
have been excluded in the fit, since they are subject to significant
higher-order corrections.  The corresponding coefficients $a_{\gamma
  Z}^{u, d, \lambda}$ can be found in Table~\ref{tab:kappas}. The
average values of the real parts of $\nu_{\gamma Z}^{u}$, $\nu_{\gamma
  Z}^{d}$, and $\nu_{\gamma Z}^{\lambda}$ obtained from the fit
formulas are $-0.70$ ($-0.31$), $0.27$ ($0.12$), and $-0.10$ ($-0.04$)
for $\Mkk = 2 \, {\rm TeV}$ ($\Mkk = 3 \, {\rm TeV}$), respectively.
The imaginary parts ${\rm Im} \, \nu_{\gamma Z}^{u,d,\lambda}$ turn
out to be tiny. The reason for this feature has already been discussed
in Section~\ref{sec:higgsproduction}.

\begin{figure}[!t]
\begin{center}
\mbox{\includegraphics[height=2.85in]{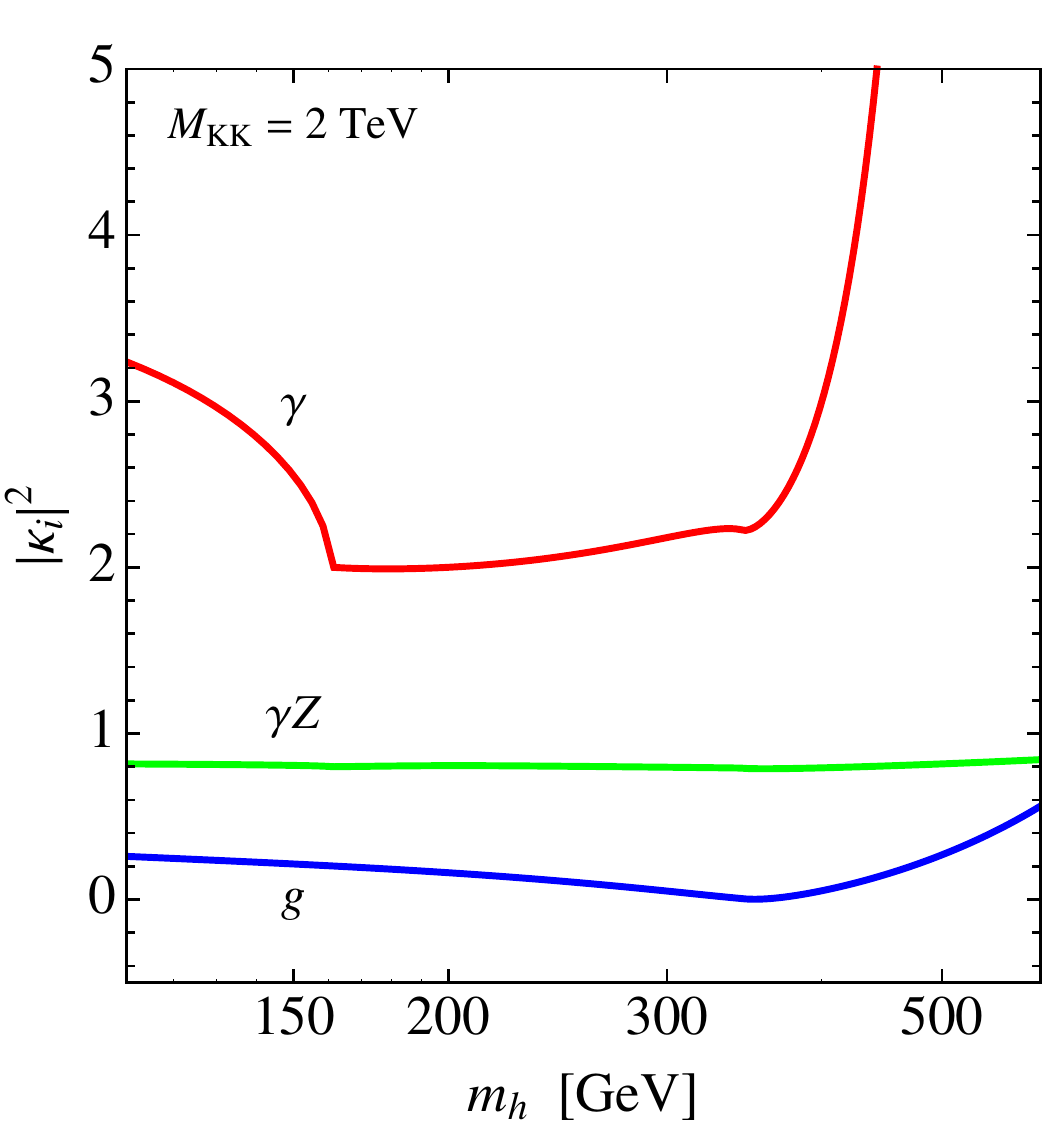}}
\hspace{6mm}
\mbox{\includegraphics[height=2.85in]{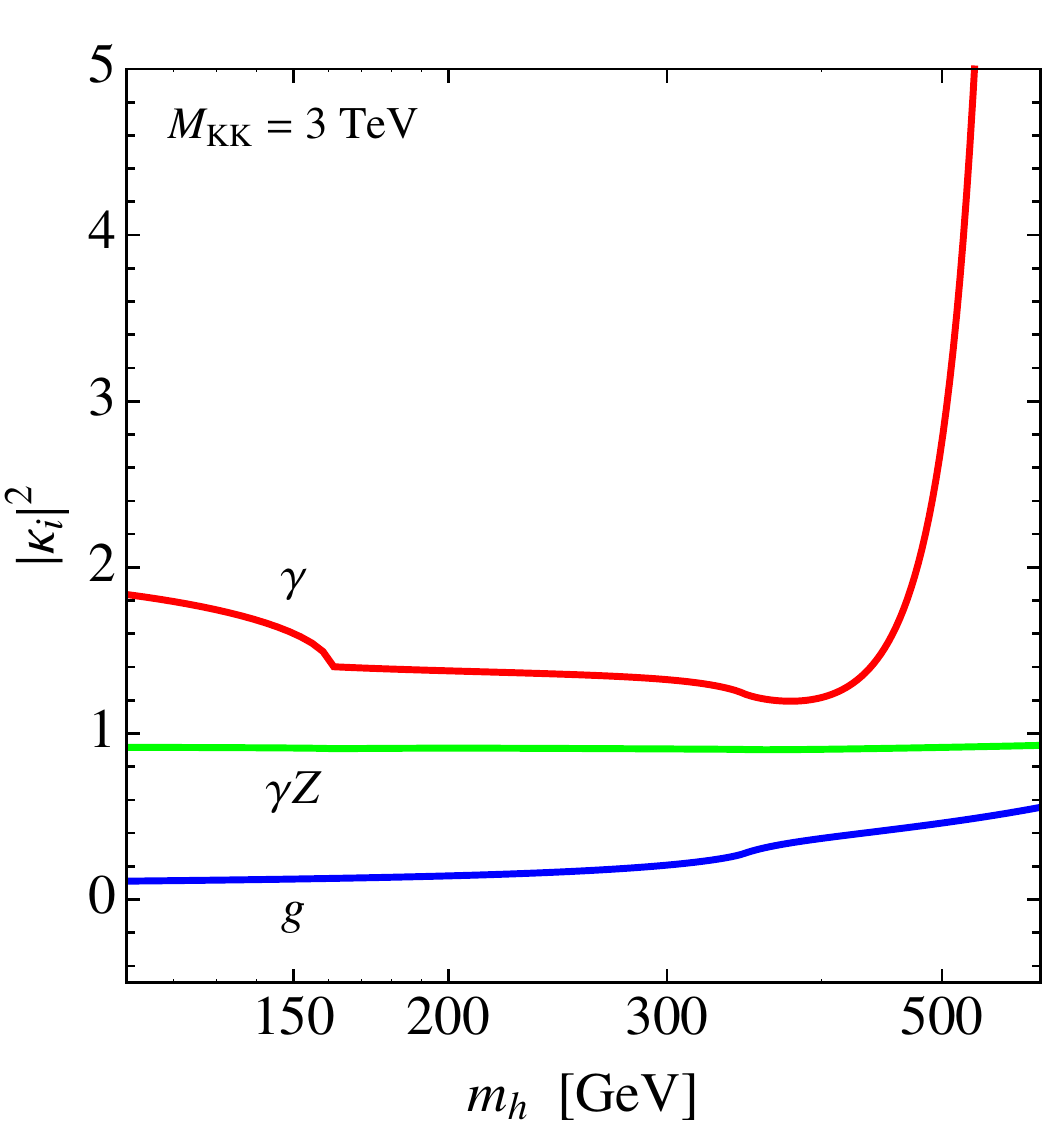}}
\parbox{15.5cm}{\caption{\label{fig:kappas} Relative corrections
    $|\kappa_g|^2$ (blue line), $|\kappa_\gamma|^2$ (red line), and
    $|\kappa_{\gamma Z}|^2$ (green line) as functions of the
    Higgs-boson mass, employing $\Mkk = 2 \, {\rm TeV}$ (left panel)
    and $\Mkk = 3 \, {\rm TeV}$ (right panel). See text for details.
  }}
\end{center}
\end{figure}

Contributions from KK-lepton triangle graphs have again not been
incorporated in (\ref{kappagammaZ}). Denoting these corrections by
$\nu_{\gamma Z}^l$, they can be included via the simple replacement
\begin{equation} \label{kkleptonsinkappaAZ} 
      {\displaystyle \sum}_{j = u, d,\lambda} \; N_c \, \displaystyle
      \frac{2 \hspace{0.5mm} Q_j \hspace{0.25mm} v_j}{c_w} \hspace{0.5mm}
      \nu^j_{\gamma Z} \; \to \; \,  {\displaystyle \sum}_{j = u,
        d,\lambda} \; N_c \, \displaystyle \frac{2 \hspace{0.5mm} Q_j
        \hspace{0.25mm} v_j}{c_w} \hspace{0.5mm} \nu^j_{\gamma Z} +
      \displaystyle \frac{2 \hspace{0.5mm} Q_l \hspace{0.25mm} v_l}{c_w}
      \hspace{0.5mm} \nu^l_{\gamma Z} \,.
\end{equation}
In order to estimate the typical size of $\nu_{\gamma Z}^l$ we need an
analytic formula for the relative strength of the vector coupling
between the $Z$ boson and fermionic KK modes appearing in
(\ref{eq:nugammaZu}). We find
\begin{equation}
  \kappa_n^{f,V} = 1 - \frac{\left({\bm \delta}_F\right)_{nn} - 
  \left({\bm \delta}_f\right)_{nn}}{v_f} +  
  {\cal O}\left(\frac{m_Z^2}{\Mkk^2}\right) , 
\end{equation}
where the expressions for ${\bm \delta}_{F,f}$ can be found in
(\ref{eq:delta2}).  In the case of extended $P_{LR}$ symmetry
(\ref{eq:extendedPLR}), it turns out that for down- and $\lambda$-type
KK quarks the result for $\kappa_n^{f,V}$ can be expressed in terms of
the electric charge and the third component of the weak isospin of the
involved fermion, while no such formula can be derived for up-type
quark KK modes. We obtain to excellent approximation ($f =d, \lambda$)
\begin{equation}
\kappa_n^{f,V} = 1 + \frac{T_L^{3 \, f_L}}{v_f} \,,
\end{equation}
which implies that all down-type ($\lambda$-type) KK-quark modes
couple with universal strength to the vector part of the $Z$-boson
coupling. It follows that in the decoupling limit, $\tau_n^f,
\lambda_n^f \to \infty$, one has
\begin{equation}
  \left ( 1 + \frac{T_L^{3 \, f_L}}{v_f}  \right ) 
  \frac{A^h_f(\tau_n^f, \lambda_n^f)}{A^h_f(\tau_n^f)} =
  \frac{a^f_{\gamma Z}}{a_f} \,.
\end{equation}
From the numbers of the fit coefficients given in
Table~\ref{tab:kappas}, we see that this relation is satisfied
to an accuracy of around $1\%$. The KK-fermion effects in the down-
and $\lambda$-type quark sectors that contribute to $h \to gg, \gamma
\gamma$, and $\gamma Z$ are thus universal, in the sense that they
can be simply obtained from each other by an appropriate replacement
of the vector couplings of the external fields.

Making now the plausible assumption that in the decoupling limit the
sums $\nu_{\gamma Z}^{d}$ and $\nu_{\gamma Z}^{l}$ differ only by the
presence of the vector couplings $\kappa_n^{d,V}$ and
$\kappa_n^{l,V}$, we obtain the following estimate for
the contribution to (\ref{kkleptonsinkappaAZ}) from leptonic relative
to down-type quark KK modes:
\begin{equation}
  \frac{Q_l \hspace{0.25mm} v_l \hspace{0.25mm} \nu_{\gamma Z}^l}
  {N_c \hspace{0.25mm} Q_d \hspace{0.25mm} v_d \hspace{0.25mm} 
    \nu_{\gamma Z}^d} \approx  \frac{Q_l \hspace{0.25mm} v_l \hspace{0.25mm} 
    \kappa_n^{l,V}} {N_c \hspace{0.25mm} Q_d \hspace{0.25mm} v_d \hspace{0.25mm} 
    \kappa_n^{d,V}} = \frac{3- 6 s_w^2}{3-2 s_w^2} \approx 0.64 \,.
\end{equation}
As a result, the sum (\ref{kkleptonsinkappaAZ}) can be approximated as
\begin{equation} \label{eq:sumlKK}
{\displaystyle \sum}_{j = u,
        d,\lambda} \; N_c \, \displaystyle \frac{2 \hspace{0.5mm} Q_j
        \hspace{0.25mm} v_j}{c_w} \hspace{0.5mm} \nu^j_{\gamma Z} +
      \displaystyle \frac{2 \hspace{0.5mm} Q_l \hspace{0.25mm} v_l}{c_w}
      \hspace{0.5mm} \nu^l_{\gamma Z} \approx 0.88 \,
      \nu_{\gamma Z}^u + 0.79 \, \nu_{\gamma Z}^d - 3.04 \, \nu_{\gamma
        Z}^\lambda +  0.50 \, \nu_{\gamma Z}^d \,,
\end{equation}
where the last term on the right-hand side encodes the effects due to
KK leptons, and in order to obtain the numerical values we have
inserted the relevant electroweak quantum numbers and used $s_w^2
\approx 0.23$. For $\Mkk = 2\, {\rm TeV}$, the real part of the
relation (\ref{eq:sumlKK}) evaluates to $-0.11$ ($0.03$) if effects
due to KK leptons are excluded (included). While these numbers imply
that an omission of KK lepton effects can change the numerical value
of the KK fermion contribution notably, it is not difficult to see
that the impact on (\ref{kappagammaZ}) itself is limited, since the
coefficient $\kappa_{\gamma Z}$ is dominated by the $W$-boson triangle
contribution. We thus conclude that the absence of KK-lepton
contributions in our prediction for $h \to \gamma Z$ (which is anyhow
difficult to study at the LHC) will not change any of the conclusions
drawn below.

\begin{figure}[!t]
\begin{center}
\mbox{\includegraphics[width=13cm]{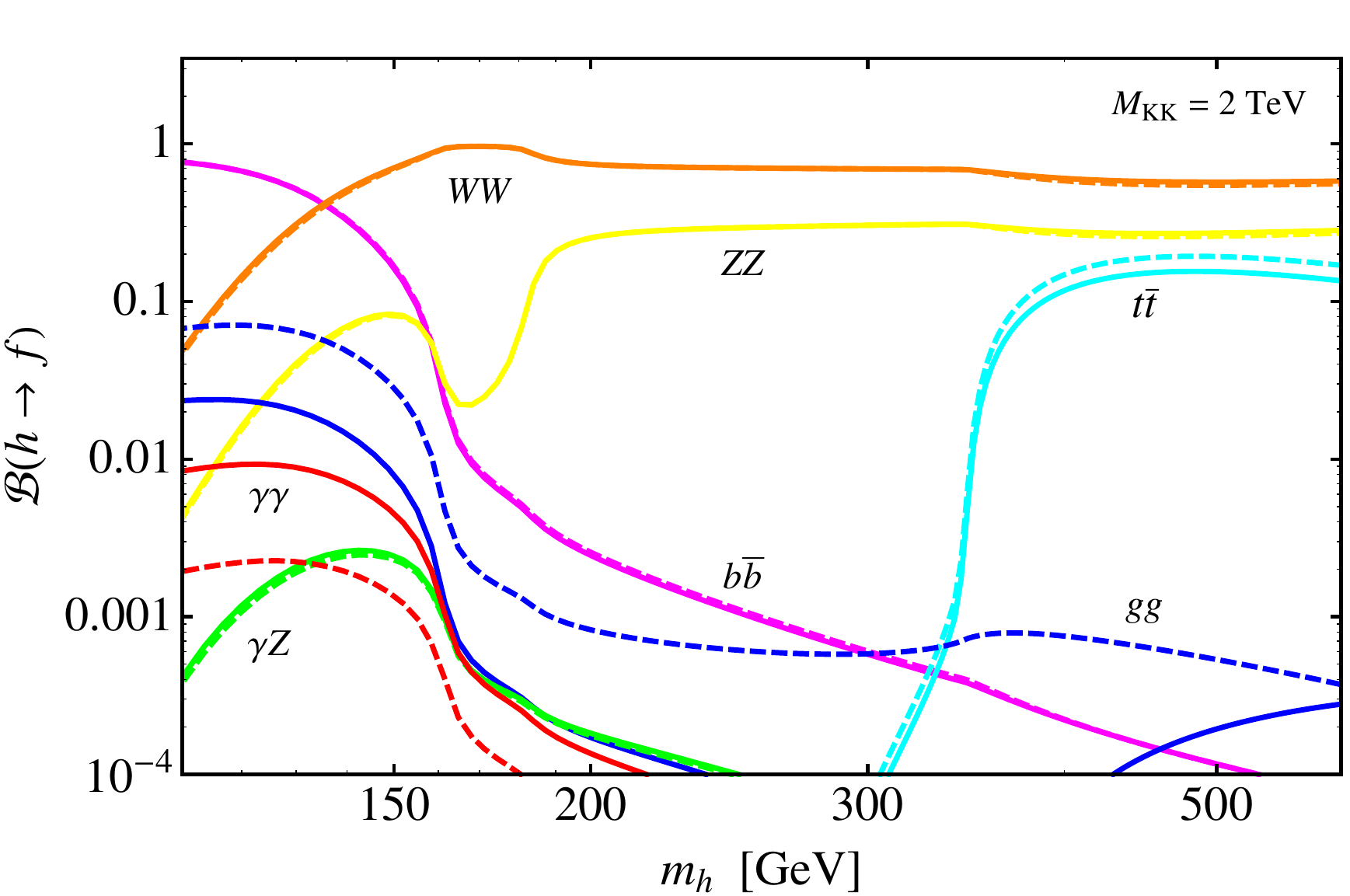}}

\mbox{\includegraphics[width=13cm]{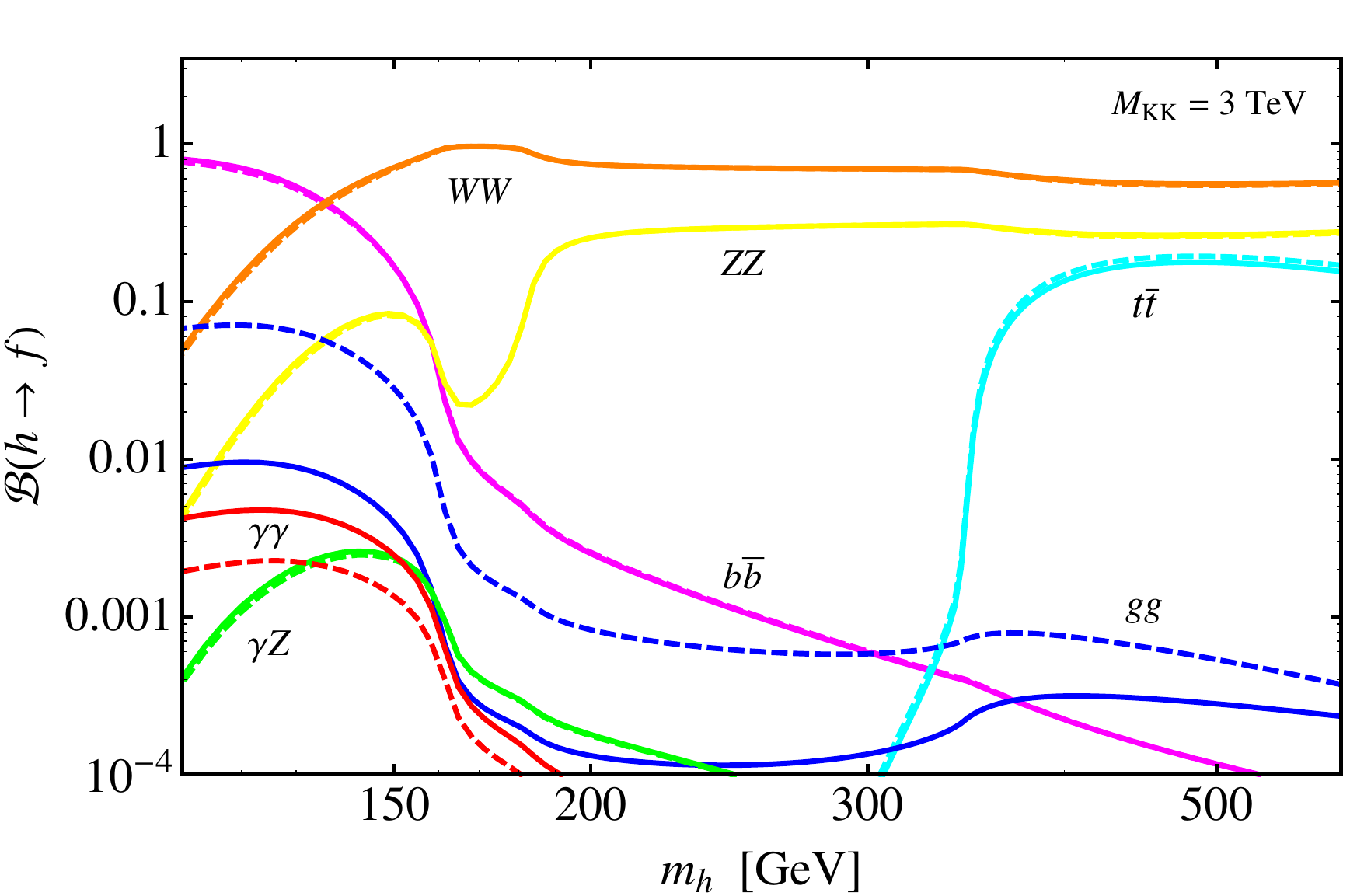}}

\vspace{4mm}

\parbox{15.5cm}{\caption{\label{fig:hXX} Branching ratios for $h \to
    f$ as functions of the Higgs-boson mass for $\Mkk =2 \, {\rm TeV}$
    (upper panel) and $\Mkk = 3 \, {\rm TeV}$ (lower panel). The
    dashed lines indicate the SM predictions, while the solid lines
    show the corresponding RS expectations. Branching fractions of
    less than $10^{-4}$ and decay channels into final states with
    muon, tau, charm-, and strange-quark pairs, which are all expected
    to remain SM-like, are not shown. See text for details.  }}
\end{center}
\end{figure}

The coefficient $\nu_{\gamma Z}^W$ in (\ref{kappagammaZ}) incorporates
the effects in $h \to \gamma Z$ due to charged KK-boson excitations in
the loop. The associated Feynman graph is displayed on the very right
in Figure~\ref{fig:hkkcontr}. This contribution can be written as
\begin{equation} \label{eq:nugammaZ} 
  \nu_{\gamma Z}^W = \frac{2 \pi
    x_W^2 \left ( g_L^2 + g_R^2 \right )}{g_L^2} \, \sum_{n =
    1}^\infty \, \frac{ \vec{d}_W^{\, T} \hspace{0.5mm}
    \vec{\chi}_n^{\, W} (1) \, \vec{\chi}_n^{\, W \, T} (1)
    \hspace{0.5mm} \vec{d}_W}{\big (x_n^W \big)^2} \; {\mathcal
    I}_{nn0}^{WWZ}\hspace{0.25mm} A_W^h(\tau_n^W,\lambda_n^W)\,,
\end{equation}
with
\begin{equation} \label{eq:IWWZ}
\begin{split}
  {\mathcal I}_{nn0}^{WWZ}=\,\frac{(2 \pi)^{3/2}}L
  \int_\epsilon^1\frac{dt}{t}\,\Big{[}\, &\chi_0^{(+)Z}{(\vec
    A_0^Z)}_1 \left( {\chi_n^{(+)W}}^2{(\vec A_n^W)}_1^{\,2} +
    \frac{g_Y^2}{g_L^2}\,
    {\chi_n^{(-)W}}^2{(\vec A_n^W)}_2^{\,2} \right)\\
  &-\,\sqrt{1-g_Y^4/g_L^4}\;\chi_0^{(-)Z}{(\vec
    A_0^Z)}_2\,{\chi_n^{(-)W}}^2{(\vec A_n^W)}_2^{\,2}\,\Big{]}\,,
\end{split}
\end{equation}
and $\lambda_n^W\equiv 4 \hspace{0.25mm}
\big(m_n^W\big)^2/m_Z^2$. Notice that the prefactor in the second line
of the above formula corresponds to the choice $g_L=g_R$.  Since the
first term in the sum of (\ref{eq:nugammaZ}) is already suppressed by
a factor of $v^2/\Mkk^2$, the computation of $\nu_{\gamma Z}^W$ to this 
order only requires the knowledge of the overlap integral
(\ref{eq:IWWZ}) to zeroth order in the ratio of the weak over the KK
scale. We obtain
\begin{equation} \label{eq:IWWZapprox}
\begin{split}
  {\mathcal I}_{nn0}^{WWZ}=\,\frac{2 \pi}{L}
  \int_\epsilon^1\frac{dt}{t}\, \left( {\chi_n^{(+)W}}^2{(\vec
      A_n^W)}_1^{\,2} + \frac{g_Y^2}{g_L^2}\, {\chi_n^{(-)W}}^2{(\vec
      A_n^W)}_2^{\,2} \right)+\ord\left(\frac{v^2}{\Mkk^2} \right) \,.
\end{split}
\end{equation}
It is again an excellent approximation to evaluate the loop function
$A_W^h(\tau_n^W,\lambda_n^W)$ in the infinite mass limit $\tau_W^h,
\lambda_n^W \to \infty$, in which the form factor approaches $31 \,
c_w/6 - 11 \, s_w^2/(6 \, c_w) \approx 4.0$.  We perform the sum in
(\ref{eq:nugammaZ}) numerically, including sufficiently many KK levels
until the series converges. In this way, we find $\nu_{\gamma
  Z}^W=0.16$ $(\nu_{\gamma Z}^W = 0.07)$ for $\Mkk=2 \, {\rm TeV}$
($\Mkk=3 \, {\rm TeV}$).  Values for $\nu_{\gamma Z}^W$ corresponding
to different KK scales can be obtained by means of the fit formula
$a_{\gamma Z}^W \hspace{0.5mm} v^2/\Mkk^2$ with the coefficient
$a_{\gamma Z}^W$ given in Table~\ref{tab:kappas}.

In the two panels of Figure~\ref{fig:kappas} we display the relative
corrections $|\kappa_g|^2$, $|\kappa_\gamma|^2$, and $|\kappa_{\gamma
  Z}|^2$ for $\Mkk = 2 \, {\rm TeV}$ (left) and $\Mkk = 3 \, {\rm
  TeV}$ (right). The depicted curves represent the RS results obtained
from (\ref{eq:Hbosoncoupl}) and (\ref{eq:nugammaW}) as well as the
relevant fit formulas with the values of the coefficients collected in
Table \ref{tab:kappas}. While the behavior of $|\kappa_g|^2$ has
already been explained in Section~\ref{sec:higgsproduction}, we see
that $|\kappa_{\gamma Z}|^2$ is close to 1 and independent of the
value of the Higgs-boson mass.  This implies that the partial decay
width $\Gamma (h \to \gamma Z)$ in the custodial RS model is
essentially unchanged with respect to the SM. The relative correction
$|\kappa_\gamma|^2$ is, on the other hand, a non-trivial function of
$m_h$. Below the $WW$ threshold, the $W$-boson amplitude dominates the
SM $h \to \gamma \gamma$ decay rate and the contributions due to KK
quarks and $W$ bosons both interfere constructively with the SM
gauge-boson triangle graph. For $m_h = 130 \, {\rm GeV}$, the
new-physics contributions amount to around $70\%$ ($30\%$) of the
total SM amplitude for $\Mkk = 2\, {\rm TeV}$ ($\Mkk = 3\, {\rm
  TeV}$), resulting in values $|\kappa_\gamma|^2 \approx 3$
($|\kappa_\gamma|^2 \approx 1.7$).  For $m_h \gtrsim 160 \, {\rm
  GeV}$, the Higgs-mass dependence of the SM amplitude becomes less
pronounced and the RS prediction stays almost constant. The strong
rise of $|\kappa_\gamma|^2$, visible at higher values of the Higgs
mass, results from the fact that for $m_h \approx 650 \, {\rm GeV}$
the top-quark loop nearly cancels the $W$-boson contribution in the
SM. In consequence, for $m_h \gtrsim 500 \, {\rm GeV}$ the partial
width $\Gamma (h \to \gamma \gamma)$ is almost entirely due to loops
involving heavy KK modes, with the contribution from KK quarks being
the dominant correction.

The various Higgs-boson branching ratios obtained using the above
results are shown in Figure~\ref{fig:hXX}. The dashed lines illustrate
the SM expectations calculated with the help of {\tt HDECAY}
\cite{Djouadi:1997yw},\footnote{Expect for the parameters listed in
  Appendix~\ref{app:masses}, the original input file of {\tt HDECAY}
  version 3.51 is used.} while the solid lines represent the RS
predictions based on the results for $\kappa_{t,b,W,Z}$ quoted above
and the curves for $|\kappa_{g, \gamma, \gamma Z}|^2$ displayed in
Figure~\ref{fig:kappas}. It is evident that in the custodial RS model
the branching ratios $h \to b \bar b$, $h \to WW$, and $h \to ZZ$
receive only insignificant corrections, not exceeding the level of
$\pm 5\%$. For $m_h\gtrsim 180 \, {\rm GeV}$ the experimentally
cleanest signature for the discovery of the Higgs boson at the LHC is
its ``golden'' decay to four leptons, $h \to Z^{(\ast)}Z^{(\ast)} \to
l^+ l^- l^+ l^-$. Since the $h \to ZZ$ branching fraction is essential
SM-like, the reduction in the $gg \to h$ production cross section will
make an observation of the Higgs boson in the golden channel more
difficult. Moderate effects occur in the non-discovery channels $h \to
\gamma Z$ and $h \to t \bar t$. In the relevant ranges for the Higgs
mass, the modifications in the branching ratios amount to around $+10
\%$ ($+10\%$) and $-25 \%$ ($-10\%$) for $\Mkk = 2 \, {\rm TeV}$
($\Mkk = 3 \, {\rm TeV}$). The most pronounced effects are found for
$h \to gg$ and $h \to \gamma \gamma$. For Higgs masses below the $WW$
threshold, the branching fraction of the former mode is reduced by a
factor of almost 4 (8), while the branching ratio of the latter
transition is enhanced by a factor of around 4 (2). The corresponding
maximal values of ${\cal B} (h \to \gamma \gamma)$ are $9.3 \cdot
10^{-3}$ ($4.8 \cdot 10^{-3}$) for $\Mkk = 2 \, {\rm TeV}$ ($\Mkk = 3
\, {\rm TeV}$) and arise at $m_h \approx 120 \, {\rm
  GeV}$. Calculating the rescaling factor $\varkappa = \left
  (\sigma_{\rm RS} (gg \to h) \hspace{1mm} {\cal B} (h \to \gamma
  \gamma)_{\rm RS} \right )/\left (\sigma_{\rm SM} (gg \to h)
  \hspace{1mm} {\cal B} (h \to \gamma \gamma)_{\rm SM} \right )$ for
$\sqrt s = 10 \, {\rm TeV}$ and the quoted maximal branching
fractions, we obtain the values $1.03$ ($0.24$). These numbers suggest
that the statistical significance for a LHC discovery of the Higgs
boson in $h \to \gamma \gamma$ can be enhanced in the custodial RS
model for low KK scales. A detailed study of how the deviations found
in the RS framework affect the searches for the Higgs boson at the LHC
will be presented elsewhere. We add that if the KK scale is lowered to
$1 \, {\rm TeV}$, the branching ratio of $h \to tc$ can reach values
above $10^{-4}$ for Higgs masses above $m_h \approx 180 \, {\rm
  GeV}$.\footnote{In the limit of vanishing charm-quark mass, $r_c =
  0$, the corresponding decay rate is simply obtained from
  (\ref{BRthc}) by multiplying the branching fraction for $t \to ch$
  with $g^2 ( 1 - r_W^2 )^2 (1 + 2 \hspace{0.25mm} r_W^2)/(2
  \hspace{0.25mm} r_W^2) \, m_h/(16 \pi)$ and replacing $r_h$ through
  $r_t$. Of course, an analogous formula applies in the case of $h\to
  bs$.} For such a low KK scale, also the decay channel $h\rightarrow
bs$ can open up below the $WW$ threshold, but typically stays below
the level of $10^{-3}$.  Note that our results for the Higgs-boson
branching fractions depend primarily on the value of the KK scale, and
are rather insensitive to the other free parameters present in the
model. For example, the final results do not strongly depend on the
precise localization pattern of the fermionic bulk fields. We also
verified that the omission of KK-lepton effects does not have a
pronounced effect. RS predictions for the various branching fractions
of the Higgs boson have been presented previously in
\cite{Azatov:2009na}. Yet a direct comparison with our results is
difficult, as the latter work only includes RS corrections affecting
the tree-level couplings of the Higgs boson to fermions.

\section{Conclusions}
\label{sec:concl}

We have performed a thorough analysis of the structure of tree-level
effects in the RS model with enlarged bulk gauge symmetry
$SU(2)_L\times SU(2)_R\times U(1)_X\times P_{LR}$ and an IR
brane-localized Higgs sector. In contrast to the existing literature,
where the Yukawa couplings have always been treated as a perturbation,
we have performed the KK decomposition of the gauge fields in a
covariant $R_\xi$ gauge within the basis of mass eigenstates, by
constructing the exact solutions to the bulk equations of motion
augmented with appropriate boundary conditions. The KK decomposition
in the matter sector has been performed employing the same formalism
and including the mixing of fermionic fields between different
representations and generations in a completely general way. By
expanding the exact results, we have derived simple analytic
expressions for the profiles and masses of the various SM particles
as well as for the sums over KK towers of gauge bosons, which include 
all terms up to second order in the ratio of the Higgs vacuum 
expectation value $v$ over the KK mass scale $\Mkk$.

We have demonstrated that our exact approach is not only more elegant,
but also offers some distinct advantages over treating the couplings
of the bulk fields to the Higgs sector perturbatively. By expanding
the low-energy spectrum as well as the gauge couplings in powers of
$v^2/\Mkk^2$, we have obtained analytic formulas which allow not only
for a numerical treatment, but for a transparent and explicit
understanding of the model-specific protection mechanisms of the
Peskin-Takeuchi parameter $T$ and the left-handed $Z$-boson vertices
involving down-type quarks. In the case of the gauge-boson corrections
to the $Z d_L^i \bar d_L^j$ couplings, we have pointed out all terms
that escape the custodial protection and identified them with the
irreducible sources of $P_{LR}$-symmetry breaking, originating from
the different boundary conditions of untwisted and twisted gauge-boson
profiles on the UV brane. Unlike in the perturbative approach, which
in general requires diagonalizing high-dimensional matrices
numerically, the interpretation of our results in physical terms is
thus very clear.  By making the dependence on the implementation of
the matter sector explicit, we were also able to address the important
question about the model-dependence of the resulting gauge-boson
interactions with SM fermions. We have shown in this context, that the
$P_{LR}$ symmetry is explicitly broken by the bulk mass parameters of
the $Z_2$-odd $SU(2)_L$ singlet fields if their values differ from the
ones of their $Z_2$-even counterparts. Turning our attention to the
charged-current interactions, we have then demonstrated that a
custodial protection in not at work in this case. We have finally
revisited the issue of the flavor-misalignment between fermion
zero-mode masses and Yukawa couplings, extending existing analyses of
the structure of the flavor-changing Higgs-boson couplings to the case
of the RS scenario with custodial protection.

Subsequently we have considered some simple applications of our
general results. A thorough discussion of the constraints imposed by
the precision measurements of the bottom-quark pseudo observables
opened our phenomenological survey. We found that, contrary to the
minimal case, the prediction for the correction to the $Z b_L \bar
b_L$ vertex in the RS model with extended $P_{LR}$ symmetry is
essentially independent of the left-handed bulk mass parameter of the
third-generation quarks. This feature relaxes the bounds that
originate from the precision measurements of the left-handed $Z$-boson
coupling significantly, giving a strong motivation to protect the
latter vertex through a suitable embedding of the bottom quarks. We
have furthermore pointed out that, irrespectively of the bulk gauge
group and barring an unnatural large value of the bulk mass parameter
of the right-handed top quark, the requirement to obtain the correct
top- and bottom-quark masses excludes large corrections to the $Z b_R
\bar b_R$ coupling. A direct explanation of the anomaly in the
forward-backward asymmetry for bottom quarks seems therefore
generically challenging in warped extra-dimension models in which the
left-handed bottom and top quark are part of the same multiplet.
Allowing for a heavy Higgs boson with a mass in the ballpark of 0.5
TeV (which is the naturally expected mass range for $m_h$ in models
with a brane-localized Higgs sector) leads however to a good agreement
between $Z \to b\bar b$ data and theory. Yet, a heavy Higgs boson
would need tuning in models with custodial symmetry, since the shifts
induced by $m_h = 0.5 \, {\rm TeV}$ in the parameters $S$ and $T$
cannot be compensated by RS tree-level effects, and thus would require
the presence of sizable oblique loop corrections in order not to spoil
the global electroweak fit.  Detailed numerical analyses of the
new-physics effects in rare top-quark decays as well as of the changes
in the production cross section and branching fractions of the Higgs
boson completed our phenomenological investigations. In the former
case, we found that due to the protection of the $Z b_L \bar b_L$
vertex, the experimental prospects for observing $t\to c Z$ and $t \to
ch$ are more favorable in the extended than in the minimal RS
scenario. In particular, for KK gauge-boson masses below $5 \, {\rm
  TeV}$ the branching fractions of both $t \to cZ$ and $t \to ch$ can
be within the reach of the LHC.  In the latter case, our study
revealed that due to the composite nature of the Higgs boson, the top
quark, and the KK modes, observable effects in Higgs physics can
naturally occur in the scenario under consideration. In order to
arrive at this conclusion, we have performed the first complete
one-loop calculation of all Higgs-boson production and decay channels
relevant at hadron colliders, incorparating all effects stemming from
the extended electroweak gauge boson and fermion sectors. Concerning
the main Higgs-boson production modes at the Tevatron and the LHC,
proceeding through $gg \to h$, $q \bar q^\prime \to Wh$, and
$qq^{(\prime)} \to qq^{(\prime)} h$, we found that they are all
suppressed in the custodial RS model relative to the SM. Since the
shifts in the production cross sections can exceed the combined
experimental and theoretical uncertainties, the reduction in Higgs
events predicted in the RS framework might be observable at the
LHC. On the other hand, the reduced $gg \to h$ production cross
section should make an observation of the Higgs boson with a mass
above the $ZZ$ threshold via the ``golden'' four-lepton channel more
difficult, because the $h \to ZZ$ branching fraction remains essential
SM-like in the custodial RS model.  The possible enhancement of the
branching ratio for $h \to \gamma \gamma$ might however lead to a
higher statistical significance and a faster LHC discovery of the
Higgs boson, if its mass is below the $WW$ threshold. We emphasize
that our findings concerning Higgs physics have to be considered
robust predictions, since they depend rather weakly on the details of
the spectrum (and thus the specific RS parameter values) once the
contributions of the entire KK towers have been included.

The analytical and numerical results obtained in this article form the
basis for general calculations of flavor-changing processes in the
custodial RS model. A detailed phenomenological analysis of the
potential new-physics effects in neutral-meson mixing and in rare
decays of kaons and $B$ mesons, including both inclusive and exclusive
processes, is left for future work.

\subsubsection*{Acknowledgments}

It is a pleasure to thank A.~Azatov, M.~Toharia, and L.~Zhu for
helpful correspondence concerning flavor-changing Higgs-boson
couplings. We are also grateful to V.~Ahrens, M.~Benzke, and D.~Dolce
for useful discussions. The Feynman diagrams shown in this work are
drawn using {\tt FeynArts} \cite{Hahn:2000kx}. The research of
S.C. is supported by the DFG cluster of excellence ``Origin and
Structure of the Universe''.  The research of F.G., U.H., M.N., and
T.P. is supported in part by the German Federal Ministry for Education
and Research grant 05H09UME (``Precision Calculations for Collider and
Flavour Physics at the LHC''), and by the Research Centre ``Elementary
Forces and Mathematical Foundations'' funded by the Excellence
Initiative of the State of Rhineland-Palatinate. U.H. thanks the
Galilo Galilei Institute for Theoretical Physics for the hospitality
and the INFN for partial support during the final stage of this work.

\begin{appendix}

\section{IR BCs and Higgs-Boson FCNCs}
\label{app:higgsstuff}

\renewcommand{\theequation}{A\arabic{equation}}
\setcounter{equation}{0}

In this appendix we rederive (\ref{eq:bcIRrescaled}) and
(\ref{eq:gtil1}) to (\ref{eq:bmh}), using the rectangular function
\begin{equation} \label{eq:rect}
    \delta^\eta (t - 1) = \begin{cases}
    \, \displaystyle \frac{1}{\eta} \,, & t \in [1 - \eta, 1]
    \,, \\[4mm]
    \, 0 \,, & {\rm otherwise} \,, \end{cases}
\end{equation}
to regularize the $\delta$-functions appearing in the EOMs
(\ref{eq:EOM}).

Keeping only terms relevant in the range $t \in [1
  -\eta, 1]$, the EOMs (\ref{eq:EOM}) close to the IR brane
    take the simpler form
  \begin{align}  \label{eq:EOM_close_to_brane}
    - \partial_t \, \bm{S}_n^Q(t)\, \vec a_n^Q &= \delta^\eta (t-1) \,
    \frac{v}{\sqrt{2} M_{\rm KK}} \, {\bm Y}_{\vec q} \hspace{1mm}
    \bm{C}_n^{\hspace{0.25mm} q}(t) \, \vec a_n^{\hspace{0.25mm} q} \,, \nonumber \\
    \partial_t \, \bm{S}_n^{\hspace{0.25mm} q}(t)\, \vec
    a_n^{\hspace{0.25mm} q} &= \delta^\eta (t-1) \, \frac{v}{\sqrt{2}
      M_{\rm KK}} \, {\bm Y}_{\vec
      q}^\dagger \hspace{1mm} \bm{C}_n^Q(t) \, \vec a_n^Q \,, \nonumber \\
    \partial_t \, \bm{C}_n^Q(t)\, \vec a_n^Q &= \delta^\eta (t-1) \,
    \frac{v}{\sqrt{2} M_{\rm KK}} \, {\bm Y}_{\vec q} \hspace{1mm}
    \bm{S}_n^q(t) \, \vec a_n^{\hspace{0.25mm} q} \,, \nonumber \\
    -\partial_t \, \bm{C}_n^{\hspace{0.25mm} q}(t)\, \vec
    a_n^{\hspace{0.25mm} q} &= \delta^\eta (t-1) \, \frac{v}{\sqrt{2}
      M_{\rm KK}} \, {\bm Y}_{\vec q}^\dagger \hspace{1mm}
    \bm{S}_n^Q(t) \, \vec a_n^Q \,.
  \end{align}
Combining the first (second) with the fourth (third) relation and
using (\ref{eq:rect}), we obtain
\begin{equation} \label{eq:SDGL}
  \begin{split}
    \left[\partial_t^2 - \bigg ( \frac{{\bm X}_{\vec q}}{\eta} \bigg
      )^2 \right] \bm{S}_n^Q(t) & = 0\,, \qquad \left [ \partial_t^2 -
      \bigg ( \frac{\bm{\tilde X}_{\vec q}}{\eta} \bigg )^2 \right
    ]\bm{S}_n^{\hspace{0.25mm} q}(t) = 0 \,,
  \end{split}
\end{equation}
where 
\begin{equation} 
  {\bm X}_{\vec q} \equiv \frac{v}{\sqrt{2} M_{\rm KK}} \,
  \sqrt{{\bm Y}_{\vec q} \, {\bm Y}_{\vec q}^\dagger} \;,\qquad
  \bm{\tilde X}_{\vec q} \equiv \frac{v}{\sqrt{2} M_{\rm KK}} \,
  \sqrt{{\bm Y}_{\vec q}^\dagger\, {\bm Y}_{\vec q}} \,.
\end{equation}
Imposing now the BCs ${\bm S}_n^{Q,q} (1) = 0$ and matching ${\bm
S}_n^{Q,q} (1 - \eta)$ onto the solutions of (\ref{eq:EOM}) evaluated
in the limit $t \to 1^-$, we find that the differential equations
(\ref{eq:SDGL}) are solved by
\begin{equation} \label{eq:Ssol}
  \begin{split}
    {\bm S}_n^{Q} (t) =\frac{\sinh \bigg (\displaystyle \frac{{\bm
          X}_{\vec q}}{\eta} \; (1-t) \bigg )}{\sinh \big ( {\bm
        X}_{\vec q} \big )} \; {\bm S}_n^{Q} (1^-) \,, \qquad {\bm
      S}_n^{\hspace{0.25mm} q} (t) = \frac{\sinh \bigg (\displaystyle
      \frac{\bm{\tilde X}_{\vec q}}{\eta} \; (1-t) \bigg )}{\sinh \big
      ( \bm{\tilde X}_{\vec q} \big )} \; {\bm S}_n^{\hspace{0.25mm}
      q} (1^-) \,.
  \end{split}
\end{equation}
This implies that in the interval $t \in [1-\eta, 1]$ the $Z_2$-even
fermion profiles take the form
\begin{equation} \label{eq:Csol}
  \begin{split}
    {\bm C}_n^{Q} (t) & =\frac{\cosh \bigg (\displaystyle \frac{{\bm
          X}_{\vec q}}{\eta} \; (1-t) \bigg )}{\cosh \big ( {\bm
        X}_{\vec q} \big )} \; {\bm C}_n^{Q} (1^-) \,, \qquad {\bm
      C}_n^{\hspace{0.25mm} q} (t) = \frac{\cosh \bigg (\displaystyle
      \frac{\bm{\tilde X}_{\vec q}}{\eta} \; (1-t) \bigg )}{\cosh \big
      ( \bm{\tilde X}_{\vec q} \big )} \; {\bm C}_n^{\hspace{0.25mm}
      q} (1^-) \,.
  \end{split}
\end{equation}
Reinserting the solutions (\ref{eq:Ssol}) and (\ref{eq:Csol}) into
(\ref{eq:EOM_close_to_brane}), allows us to determine the IR BCs which
relate the $Z_2$-even profiles with the -odd ones at $t = 1^-$. The
resulting expressions read
\begin{equation} \label{eq:IRBCsagain}
  \begin{split} {\bm S}_n^Q (1^-) \, \vec a_n^Q & = \frac{v}{\sqrt{2}
      M_{\rm KK}} \, {\bm Y}_{\vec q} \, \big ( \bm{\tilde X}_{\vec q}
    \big)^{-1} \, \tanh \big ( \bm{\tilde X}_{\vec q} \big ) \, {\bm
      C}_n^{\hspace{0.25mm} q} (1^-) \, \vec a_n^{\hspace{0.25mm} q} \,,
    \\[1mm] -{\bm S}_n^{\hspace{0.25mm} q} (1^-) \, \vec
    a_n^{\hspace{0.25mm} q} & = \frac{v}{\sqrt{2} M_{\rm KK}} \, {\bm
      Y}_{\vec q}^\dagger \, \big (\bm{X}_{\vec q} \big )^{-1} \, \tanh \big
    ( \bm{X}_{\vec q} \big ) \, {\bm C}_n^{Q} (1^-) \, \vec a_n^{Q} \,,
  \end{split}
\end{equation}
which, after introducing the rescaled Yukawa couplings ${\bm{\tilde
Y}}_{\vec q}$, resembles (\ref{eq:bcIRrescaled})

Employing the regularization (\ref{eq:rect}) for the
$\delta$-function, the flavor-changing Higgs-boson couplings
(\ref{eq:Deltagtilde}) become
\begin{equation} \label{eq:Deltagtildeagain} 
  (\Delta \tilde{g}_h^q
  )_{mn} = - \sqrt{2} \, \frac{2\pi}{L\epsilon}\int_{1-\eta}^1 \!dt \;
  \frac{1}{\eta} \; \vec a_m^{\hspace{0.25mm} q\,\dagger}\,
  \bm{S}_m^{\hspace{0.25mm} q} (t) \, \bm{Y}_{\vec q}^{\dagger} \,
  \bm{S}_n^Q(t)\, \vec a_n^Q \,.
\end{equation}
Combining (\ref{eq:Ssol}), (\ref{eq:Csol}), and (\ref{eq:IRBCsagain})
and using
\begin{equation}
  \int_{1-\eta}^1 \! dt \; \frac{1}{\eta} \, \sinh^2 
  \left ( \frac{{\bm A}}{\eta}{\,(1-t)} \right ) = \frac{1}{2} \, 
  \Big (\sinh \big (2 {\bm A} \big ) \big (2 {\bm A})^{-1}
  - {\bm 1} \Big ) \,,
\end{equation}
valid for any arbitrary invertible matrix ${\bm A}$, we then obtain
(\ref{eq:gtil1}) to (\ref{eq:bmh}).

\section{Reference Values for the SM Parameters}
\label{app:masses}

\renewcommand{\theequation}{B\arabic{equation}}
\setcounter{equation}{0}

The central values and errors of the quark masses used in our analysis
are
\begin{equation}\label{eq:fitmasses}
  \begin{aligned}
    m_u &= (1.5\pm 1.0)\,\mbox{MeV} \,, & \qquad m_c &= (520\pm
    40)\,\mbox{MeV} \,, & \qquad
    m_t &= (144\pm 5)\,\mbox{GeV} \,, \\
    m_d &= (3.0\pm 2.0)\,\mbox{MeV} \,, & \qquad m_s &= (50\pm
    15)\,\mbox{MeV} \,, & \qquad m_b &= (2.4\pm 0.1)\,\mbox{GeV} \,.
  \end{aligned}
\end{equation} 
They correspond to $\overline{\rm MS}$ masses evaluated at the scale
$\Mkk=1$\,TeV, obtained by using the low-energy values as compiled in
\cite{Amsler:2008zz}. The central values and errors of the Wolfenstein
parameters are taken from \cite{Charles:2004jd} and read
\begin{equation}\label{eq:fitwolf}
  \lambda = 0.2265\pm 0.0008 \,, \quad 
  A = 0.807\pm 0.018 \,, \quad
  \bar{\rho} = 0.141\,_{-0.017}^{+0.029} \,, \quad 
  \bar{\eta} = 0.343\pm 0.016 \,.
\end{equation}
The central values and errors for the parameters entering our analysis
of the bottom-quark pseudo observables are \cite{LEPEWWG:2005ema,
Group:2008nq}
\begin{equation}
  \begin{aligned}
    \Delta\alpha^{(5)}_{\rm had}(m_Z) &= 0.02758\pm 0.00035 \,, \qquad
    & m_Z &= (91.1875\pm 0.0021)\,\mbox{GeV} \,, \\
    \alpha_s(m_Z) &= 0.118\pm 0.003 \,, \qquad & m_t &= (172.6\pm
    1.4)\,\mbox{GeV} \,.
  \end{aligned}
\end{equation}
We refer to the central values for these quantities as SM reference
values. Unless noted otherwise, the reference value for the
Higgs-boson mass is $m_h=150$\,GeV.

\section{Form Factors for Higgs-Boson Production and Decay}
\label{app:formfactors}

\renewcommand{\theequation}{C\arabic{equation}}
\setcounter{equation}{0}

The form factors $A_{q,W}^h (\tau)$ and $A_{q,W}^h (\tau, \lambda)$
describing the effects of quark and $W$-boson loops in the production
and the decay of the Higgs boson are given by \cite{Djouadi:2005gi}
\begin{equation}
  \begin{split}
    A_{q}^h (\tau) & = \frac{3 \hspace{0.25mm} \tau}{2} \left [
      \hspace{0.25mm} 1 + \left ( 1 - \tau \right ) f (\tau)
      \hspace{0.25mm} \right ] \,, \\
    A_{W}^h (\tau) & = -\frac{3}{4} \left [ \hspace{0.25mm} 2 + 3 \tau
      + 3 \tau \left ( 2 - \tau \right ) f (\tau) \hspace{0.25mm}
    \right ]
    \,, \\[2mm]
    A_{q}^h (\tau, \lambda) & = - I (\tau, \lambda) + J (\tau,
    \lambda) \,, \\[1mm]
    A_{W}^h (\tau, \lambda) & = c_w \hspace{0.5mm} \left \{ 4 \left (
        3 - \frac{s_w^2}{c_w^2} \right ) I (\tau, \lambda) + \left [
        \left ( 1 + \frac{2}{\tau} \right ) \frac{s_w^2}{c_w^2} -
        \left ( 5 + \frac{2}{\tau} \right ) \right ] J (\tau, \lambda)
    \right \} \,.
  \end{split}
\end{equation}
The functions $I(\tau, \lambda)$ and $J(\tau, \lambda)$ take the form
\begin{equation}
  \begin{split}
    I (\tau, \lambda) & = -\frac{\tau \lambda}{2 (\tau - \lambda)} \,
    \big [ f(\tau) - f(\lambda) \big ] \,, \\
    J (\tau, \lambda) & = \frac{\tau \lambda}{2 \left (\tau - \lambda
      \right )} + \frac{\tau^2 \lambda^2}{2 \left (\tau - \lambda
      \right )^2} \, \big [ f(\tau) - f(\lambda) \big ] + \frac{\tau^2
      \lambda}{(\tau - \lambda)^2} \, \big [ g(\tau) - g(\lambda) \big
    ] \,,
  \end{split}
\end{equation}
while the functions $f(\tau)$ and $g(\tau)$ read 
\begin{align}
    f(\tau) & = \begin{cases} - \displaystyle \frac{1}{4} \left [ \,
        \ln \left ( \displaystyle \frac{1 + \sqrt{1 - \tau}}{1 -
            \sqrt{1 - \tau}} \right ) - i \pi \, \right ]^2 \,, & \tau
      \leq 1 \,, \\[4mm] \arcsin^2 \left ( \displaystyle
        \frac{1}{\sqrt{\tau}}
      \right ) \,, & \tau > 1 \,, \end{cases} \\[2mm]
    g(\tau) & = \begin{cases} \sqrt{\tau - 1} \hspace{0.5mm} \arcsin
      \left ( \displaystyle
        \frac{1}{\sqrt{\tau}} \right ) \,, & \tau \leq 1 \,,\\[6mm]
      \displaystyle \frac{1}{2} \, \sqrt{1 - \tau} \, \left [ \, \ln
        \left ( \displaystyle \frac{1 + \sqrt{1 - \tau}}{1 - \sqrt{1 -
              \tau}} \right ) - i \pi \, \right ] \,, & \tau > 1
      \,. \end{cases}
\end{align}

\end{appendix}

\end{document}